\documentclass[12pt,a4paper]{article}
\usepackage[english, USenglish]{babel}
\usepackage{amsmath}
\usepackage{amsfonts}
\usepackage{latexsym}
\usepackage[dvips]{graphicx}
\usepackage{bm}
\usepackage[bookmarksnumbered=true,breaklinks=true]{hyperref}
\usepackage{epsfig}
\usepackage{a4wide}
\catcode`\@=11
\@addtoreset{equation}{section}

\catcode`\@=11

\def\section{\@startsection{section}{1}{\z@}{3.5ex plus 1ex minus
   .2ex}{2.3ex plus .2ex}{\large\bf}}
\def\thesubsection{\arabic{section}.\arabic{subsection}}

\author{\\[-0.3cm]\Large I.~Antoniadis\footnote{{\tt ignatios.antoniadis@cern.ch},\newline 
~~On leave from CPHT (UMR CNRS 7644) Ecole Polytechnique, F-91128 Palaiseau}~, S.~Hohenegger\footnote{{\tt stefan.hohenegger@cern.ch}}~, K.S.~Narain\footnote{{\tt narain@ictp.trieste.it}}
}
\title{\begin{flushright}{\small CERN-PH-TH/2006-212}\end{flushright}
\vspace{1cm}\bf{$N=4$ Topological Amplitudes and String Effective Action}}
\dateUSenglish
\date{\today}
\graphicspath{{figures/}}
\begin{document}
\begin{titlepage}
\maketitle
\begin{center}
\renewcommand{\thefootnote}{\fnsymbol{footnote}}\vspace{-0.5cm}
\footnotemark[1]\footnotemark[2]Department of Physics, CERN -- Theory Division\\ CH-1211 Geneva 23, Switzerland\\[0.5cm]
\vspace{-0.3cm}\footnotemark[3]High Energy Section,\\The Abdus Salam International Center for Theoretical Physics,\\Strada Costiera, 11-34014 Trieste, Italy\\[0.5cm]
\end{center}
\begin{abstract}
Certain scattering amplitudes in the gravitational sector of type II string theory on $K3\times T^2$ are 
found to be computed by correlation functions of the  $N=4$ topological string. 
This analysis extends the already known results for $K3$ by Berkovits and Vafa, 
which correspond to six-dimensional terms in the effective action, involving four 
Riemann tensors and $4g-4$ graviphotons, $R^4T^{4g-4}$, at genus $g$. We find two additional classes of
topological amplitudes that use the full internal SCFT of $K3 \times T^2$.
One of these string amplitudes is mapped to a 1-loop contribution in the heterotic theory, 
and is studied explicitly. It corresponds to the four-dimensional term 
$R^2(dd\Phi)^2T^{2g-4}$, with $\Phi$ a Kaluza-Klein graviscalar from $T^2$.
Finally, the generalization of the harmonicity relation for its moduli dependent
coupling coefficient is obtained and shown to contain an anomaly, generalizing
the holomorphic anomaly of the $N=2$ topological partition function $F_g$.\\[5pt]
\end{abstract}
\thispagestyle{empty}
\end{titlepage}

\tableofcontents
\pagebreak
\section{Introduction}
It is now known that certain physical superstring amplitudes corresponding to BPS-type couplings can be 
computed within a topological theory, obtained by twisting the underlying $N=2$ superconformal field theory 
(SCFT) that describes the internal compactification space~
\cite{Witten:1992fb, Antoniadis:1993ze, Bershadsky:1993cx}. The better studied example is the topological 
Calabi-Yau (CY) $\sigma$-model describing $N=2$ supersymmetric compactifications of type II string in four 
dimensions. Its partition function computes a series of higher derivative F-terms of the form $W^{2g}$, 
where $W$ is the $N=2$ Weyl superfield; its lowest component is the self-dual graviphoton field strength 
$T_+$ while its next upper bosonic component is the self-dual Riemann tensor, so that $W^{2g}$ generates in 
particular the amplitude $R^2T^{2g-2}$, receiving contributions only from the $g$-loop order~
\cite{Antoniadis:1993ze}. The corresponding coupling coefficients $F_g$ are functions of the moduli that 
belong to $N=2$ vector multiplets. Although na\"ively these functions should be analytic, there is a 
holomorphic anomaly expressed by a differential equation providing a recursion relation for the 
non-analytic part~\cite{Cecotti:1992qh, Bershadsky:1993ta, Bershadsky:1993cx}.

The topological partition function has several physical applications. It provides the corrections to the 
entropy formula of $N=2$ black holes~\cite{deWit:1996ix,LopesCardoso:1998wt,LopesCardoso:1999cv, Ooguri:2004zv}. 
Moreover, by applying the $N=1$ involution that introduces boundaries along the $a$-cycles, one obtains a 
similar series of higher dimensional $N=1$ F-terms involving powers of the gauge superfield ${\cal W}$~
\cite{Antoniadis:1996qg, Beasley:2005iu, Antoniadis:2005sd}. In particular, $({\rm Tr}{\cal W}^2)^2$ 
gives rise to gaugino masses upon turning on expectation values for the D-auxiliary component, generated 
for instance by internal magnetic fields, or equivalently by D-branes at angles~\cite{Antoniadis:2005sd, 
Antoniadis:2004qn}.

The $N=2$ $W^{2g}$ F-terms appear also in the heterotic string compactified in four dimensions on 
$K3\times T^2$. Under string duality, all $F_g$'s are generated already at one-loop order and can be 
explicitly computed in the appropriate perturbative limit~\cite{Antoniadis:1995zn,Marino:1998pg}. 
This allows in particular the study of several properties, such as their exact behavior around the 
conifold singularity (\cite{Antoniadis:1995zn,Strominger:1995cz,Ghoshal:1995wm,Vafa:1995ta,
Gopakumar:1998ii,Gopakumar:1998jq}).

In this work, we perform a systematic study of $N=4$ topological amplitudes, associated to type II string 
compactifications on $K3\times T^2$, or heterotic compactifications on $T^6$. The $N=4$ topological 
$\sigma$-model on $K3$, describing the six-dimensional (6d) type II compactifications was studied in 
Refs.~\cite{Berkovits:1994vy, Berkovits:1998ex}. Although in principle simpler than the $N=2$ case, 
in practice the analysis presents a serious complication due to the trivial vanishing of the corresponding 
partition function.

The topological twist consists of redefining the energy momentum tensor $T_B$ of the internal SCFT by 
$T_B-\frac{1}{2}\partial J$, with $J$ the $U(1)$ current of the $N=2$ subalgebra of $N=4$. The conformal 
weights $h$ are then shifted by the $U(1)$ charges $q$ according to $h\to h-q/2$. As a result, the $N=2$ 
supercurrents $G^+$ and $G^-$, of $U(1)$ charge $q=+1$ and $q=-1$, acquire conformal dimensions one and two, respectively; $G^+$ becomes the 
BRST operator of the topological theory, while $G^-$ plays the role of the reparametrization anti-ghost 
$b$ used to be folded with the Beltrami differentials (when combined from left and right-moving sectors) 
and define the integration measure over the moduli space of the Riemann surface. Thus, the genus $g$ 
partition function of the topological theory involves $(3g-3)$ $G^-$ insertions. On the other hand, 
there is a $U(1)$ background charge given by ${\hat c}(g-1)$, with $\hat c$ the central charge of the 
$N=2$ SCFT before the topological twist. In the $N=2$ case, the ``critical" (complex) dimension is 
${\hat c}=3$, corresponding to compactifications on a Calabi-Yau threefold, and the total charge of 
the $(3g-3)$ $G^-$ insertions matches precisely with the required $U(1)$ charge to give a non-vanishing 
result. In the $N=4$ case however, the critical dimension is ${\hat c}=2$, corresponding to $K3$ 
compactifications, and the $U(1)$ charge is not balanced. In Ref.~\cite{Berkovits:1994vy}, it was then 
proposed to consider correlation functions by adding $(g-1)$ (integrated) ${\tilde G}^+$ insertions, 
where ${\tilde G}^\pm$ are the other two supercurrents of the $N=4$ SCFT.\footnote{Actually, an additional 
insertion of the (integrated) $U(1)$ current $J$ is required in order to obtain a non-vanishing result.}

It was also shown \cite{Berkovits:1994vy}, using the Green-Schwarz (GS) formalism, that the above 
topological correlation functions 
compute a particular class of higher derivative interactions in the 6d effective field theory of the form 
$W_{++}^{4g}$, where $W_{++}$ is an $N=4$ chiral superfield ($N=2$ in six dimensions) containing a 
graviphoton field strength as lower component and the Riemann tensor as the next bosonic upper component. 
Moreover, an effective 6d self-duality is defined by projecting antisymmetric vector indices, say 
$A_{\mu\nu}$, using the Lorentz generators in the reduced (chiral) spinor representation 
${(\sigma^{\mu\nu})_{a}}^b$, with $a,b=1,\dots 4$. As a result, $W_{++}^{4g}$ generates in particular 
an effective action term of the form $R^4T^{4g-4}$, receiving contributions only from $g$-loop order. 
These terms can be covariantized in terms of the full $SU(2)$ R-symmetry of the extended supersymmetry 
algebra and the corresponding coefficients were shown to satisfy a harmonicity relation, which is a 
generalization of the holomorphicity to the harmonic superspace and follows as a consequence of the 
BPS character of the effective action term integrated over half of the full $N=2$ 6d superspace.

It turns out that upon string duality, these amplitudes are 
mapped on the heterotic side to contributions starting from genus $g+1$, and thus they cannot be studied 
similar to the $N=2$ 4d couplings $F_g$ at one loop.
In this work, we propose and study alternative definitions of the $N=4$ topological amplitudes and their 
string theory interpretations, that avoid multiple point insertions and have tractable heterotic 
representations.

Our starting point is to consider type II compactifications in four dimensions on $K3\times T^2$, 
associated to a SCFT which in addition to the $N=4$ part of $K3$ has an $N=2$ piece describing $T^2$. 
Since the central charge is now ${\hat c}=3$, the $U(1)$ charge of the $(3g-3)$ $G^-$ insertions, with
$G^-$ being the sum of $K3$ and $T^2$ parts, defining 
the genus $g$ topological partition function, balances the corresponding background charge deficit, 
avoiding the trivial vanishing of the critical case. However, the partition function still vanishes due 
to the $N=4$ extended supersymmetry of $K3$. In order to obtain a non-vanishing result, one needs to add 
just two insertions. One possibility is to insert the (integrated) $U(1)$ current of $K3$, $J_{K3}$, and 
the (integrated) $U(1)$ current of $T^2$, $J_{T^2}$: 
\begin{equation}
\int_{\mathcal{M}_g}\langle |\prod_{a=1}^{3g-3} G^-(\mu_a) \int J_{K3} \int J_{T^2}|^2\rangle_{K3\times T^2} 
\end{equation}
where $\mu_a$ are the Beltrami differentials. Another possible definition of a non-trivial topological 
quantity is  
\begin{equation}
\int_{\mathcal{M}_g}\langle |\prod_{a=1}^{3g-4} G^-(\mu_a) J^{--}_{K3}(\mu_{3g-3})~ \psi(z)|^2
\rangle_{K3\times T^2}
\end{equation} 
where $J^{--}$ is part of the $SU(2)$ currents of the $N=4$ superconformal algebra $(J^{--},J,J^{++})$, $\psi$ is the fermionic partner of the (complex) $T^2$ coordinate $(J_{T^2}=\psi\bar\psi)$, 
whose dimension is zero and charge +1
in the twisted theory, and $z$ is an arbitrary point on the genus $g$ Riemann surface. Furthermore  $g \geq 2$ and a complete
antisymmetrization of the Beltrami differentials is understood. It turns out that both 
definitions of the topological partition function correspond to actual physical string amplitudes in four dimensions. The former corresponds in particular to $R_+^2R_-^2T_+^{2g-2}$, with the subscripts $(-)$ $+$ denoting the (anti) self-dual part of the corresponding field strength\footnote{The $\pm$ subscript will be mostly dropped in the following, for notational simplicity.}, while the latter corresponds to the term $R_+^2(dd\Phi_+)^2T_+^{2g-4}$, where $\Phi_+$ is the complex graviscalar from the 6d `self-dual' graviton with positive charge under the $N=2$ $U(1)$ current $J_{T^2}$; it corresponds to the K\"ahler class modulus of $T^2$. Moreover, on the heterotic side, the first starts appearing at two loops for every $g$, while the second at one loop and can therefore be studied in a way similar to the $N=2$ couplings $F_g$'s.

We also obtain the generalization of the harmonicity relation for the full $SU(4)$ R-symmetry using known results of 4d $N=4$ supersymmetry. The moduli in this case belong to vector multiplets and transform in the two-index antisymmetric representation of $SU(4)$. Although we cannot check this relation on the type II side because it involves RR (Ramond-Ramond) field backgrounds, we can also derive it in the heterotic theory at the string level, where the whole $SO(6)$ is manifest, for the minimal series that appears already at one loop. The resulting equation reveals the presence of an anomalous term, in analogy with the holomorphic anomaly of the $N=2$ $F_g$'s.

The structure of this paper is the following. We start in Section \ref{SectN4Top} by a brief review of 
the $N=4$ Superconformal Algebra and its topological twist. We then proceed with
our strategy in finding topological amplitudes, depicted in Figure \ref{fig:overview}:
\begin{figure}[htb]
\begin{center}
\epsfig{file=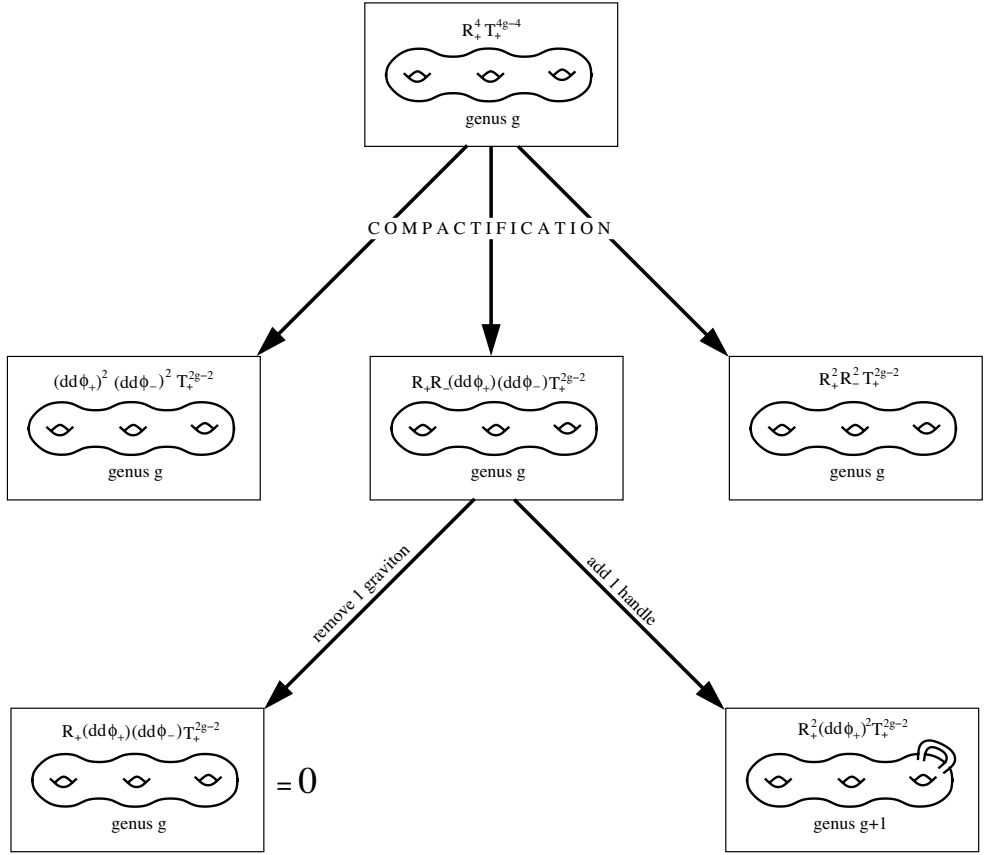, width=13cm}
\caption{Overview over the topological $N=4$ amplitudes. $R_+$ ($R_-$) denotes the (anti-) self-dual part of the Riemann tensor, $T_+$ the self-dual part of the graviphoton field strength and $\Phi_+$ ($\Phi_-$) stands for a graviscalar with one positive (negative) unit of $U(1)$-charge with respect to the current $J_{T^2}$.}
\label{fig:overview}
\end{center}
\end{figure} 
From the known $R^4T^{4g-4}$ coupling of 4 gravitons and $(4g-4)$ graviphotons on $K3$, 
we redo the computation (which was initially performed in \cite{Berkovits:1994vy}) in Section 
\ref{SectTopAmpK3} using the RNS formalism. In Section \ref{Section4dscattering}, we 
compactify two additional dimensions on a 2-torus and study various couplings of $(2g-2)$ 
graviphotons and a number of gravitons and Kaluza-Klein (KK) graviscalars $\Phi$ from the four 6d Riemann tensors. In Section \ref{ChangeAmpl}, we consider further modifications by looking at an odd number of vertex insertions and modifying the genus of the corresponding world-sheet. In Section \ref{sect:dual}, we translate the type IIA scattering amplitudes to their duals in the heterotic theory compactified on $T^4$ and $T^6$, respectively, and in Section \ref{sect:hetT6} we explicitly compute one of them as a 1-loop amplitude. In Section \ref{harmoIIK3}, we give a brief review of the holomorphicity of $N=2$ $F_g$'s and we present a qualitative derivation of the harmonicity relation for 
the topological amplitudes on $K3$. This allows us in Section \ref{sect:harmrelIIA} to generalize it in 
four dimensions for the new topological amplitudes on $K3\times T^2$, by introducing appropriate ``harmonic"-like variables that restore the $SU(4)$ symmetry. Indeed, in Section \ref{sect:harmHet}, we derive this relation on the heterotic side on $T^6$ and find also an anomaly arising from boundary terms. Finally, in Section \ref{sect:conclusion}, we present our conclusions. 
Some basic material and part of our notation is presented in a number of appendices.
Appendix \ref{AppendixGamma} contains a representation of $\gamma$-matrices and Lorentz generators, while in Appendix \ref{AppendixSelfDual} we define the notion of self-duality in six dimensions. Appendix \ref{AppendixRiemanntot} contains the definition of $\vartheta$-functions, prime forms and the Riemann vanishing theorem. In Appendix \ref{AppendixSuperconformal}, we review the $N=4$ superconformal algebra for $K3$ orbifolds and $K3\times T^2$. Finally, in Appendix \ref{AppendixLeftRightAsym}, we present some new amplitudes with insertions of Neveu-Schwarz (NS) graviphotons. 

\section{$N=4$ Superconformal Algebra and its Topological Twist}\label{SectN4Top}
A detailed review can be found in Appendix \ref{AppendixSuperconformal}. Here, we summarize the main expressions and properties of its generators, the algebra and the topological twist, that will be used throughout this paper. The $N=4$ algebra associated to type II compactifications on $K3$ involves besides the energy momentum tensor $T^B_{K3}$, two complex spin-3/2 super currents $G^+_{K3}$, ${\tilde G}^+_{K3}$ (and their complex conjugates $G^-_{K3}$, ${\tilde G}^-_{K3}$), and three spin-1 currents forming an $SU(2)$ R-symmetry algebra $J^{++}_{K3}$, its complex conjugate $J^{--}_{K3}$ and $J_{K3}$. In this notation, the superscripts $+/-$ count the units of charge with respect to the $U(1)$ generator $J_{K3}$, while the supercurrents form two $SU(2)$ doublets $(G^+_{K3},{\tilde G}^-_{K3})$ and $({\tilde G}^+_{K3},G^-_{K3})$. 
It follows that the non-trivial operator product expansions (OPE's) among them are:
\begin{align}
&J_{K3}(z)J^{\pm\pm}_{K3}(0)\sim\pm 2\frac{J^{\pm\pm}_{K3}(0)}{z}, 
&&J^{--}_{K3}(z)J^{++}_{K3}(0)\sim\frac{J_{K3}(0)}{z},\nonumber\\
&J_{K3}(z)G_{K3}^\pm(0)\sim\pm \frac{G_{K3}^\pm(0)}{z}, 
&&J_{K3}(z){\tilde G}_{K3}^\pm(0)\sim\pm \frac{{\tilde G}_{K3}^\pm(0)}{z},\nonumber\\
&J^{--}_{K3}(z)G_{K3}^+(0)\sim\frac{\tilde{G}^-_{K3}(0)}{z}, &&J_{K3}^{++}(z)\tilde{G}_{K3}^-(0)
\sim-\frac{G^+_{K3}(0)}{z},\label{N4K3}\\ 
&J_{K3}^{++}(z)G^-_{K3}(0)\sim \frac{\tilde{G}^+_{K3}(0)}{z}, &&J_{K3}^{--}(z)
\tilde{G}^+_{K3}(0)\sim-\frac{G^-_{K3}(0)}{z},\nonumber\\
&\tilde{G}^+_{K3}(z)G^+_{K3}(0)\sim 2\frac{J^{++}_{K3}(0)}{z^2}+\frac{\partial J^{++}_{K3}(0)}{z},
&&\tilde{G}^-_{K3}(z)G^-_{K3}(0)\sim 2\frac{J^{--}_{K3}(0)}{z^2}+\frac{\partial J^{--}_{K3}(0)}{z},
\nonumber
\end{align}
\begin{align}
&G^-_{K3}(z)G^+_{K3}(0)\sim\tilde{G}^-_{K3}(z)\tilde{G}^+_{K3}(0)\sim
-\frac{J_{K3}(0)}{z^2}+\frac{T^B_{K3}(0)-{1\over 2}\partial J_{K3}(0)}{z}.\nonumber
\end{align}

In the case of $K3\times T^2$, the R-symmetry group of the $N=4$ algebra is extended to $SU(2)\times U(1)$ and 
the above operators are supplemented by those referring to the torus: $G^\pm_{T^2}$, $\tilde{G}^\pm_{T^2}$ and 
$J_{T^2}$ satisfying the non-trivial OPE's:
\begin{align}
&J_{T^2}(z)G_{T^2}^\pm(0)\sim\pm \frac{G_{T^2}^\pm(0)}{z}, 
&&J_{T^2}(z){\tilde G}_{T^2}^\pm(0)\sim\pm \frac{{\tilde G}_{T^2}^\pm(0)}{z},
\end{align}
\begin{align}
&G^-_{T^2}(z)G^+_{T^2}(0)\sim\tilde{G}^-_{T^2}(z)\tilde{G}^+_{T^2}(0)\sim
-\frac{J_{T^2}(0)}{z^2}+\frac{T^B_{T^2}(0)-{1\over 2}\partial J_{T^2}(0)}{z}.\label{N4T2}
\end{align}
Note that $G^+=G^+_{K3}+G^+_{T^2}$ and $J=J_{K3}+J_{T^2}$ are the supercurrent and the $U(1)$ current of the 
$N=2$ subalgebra, while $T_F=G^++G^-$ is the internal part of the $N=1$ world-sheet supercurrent.

In this work for simplicity, we will consider only $K3$ orbifolds, but the final results will be true for 
generic $K3$.  In the orbifold case, the $N=4$ generators are given in terms of 
free fields. Starting with the ten real bosonic and fermionic space-time coordinates, $Z^M$ and $\chi^M$ 
respectively, with $M=0,1,\dots 9$, we introduce a (Euclidean) complex basis:\footnote{In order to avoid 
confusion with the indices, we adopt the following convention:
\begin{align}
&\text{real space-time indices:}\ \mu,\nu=0,\ldots,5,\nonumber\\
&\text{complex space-time indices:}\ A,B=1,\bar{1},2,\bar{2},3,\bar{3},\nonumber
\end{align}
where a bar means complex conjugation.}
\begin{align}
&X^1=\frac{1}{\sqrt{2}}(Z^0-iZ^1), &&\psi^1=\frac{1}{\sqrt{2}}(\chi^0-i\chi^1),\label{compbas1}\\
&X^2=\frac{1}{\sqrt{2}}(Z^2-iZ^3), &&\psi^2=\frac{1}{\sqrt{2}}(\chi^2-i\chi^3),\label{compbas2}\\
&X^3=\frac{1}{\sqrt{2}}(Z^4-iZ^5), &&\psi^3=\frac{1}{\sqrt{2}}(\chi^4-i\chi^5).\label{compbas3}\\
&X^4=\frac{1}{\sqrt{2}}(Z^6-iZ^7), &&\psi^4=\frac{1}{\sqrt{2}}(\chi^6-i\chi^7),\label{compbas4}\\
&X^5=\frac{1}{\sqrt{2}}(Z^8-iZ^9), &&\psi^5=\frac{1}{\sqrt{2}}(\chi^8-i\chi^9).\label{compbas5}
\end{align}
In our conventions, $X^{1,2}$, $X^3$ and $X^{4,5}$ correspond to the (complex) coordinates of the non-compact 
4d space-time, the 2-torus $T^2$, and $K3$ orbifold, respectively. In this basis, the expressions of the $N=4$ 
generators are:
\begin{align}
&G^+_{K3}=\psi_4\partial \bar{X}_4+\psi_5\partial\bar{X}_5, &&\tilde{G}^+_{K3}=-\psi_5\partial X_4+\psi_4
\partial X_5,\nonumber\\
&G^{-}_{K3}=\bar{\psi}_4\partial X_4+\bar{\psi}_5\partial X_5, &&\tilde{G}^-_{K3}=-\bar{\psi}_5\partial 
\bar{X}_4+\bar{\psi}_4\partial \bar{X}_5,\nonumber
\end{align}
\begin{align}
&J_{K3}=\psi_4\bar{\psi}_4+\psi_5\bar{\psi_5}, &&J^{++}_{K3}=\psi_4\psi_5, &&J^{--}_{K3}=\bar{\psi}_4
\bar{\psi}_5.\label{N4K3expr}
\end{align}
\begin{align}
&G^+_{T^2}=\psi_3\partial \bar{X}_3, &&\tilde{G}^+_{T^2}=\psi_3\partial X_3,\nonumber\\
&G^{-}_{T^2}=\bar{\psi}_3\partial X_3, &&\tilde{G}^-_{T^2}=\bar{\psi}_3\partial \bar{X}_3,\nonumber
\end{align}
\begin{align}
J_{T^2}=\psi_3\bar{\psi}_3.\label{N4T2expr}
\end{align}

The topological twist is defined as usual by shifting the energy momentum tensor $T_B$ with (half of) the 
derivative of the $U(1)$ current of the $N=2$ subalgebra of the full $N=4$ SCFT:
\begin{align}
T_B\to T_B-\frac{1}{2}\partial J.
\end{align}
As a result, the conformal dimensions $h$ of the operators are shifted by (half of) their $U(1)$ charges, 
$h\to h-q/2$. Thus, the new conformal weights are:
\begin{align}
&\text{dim}[G_{T^2,K3}^+]=\text{dim}[\tilde{G}_{T^2,K3}^+]=1, &&\text{dim}[G_{T^2,K3}^-]=\text{dim}
[\tilde{G}_{T^2,K3}^-]=2,\nonumber\\
&\text{dim}[J^{++}]=0, &&\text{dim}[J^{--}]=2,\label{twistweights}\\
&\text{dim}[\psi_A]=0, &&\text{dim}[\bar{\psi}_A]=1.\nonumber
\end{align}
Moreover, the last of the OPE relations (\ref{N4K3}) and (\ref{N4T2}) is changed to:
\begin{align}
&G^-(z)G^+(0)\sim\tilde{G}^-(z)\tilde{G}^+(0)\sim
-\frac{J(0)}{z^2}+\frac{T_B(0)}{z}.\label{N4top}
\end{align}
The physical Hilbert space of the topological theory is defined by the cohomology of the operators $G^+$ and 
${\tilde G}^+$, while $G^-$ (or ${\tilde G}^-$) can be used to define the integration measure over the Riemann 
surfaces. Thus, the physical states are all primary chiral states of the $N=2$ SCFT satisfying $h=q/2$, while 
antichiral fields are BRST exact and should decouple from physical amplitudes (in the absence of anomalies). 
Indeed, they can be written as contour integrals of $G^+$ and upon contour deformation and the OPE relation 
(\ref{N4top}), one gets insertions of the energy momentum tensor $T_B$ leading to total derivatives (see e.g. \cite{Bershadsky:1993cx}).

\section{Review of $N=4$ Topological Amplitudes in 6 Dimensions}\label{SectTopAmpK3}
In \cite{Berkovits:1994vy} g-loop scattering amplitudes of type II string theory compactified on 
$K3$ involving four gravitons and $(4g-4)$ graviphotons were found to be topological by 
using the Green Schwarz formalism. The goal of this section is to repeat this computation 
using the RNS formalism for orbifold compactifications in 6 dimensions.

\subsection{The Supergravity Setup}
The superfield used to write down the 6-dimensional couplings is the following subsector of the 6d, $N=2$ 
Weyl superfield
\begin{align}
({W_a}^b)^{ij}={(\sigma^{\mu\nu})_a}^bT^{ij}_{\mu\nu}+{(\sigma^{\mu\nu})_a}^b(\theta_L^i\sigma^{\rho\tau}
\theta_R^j)R_{\mu\nu\rho\tau}+\ldots,
\label{Wdef}
\end{align}
with $T$ the graviphoton field strength and $R_{\mu\nu\rho\tau}$ the Riemann tensor in 6 
dimensions. The space-time indices run over $\mu,\nu,\rho,\tau=0,1,2,3,4,5$ and the spinor 
indices are Weyl of the same 6-dimensional chirality and take values $a,b=1,2,3,4$. For the 
$SU(2)_L\times SU(2)_R$ R-symmetry indices $i,j$, we will in most applications choose one special component 
(say $i=j=1$ as in \cite{Berkovits:1994vy}) and hence neglect them in the following for notational simplicity. 
Finally, the matrices ${(\sigma^{\mu\nu})_a}^b$ are in the reduced spinor representation of the the 6-dimensional 
Lorentz group (for a precise definition and useful identities see Appendix \ref{AppendixGamma}).

This multiplet has been formally introduced in \cite{Berkovits:1994vy} to reproduce the topological amplitudes, 
and its full expression must contain additional fields of the 6d $N=2$ gravity multiplet, such as the two-form 
$B_{\mu\nu}$, which are not relevant for our analysis. Moreover in \cite{Berkovits:1994vy}, the superspace 
Lagrangian
\begin{align}
\int d^4\theta^1_L\int d^4\theta^1_R {\cal F}_g^{(6d)}({W_{a_1}}^{b_1}{W_{a_2}}^{b_2}{W_{a_3}}^{b_3}{W_{a_4}}^{b_4}
\epsilon^{a_1a_2a_3a_4}\epsilon_{b_1b_2b_3b_4})^{g},\label{SuperExp}
\end{align}
was considered,
with $\theta^1_{L/R}$ a special choice for the $SU(2)_{L/R}$ indices, and was shown using the GS formalism that 
correlation functions computed from these couplings are 
topological. Note that the integral is extended only over half the superspace and thus 
corresponds to an $N=4$ F-type (BPS) term.\footnote{In this paper, we count supersymmetries in 4 dimensions, 
unless otherwise stated.}

In order to repeat the computation of \cite{Berkovits:1994vy} in the RNS formalism the precise 
knowledge of the helicities and kinematics of the involved fields is essential. As far as 
the gravitons are concerned they can easily be extracted by considering the case $g=1$ 
and performing the superspace integrations (we take the lowest
component of ${\cal F}_g^{(6d)}$ which is just a function of moduli scalars)
\begin{align}
\int d^4\theta^1_L&\int d^4\theta^1_R {W_{a_1}}^{b_1}{W_{a_2}}^{b_2}{W_{a_3}}^{b_3}{W_{a_4}}^{b_4}
\epsilon^{a_1a_2a_3a_4}\epsilon_{b_1b_2b_3b_4}=\nonumber\\
=\ &{(\sigma^{\mu_1\nu_1})_{a_1}}^{b_1}{(\sigma^{\mu_2\nu_2})_{a_2}}^{b_2}
{(\sigma^{\mu_3\nu_3})_{a_3}}^{b_3}{(\sigma^{\mu_4\nu_4})_{a_4}}^{b_4}\epsilon^{a_1a_2a_3a_4}
\epsilon_{b_1b_2b_3b_4}\epsilon_{c_1c_2c_3c_4}\epsilon^{d_1d_2d_3d_4}\cdot\nonumber\\
&{(\sigma^{\rho_1\tau_1})_{d_1}}^{c_1}{(\sigma^{\rho_2\tau_2})_{d_2}}^{c_2}
{(\sigma^{\rho_3\tau_3})_{d_3}}^{c_3}{(\sigma^{\rho_4\tau_4})_{d_4}}^{c_4}\cdot 
R_{\mu_1\nu_1\rho_1\tau_1}R_{\mu_2\nu_2\rho_2\tau_2}R_{\mu_3\nu_3\rho_3\tau_3}
R_{\mu_4\nu_4\rho_4\tau_4}.
\end{align}
The antisymmetrization identity
\begin{align}
&\epsilon_{c_1c_2c_3c_4}\epsilon^{d_1d_2d_3d_4}=\delta^{[d_1}_{[c_1}\delta^{d_2}_{c_2}
\delta^{d_3}_{c_3}\delta^{d_4]}_{c_4]},
\end{align}
finally leads to contractions of $\sigma$-matrices of the form
\begin{align}
=24\big[&\text{tr}(\sigma^{\rho_1\tau_1}\sigma^{\rho_2\tau_2})\ \text{tr}
(\sigma^{\rho_3\tau_3}\sigma^{\rho_4\tau_4})-\text{tr}(\sigma^{\rho_1\tau_1}
\sigma^{\rho_2\tau_2}\sigma^{\rho_3\tau_3}\sigma^{\rho_4\tau_4})-\text{tr}
(\sigma^{\rho_4\tau_4}\sigma^{\rho_3\tau_3}\sigma^{\rho_2\tau_2}\sigma^{\rho_1\tau_1})+\nonumber\\
+&\text{tr}(\sigma^{\rho_1\tau_1}\sigma^{\rho_3\tau_3})\ \text{tr}(\sigma^{\rho_2\tau_2}
\sigma^{\rho_4\tau_4})-\text{tr}(\sigma^{\rho_1\tau_1}\sigma^{\rho_3\tau_3}\sigma^{\rho_2\tau_2}
\sigma^{\rho_4\tau_4})-\text{tr}(\sigma^{\rho_4\tau_4}\sigma^{\rho_2\tau_2}\sigma^{\rho_3\tau_3}
\sigma^{\rho_1\tau_1})+\nonumber\\
+&\text{tr}(\sigma^{\rho_1\tau_1}\sigma^{\rho_4\tau_4})\ \text{tr}(\sigma^{\rho_3\tau_3}
\sigma^{\rho_2\tau_2})-\text{tr}(\sigma^{\rho_1\tau_1}\sigma^{\rho_4\tau_4}
\sigma^{\rho_3\tau_3}\sigma^{\rho_2\tau_2})-\text{tr}(\sigma^{\rho_2\tau_2}
\sigma^{\rho_3\tau_3}\sigma^{\rho_4\tau_4}\sigma^{\rho_1\tau_1})\big]\cdot\nonumber\\
\cdot& {(\sigma^{\mu_1\nu_1})_{a_1}}^{b_1}{(\sigma^{\mu_2\nu_2})_{a_2}}^{b_2}
{(\sigma^{\mu_3\nu_3})_{a_3}}^{b_3}{(\sigma^{\mu_4\nu_4})_{a_4}}^{b_4}\ 
\epsilon_{b_1b_2b_3b_4}\epsilon^{a_1a_2a_3a_4}\cdot\nonumber\\
\cdot&  R_{\mu_1\nu_1\rho_1\tau_1}R_{\mu_2\nu_2\rho_2\tau_2}R_{\mu_3\nu_3\rho_3\tau_3}
R_{\mu_4\nu_4\rho_4\tau_4}.
\end{align}
Using the relations (\ref{sigsig2}) and  (\ref{sigsig4}) of Appendix \ref{AppendixGamma}, this may also be 
written in terms of the full Lorentz generators $\Sigma$, as:
\begin{align}
=24\big[&4\ \text{tr}(\Sigma^{\rho_1\tau_1}\Sigma^{\rho_2\tau_2})\ \text{tr}
(\Sigma^{\rho_3\tau_3}\Sigma^{\rho_4\tau_4})-\text{tr}(\Sigma^{\rho_1\tau_1}
\Sigma^{\rho_2\tau_2}\Sigma^{\rho_3\tau_3}\Sigma^{\rho_4\tau_4})+\nonumber\\
+&4\ \text{tr}(\Sigma^{\rho_1\tau_1}\Sigma^{\rho_3\tau_3})\ \text{tr}(\Sigma^{\rho_2\tau_2}
\Sigma^{\rho_4\tau_4})-\text{tr}(\Sigma^{\rho_1\tau_1}\Sigma^{\rho_3\tau_3}\Sigma^{\rho_2\tau_2}
\Sigma^{\rho_4\tau_4})+\nonumber\\
+&4\ \text{tr}(\Sigma^{\rho_1\tau_1}\Sigma^{\rho_4\tau_4})\ \text{tr}(\Sigma^{\rho_3\tau_3}
\Sigma^{\rho_2\tau_2})-\text{tr}(\Sigma^{\rho_1\tau_1}\Sigma^{\rho_4\tau_4}\Sigma^{\rho_3\tau_3}
\Sigma^{\rho_2\tau_2})\big]\cdot\nonumber\\
\cdot &{(\sigma^{\mu_1\nu_1})_{a_1}}^{b_1}{(\sigma^{\mu_2\nu_2})_{a_2}}^{b_2}
{(\sigma^{\mu_3\nu_3})_{a_3}}^{b_3}{(\sigma^{\mu_4\nu_4})_{a_4}}^{b_4}\ 
\epsilon_{b_1b_2b_3b_4}\epsilon^{a_1a_2a_3a_4}\cdot\nonumber\\
\cdot &  R_{\mu_1\nu_1\rho_1\tau_1}R_{\mu_2\nu_2\rho_2\tau_2}R_{\mu_3\nu_3\rho_3\tau_3}
R_{\mu_4\nu_4\rho_4\tau_4}.\label{SuperTrace}
\end{align}
Defining the trace over the self-dual\footnote{For a more detailed view on the definition
of generalized self duality in 6 dimensions, see Appendix \ref{AppendixSelfDual}.} part of the 
Riemann tensor in 6 dimensions as (for an analog see for example \cite{Foerger:1998kw})
\begin{align}
\text{tr}R_+^4=&\ R_{\mu_1\nu_1\rho_1\tau_1}R_{\mu_2\nu_2\rho_2\tau_2}R_{\mu_3\nu_3\rho_3\tau_3}
R_{\mu_4\nu_4\rho_4\tau_4}\text{tr}\left(\Sigma^{\mu_1\nu_1}\Sigma^{\mu_2\nu_2}
\Sigma^{\mu_3\nu_3}\Sigma^{\mu_4\nu_4}\right)\cdot\nonumber\\
&\cdot{(\sigma^{\rho_1\tau_1})_{a_1}}^{b_1}{(\sigma^{\rho_2\tau_2})_{a_2}}^{b_2}
{(\sigma^{\rho_3\tau_3})_{a_3}}^{b_3}{(\sigma^{\rho_4\tau_4})_{a_4}}^{b_4}
\epsilon^{a_1a_2a_3a_4}\epsilon_{b_1b_2b_3b_4},\label{INV1}\\
(\text{tr}R_+^2)^2=&\ R_{\mu_1\nu_1\rho_1\tau_1}R_{\mu_2\nu_2\rho_2\tau_2}
R_{\mu_3\nu_3\rho_3\tau_3}R_{\mu_4\nu_4\rho_4\tau_4}\text{tr} 
\left(\Sigma^{\mu_1\nu_1}\Sigma^{\mu_2\nu_2}\right)\text{tr}
\left(\Sigma^{\mu_3\nu_3}\Sigma^{\mu_4\nu_4}\right)\cdot\nonumber\\
&\cdot{(\sigma^{\rho_1\tau_1})_{a_1}}^{b_1}{(\sigma^{\rho_2\tau_2})_{a_2}}^{b_2}
{(\sigma^{\rho_3\tau_3})_{a_3}}^{b_3}{(\sigma^{\rho_4\tau_4})_{a_4}}^{b_4}
\epsilon^{a_1a_2a_3a_4}\epsilon_{b_1b_2b_3b_4},\label{INV2}
\end{align}
the result of the superspace integration can be written as
\begin{align}
\int d^4\theta^1_L\int d^4\theta^1_R{W_{a_1}}^{b_1}{W_{a_2}}^{b_2}{W_{a_3}}^{b_3}{W_{a_4}}^{b_4}
\epsilon^{a_1a_2a_3a_4}\epsilon_{b_1b_2b_3b_4}=24\left(4(\text{tr}R_+^2)^2-\text{tr} 
R_+^4\right).\label{Trace}
\end{align}
In order to uncover the index structure of the Riemann tensors involved in these traces, we use the 
complex basis for the bosonic and fermionic coordinates (\ref{compbas1})-(\ref{compbas5}).
With these definitions, eq.~(\ref{Trace}) becomes 
\begin{align}
\frac{1}{8}&\left(4(\text{tr}R_+^2)^2-\text{tr} R_+^4\right)=\nonumber\\
32(&R_{1212}R_{1\bar{2}1\bar{2}}R_{\bar{1}2\bar{1}2}R_{\bar{1}\bar{2}\bar{1}\bar{2}}+R_{1313}
R_{1\bar{3}1\bar{3}}R_{\bar{1}3\bar{1}3}R_{\bar{1}\bar{3}\bar{1}\bar{3}}+R_{2323}R_{2\bar{3}2
\bar{3}}R_{\bar{2}3\bar{2}3}R_{\bar{2}\bar{3}\bar{2}\bar{3}})+\nonumber\\
+2(&R_{1313}R_{1\bar{3}1\bar{3}}R_{\bar{1}2\bar{1}2}R_{\bar{1}\bar{2}\bar{1}\bar{2}}+R_{1212}
R_{1\bar{2}1\bar{2}}R_{\bar{1}3\bar{1}3}R_{\bar{1}\bar{3}\bar{1}\bar{3}}+R_{2323}
R_{1\bar{2}1\bar{2}}R_{2\bar{3}2\bar{3}}R_{\bar{1}\bar{2}\bar{1}\bar{2}}+\nonumber\\
+&R_{2323}R_{1\bar{3}1\bar{3}}R_{\bar{2}3\bar{2}3}R_{\bar{1}\bar{3}\bar{1}\bar{3}}+R_{1313}
R_{2\bar{3}2\bar{3}}R_{\bar{1}3\bar{1}3}R_{\bar{2}\bar{3}\bar{2}\bar{3}}+R_{1212}R_{\bar{1}2
\bar{1}2}R_{\bar{2}3\bar{2}3}R_{\bar{2}\bar{3}\bar{2}\bar{3}})+\nonumber\\
&\hspace{-0.7cm}+\ldots,\label{fullRtensor}
\end{align}
where the dots indicate terms where the two pairs of indices are not the same. Two examples for 
such terms are
\begin{align}
R_{2312}R_{1\bar{2}1\bar{2}}R_{2\bar{3}\bar{1}2}R_{\bar{1}\bar{2}\bar{1}\bar{2}},
\label{exception1}
\end{align}
or
\begin{align}
R_{1223}R_{1\bar{2}\bar{2}3}R_{\bar{1}22\bar{3}}R_{\bar{1}\bar{2}\bar{2}\bar{3}},
\label{exception2}
\end{align}
and will be neglected, since they do not lead to any different results (see for 
example Appendix \ref{AppendixLeftRightAsym} for a later application).
\subsection{Computation of the Scattering Amplitude}\label{K3compamp}
\subsubsection{Basic Tools}
\begin{figure}[htbp]
\begin{center}
\epsfig{file=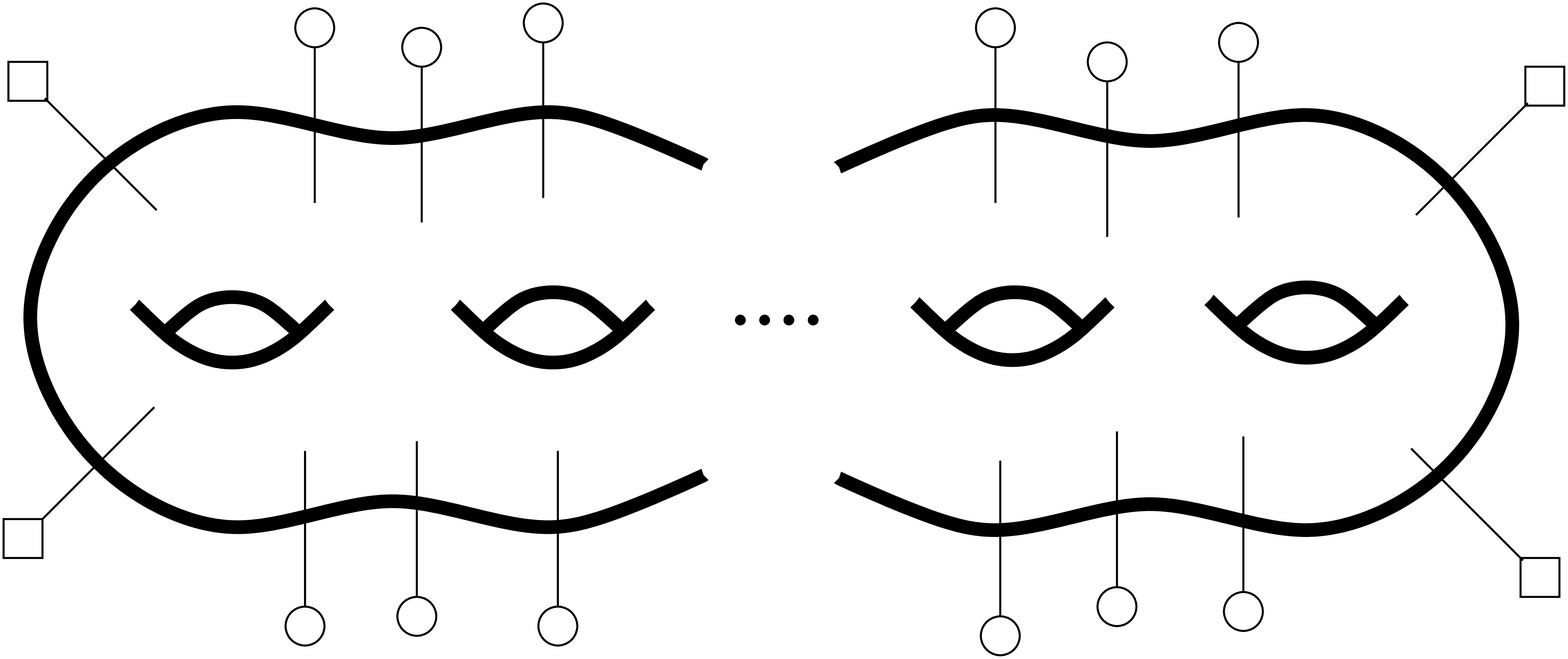, width=9cm}
\caption{Scattering of $4$ gravitons ({\large$\Box$}\normalsize) and $(4g-4)$ graviphotons 
(\huge$\circ$\normalsize) on a genus $g$ compact Riemann surface.}
\label{fig:Riemannscat}
\end{center}
\end{figure}

The analysis of the superspace expression (\ref{SuperExp}) entails computing a scattering 
amplitude on a compact genus $g$ Riemann surface with the insertion of $4$ gravitons with 
helicities corresponding to one of the terms of eq.~(\ref{fullRtensor}), and $(4g-4)$ graviphotons (see Figure \ref{fig:Riemannscat}).

We choose the gravitons to be inserted in the 0-ghost picture and the graviphotons in 
the $\left(-1/2\right)$-ghost picture. Therefore, the graviton vertex is given by
\begin{align}
V_{g}^{(0)}(p,h)=:h_{\mu\nu}\bigg(\partial Z^\mu+ip\cdot\chi\chi^\mu\bigg)\bigg(\bar{\partial} 
Z^\nu+ip\cdot\tilde{\chi}\tilde{\chi}^\nu\bigg)e^{ip\cdot Z}:,\label{vertexgraviton}
\end{align} 
with the convention, that a tilde denotes a field from the right-moving sector. Furthermore,
$h_{\mu\nu}$ is symmetric, traceless and obeys $p^\mu h_{\mu\nu}=0$. The graviphoton is a RR 
state and its vertex reads
\begin{align}
V_{T}^{\left(-1/2\right)}(p,\epsilon)=:e^{-\frac{1}{2}(\varphi+\tilde{\varphi})}
p_\nu\epsilon_{\mu}\left[S^a{(\sigma^{\mu\nu})_a}^b\tilde{S}_b\mathcal{S}\tilde{\mathcal{S}}+
S_{\dot{a}}{(\bar{\sigma}^{\mu\nu})^{\dot{a}}}_{\dot{b}}\tilde{S}^{\dot{b}}\bar{\mathcal{S}}
\bar{\tilde{\mathcal{S}}}
\right]e^{ip\cdot Z}:,\label{vertexgraviphoton}
\end{align}
where the polarization fulfills $p\cdot \epsilon=0$. $\varphi$ is a free 2d scalar bosonizing 
the super-ghost system, $S^a$ and $S_{\dot{a}}$ are (6-dimensional) space-time spin-fields 
of opposite chirality and $\mathcal{S}$ is a spin-field of the internal $N=4$ sector (and 
$\tilde{\mathcal{S}}$ its right-moving counterpart). In 
the present case of chiral graviphotons, only the first term in (\ref{vertexgraviphoton}) 
is taken into account (see also Appendix \ref{AppendixSelfDual} in this respect). 

Counting the charges of the super-ghost system, there is a surplus of $-(2g-2)$ which has to be 
canceled by inserting $(2g-2)$ picture changing operators (PCO) at random points of the surface. 
Another $(2g-2)$ insertions are due to the integration of super-moduli making a total of $(4g-4)$ 
PCOs. 

The first question is, which spin fields to choose for the graviphotons and which parts of 
the PCO to contract them with. For the moment we only discuss the left-moving sector, since the 
right movers follow exactly in the same way. Concerning the internal part we adopt the approach 
of \cite{Berkovits:1994vy,Berkovits:1998ex} and choose all graviphotons to have the same internal spin-field, 
namely
\begin{align}
\mathcal{S}=e^{\frac{i}{2}(\phi_4+\phi_5)},
\end{align}
where $\phi_I$ bosonize the fermionic coordinates $\psi_I$ defined in eqs.(\ref{compbas1})-(\ref{compbas5}), 
$\psi_I=e^{i\phi_I}$.
This creates a charge surplus of $(2g-2)$ in the 4 and 5 plane which can only be balanced by 
the picture changing operators. In other words we have $(2g-2)$ PCO contributing
\begin{align}
e^{\varphi}e^{-i\phi_4}\partial X_4,
\end{align}
and another $(2g-2)$ providing\footnote{An antisymmetric sum over all insertions is implicitly 
assumed.}
\begin{align}
e^{\varphi}e^{-i\phi_5}\partial X_5.
\end{align}

Finally we are left to deal with the space-time part and since we are considering chiral 
graviphotons, the following spin-fields are possible
\begin{align}
&S_1=e^{\frac{i}{2}(\phi_1+\phi_2+\phi_3)},\\
&S_2=e^{\frac{i}{2}(\phi_1-\phi_2-\phi_3)},\\
&S_3=e^{\frac{i}{2}(-\phi_1+\phi_2-\phi_3)},\\
&S_4=e^{\frac{i}{2}(-\phi_1-\phi_2+\phi_3)}.
\end{align}
Let us assume that $a_i$ graviphotons are inserted with the spin field 
$S_i$, where the $a_i$ are subject to the set of linear equations
\begin{align}
&a_1+a_2-a_3-a_4=0,\nonumber\\
&a_1-a_2+a_3-a_4=0,\nonumber\\
&a_1-a_2-a_3+a_4=0,\nonumber\\
&a_1+a_2+a_3+a_4=4g-4,\nonumber
\end{align}
which reflect the constraints of charge cancellation in each of the three planes and the fact 
that there is a total of $(4g-4)$ graviphotons. The unique solution of this system is $a_1=a_2=a_3=a_4=g-1$.\footnote{Of course there would 
still be the possibility of leaving a surplus charge from the graviphotons, which is then 
canceled by the graviton insertions. However these amplitudes would stem from terms of the 
form $(TRT)^4T^{4g-12}$ in the effective action which also arise when performing 
the full superspace integration in (\ref{SuperExp}) and we will thus neglect them in the 
remainder of the paper.}

Summarizing, until now we have inserted the following fields, where the columns $\phi_i$ 
denote the charges of the given field with respect to the various planes\\[10pt]
\begin{center}
\begin{tabular}{|c|c|c||c|c|c||c|c|}\hline
\textbf{insertion} & \textbf{number} & \textbf{position} & $\phi_1$ & $\phi_2$ & $\phi_3$ & 
$\phi_4$ & $\phi_5$\\\hline\hline
graviphoton & $g-1$ & $x_i$ & \parbox{0.7cm}{\vspace{0.2cm}$+\frac{1}{2}$\vspace{0.2cm}} & 
\parbox{0.7cm}{\vspace{0.2cm}$+\frac{1}{2}$\vspace{0.2cm}} & \parbox{0.7cm}{\vspace{0.2cm}
$+\frac{1}{2}$\vspace{0.2cm}} & \parbox{0.7cm}{\vspace{0.2cm}$+\frac{1}{2}$\vspace{0.2cm}} 
& \parbox{0.7cm}{\vspace{0.2cm}$+\frac{1}{2}$\vspace{0.2cm}} \\\hline 
 & $g-1$ & $y_i$ & \parbox{0.7cm}{\vspace{0.2cm}$+\frac{1}{2}$\vspace{0.2cm}} & \parbox{0.7cm}
{\vspace{0.2cm}$-\frac{1}{2}$\vspace{0.2cm}} & \parbox{0.7cm}{\vspace{0.2cm}$-\frac{1}{2}$
\vspace{0.2cm}} & \parbox{0.7cm}{\vspace{0.2cm}$+\frac{1}{2}$\vspace{0.2cm}} & 
\parbox{0.7cm}{\vspace{0.2cm}$+\frac{1}{2}$\vspace{0.2cm}} \\\hline 
 & $g-1$ & $u_i$ & \parbox{0.7cm}{\vspace{0.2cm}$-\frac{1}{2}$\vspace{0.2cm}} & 
\parbox{0.7cm}{\vspace{0.2cm}$+\frac{1}{2}$\vspace{0.2cm}} & \parbox{0.7cm}
{\vspace{0.2cm}$-\frac{1}{2}$\vspace{0.2cm}} & \parbox{0.7cm}{\vspace{0.2cm}$+
\frac{1}{2}$\vspace{0.2cm}} & \parbox{0.7cm}{\vspace{0.2cm}$+\frac{1}{2}
$\vspace{0.2cm}} \\\hline 
 & $g-1$ & $v_i$ & \parbox{0.7cm}{\vspace{0.2cm}$-\frac{1}{2}$\vspace{0.2cm}} & 
\parbox{0.7cm}{\vspace{0.2cm}$-\frac{1}{2}$\vspace{0.2cm}} & \parbox{0.7cm}
{\vspace{0.2cm}$+\frac{1}{2}$\vspace{0.2cm}} & \parbox{0.7cm}{\vspace{0.2cm}$+
\frac{1}{2}$\vspace{0.2cm}} & \parbox{0.7cm}{\vspace{0.2cm}$+\frac{1}{2}$
\vspace{0.2cm}} \\\hline\hline
PCO & $2g-2$ & $\{s_1\}$ & 0 & 0 & 0 &\parbox{0.7cm}{\vspace{0.2cm}$-1$
\vspace{0.2cm}} & 0 \\\hline 
 & $2g-2$ & $\{s_2\}$ & 0 & 0 & 0 & 0 &\parbox{0.7cm}{\vspace{0.2cm}$-1$
\vspace{0.2cm}} \\\hline 
\end{tabular}
\end{center}
${}$\\[10pt]
and $\{s_i\}$ stands for a collection of (apriori arbitrary) points on the Riemann surface.

In order to completely determine the amplitude still the graviton helicities are missing. To 
this end, one has to pick one of the terms in (\ref{fullRtensor}). Since most of them are 
related by simply exchanging two planes, it suffices to look at two generic cases:
\subsubsection{Graviton Combination I}\label{GravComb1}
When focusing on the fermionic part of the vertex operator in (\ref{vertexgraviton}), one 
possible setup of charges is (only the left-moving part is displayed):
\begin{center}
\begin{tabular}{|c|c|c||c|c|c||c|c|}\hline
\textbf{insertion} & \textbf{number} & \textbf{position} & $\phi_1$ & $\phi_2$ & $\phi_3$ & 
$\phi_4$ & $\phi_5$\\\hline\hline
graviton & $1$ & $z_1$ & 0 & \parbox{0.7cm}{\vspace{0.2cm}$+1$\vspace{0.2cm}} & \parbox{0.7cm}
{\vspace{0.2cm}$+1$\vspace{0.2cm}} & 0 & 0 \\\hline 
 & $1$ & $z_2$ & \parbox{0.7cm}{\vspace{0.2cm}$+1$\vspace{0.2cm}} & \parbox{0.7cm}
{\vspace{0.2cm}$-1$\vspace{0.2cm}} & 0 & 0 & 0 \\\hline 
 & $1$ & $z_3$ & 0 & \parbox{0.7cm}{\vspace{0.2cm}$+1$\vspace{0.2cm}} & \parbox{0.7cm}
{\vspace{0.2cm}$-1$\vspace{0.2cm}} & 0 & 0 \\\hline 
 & $1$ & $z_4$ & \parbox{0.7cm}{\vspace{0.2cm}$-1$\vspace{0.2cm}} & \parbox{0.7cm}
{\vspace{0.2cm}$-1$\vspace{0.2cm}} & 0 & 0 & 0 \\\hline
\end{tabular}
\end{center}
${}$\\[10pt]
The correlation function can now be written in the form\footnote{In the following, we drop $g$-dependent overall 
factors, which can be easily restored as in Ref.~\cite{Antoniadis:1993ze}.}
\begin{align}
\mathcal{F}_g^{(\text{6d})}=\langle&\prod_{i=1}^{g-1}e^{-\frac{\varphi}{2}}e^{\frac{i}{2}(\phi_1+\phi_2+\phi_3+
\phi_4+
\phi_5)}(x_i)\prod_{i=1}^{g-1}e^{-\frac{\varphi}{2}}e^{\frac{i}{2}(\phi_1-\phi_2-\phi_3+\phi_4+
\phi_5)}(y_i)\cdot\nonumber\\
\cdot&\prod_{i=1}^{g-1}e^{-\frac{\varphi}{2}}e^{\frac{i}{2}(-\phi_1+\phi_2-\phi_3+\phi_4+\phi_5)}
(u_i)\prod_{i=1}^{g-1}e^{-\frac{\varphi}{2}}e^{\frac{i}{2}(-\phi_1-\phi_2+\phi_3+\phi_4+\phi_5)}
(v_i)\cdot\nonumber\\
\cdot&e^{i(\phi_2+\phi_3)}(z_1)\cdot e^{i(\phi_1-\phi_2)}(z_2)\cdot e^{i(\phi_2-\phi_3)}(z_3)
\cdot e^{-i(\phi_1+\phi_2)}(z_4)\cdot\nonumber\\
\cdot&\prod_a^{\{s_1\}}e^{\varphi}e^{-i\phi_4}\partial X_4(r_a)\prod_a^{\{s_2\}}e^{\varphi}
e^{-i\phi_5}\partial X_5(r_a)\rangle.
\label{harmL}
\end{align} 
Computing all possible contractions for a fixed spin structure, the result for the amplitude is
\begin{align}
&\mathcal{F}_g^{(\text{6d})}=\frac{\vartheta_s\!\!\left(\frac{1}{2}\sum_{i=1}^{g-1}
(x_i+y_i-u_i-v_i)+z_2-z_4\right)}
{\vartheta_s\!\!\left(\frac{1}{2}\sum_{i=1}^{g-1}(x_i+y_i+u_i+v_i)-\sum_{a=1}^{4g-4}r_a+
2\Delta\right)\prod_{a<b=1}^{4g-4}E(r_a,r_b)\prod_{a=1}^{4g-4}\sigma^2(r_a)}\cdot\nonumber\\
&\cdot\frac{\vartheta_s\!\!\left(\frac{1}{2}\sum_{i=1}^{g-1}
(x_i-y_i+u_i-v_i)+z_1-z_2+z_3-z_4\right)\!\vartheta_s\!\!\left(\frac{1}{2}
\sum_{i=1}^{g-1}(x_i-y_i-u_i+v_i)+z_1-z_3\right)}{\prod_{j=1}^{g-1}E(x_i,z_4)
E(y_i,z_1)E(u_i,z_2)E(v_i,z_3)E(z_1,z_2)E(z_1,z_4)E(z_2,z_3)E(z_3,z_4)}\cdot\nonumber\\
&\cdot\vartheta_{h_4,s}\!\!\left(\frac{1}{2}\sum_{i=1}^{g-1}(x_i+y_i+u_i+v_i)-
\sum_{a}^{\{s_1\}}r_a\right)\vartheta_{h_5,s}\!\!\left(\frac{1}{2}
\sum_{i=1}^{g-1}(x_i+y_i+u_i+v_i)-\sum_{a}^{\{s_2\}}r_a\right)\cdot\nonumber\\
&\cdot\prod_{i=1}^{g-1}\sigma(x_i)\sigma(y_i)\sigma(u_i)\sigma(v_i)
\prod_{i<j}^{g-1}E(x_i,x_j)E(y_i,y_j)E(u_i,u_j)E(v_i,v_j)\prod_{a<b}^{\{s_1\}}E(r_a,r_b)\cdot\nonumber\\
&\cdot\prod_{a<b}^{\{s_2\}}E(r_a,r_b)
\prod_{j=1}^{g-1}E(x_i,z_1)E(y_i,z_2)E(u_i,z_3)E(v_i,z_4)\prod_{a}^{\{s_1\}}
\partial X_4(r_a)\prod_{a}^{\{s_2\}}\partial X_5(r_a).\nonumber
\end{align}
Here $\vartheta_{h_I,s}$ stands for the Jacobi theta-function~\cite{Fay} with the spin structure $s$ 
and twist $h_I$, $E(x,y)$ is the prime form, $\sigma$ is the $\frac{g}{2}$-differential 
with no zeros or poles and $\Delta$ stands for the Riemann theta constant (see also Appendix 
\ref{AppendixRiemanntot}). We should also mention again, that in this computation (and also 
throughout the paper), phase factors and numerical constants of the amplitude, as well 
as factors of the chiral determinant $Z_1$ of the $(1,0)$ system are dropped. The latter 
can most easily be restored by comparing with the $N=2$ case at any intermediate step 
\cite{Antoniadis:1993ze} and disappear completely in the final result. 

In order to simplify the expression the apriori arbitrary PCO insertion points 
$r_a$ are fixed in the following way\footnote{In the remainder of this section we will refer to 
this 'choice' as the 'gauge choice' or 'gauge condition'.}
\begin{align}
&\frac{1}{2}\sum_{i=1}^{g-1}(x_i+y_i-u_i-v_i)+z_2-z_4=\frac{1}{2}\sum_{i=1}^{g-1}
(x_i+y_i+u_i+v_i)-\sum_{a=1}^{4g-4}r_a+2\Delta\nonumber\\
&\Rightarrow \sum_{a=1}^{4g-4}r_a=\sum_{i=1}^{g-1}(u_i+v_i)-z_2+z_4+2\Delta,\label{gaugecond3}
\end{align}
which results in a cancellation of one $\vartheta$-function in the numerator against the one in 
the denominator. The next step is the sum over the different spin structures, for which the relevant terms 
are the four remaining $\vartheta$-functions:
\begin{align}
&\vartheta_s\!\!\left(\frac{1}{2}\sum_{i=1}^{g-1}(x_i-y_i+u_i-v_i)+z_1-z_2+z_3-z_4\right)
\vartheta_s\!\!\left(\frac{1}{2}\sum_{i=1}^{g-1}(x_i-y_i-u_i+v_i)+z_1-z_3\right)\nonumber\\
&\vartheta_{h_4,s}\!\!\left(\frac{1}{2}\sum_{i=1}^{g-1}(x_i+y_i+u_i+v_i)-\sum_{a}^{\{s_1\}}r_a
\right)\vartheta_{h_5,s}\!\!\left(\frac{1}{2}\sum_{i=1}^{g-1}(x_i+y_i+u_i+v_i)-
\sum_{a}^{\{s_2\}}r_a\right)\label{spinsum3}
\end{align}
To this end, the Riemann identity is used (see Appendix \ref{AppendixRiemann}). In the special 
case of (\ref{spinsum3}) the summed arguments read respectively
\begin{itemize}
\item $T_{++++}$:
\begin{align}
\sum_{i=1}^{g-1}x_i+\frac{1}{2}\sum_{i=1}^{g-1}(u_i+v_i)+z_1-\frac{1}{2}(z_2+z_4)-\frac{1}{2}
\sum_{a=1}^{4g-4}r_a=\sum_{i=1}^{g-1}x_i+z_1-z_4-\Delta,\nonumber
\end{align}
\item $T_{--++}$:
\begin{align}
\sum_{i=1}^{g-1}y_i+\frac{1}{2}\sum_{i=1}^{g-1}(u_i+v_i)-z_1+\frac{1}{2}(z_2+z_4)-\frac{1}{2}
\sum_{a=1}^{4g-4}r_a=\sum_{i=1}^{g-1}y_i-z_1+z_2-\Delta,\nonumber
\end{align}
\item $T_{-+-+}$:
\begin{align}
&-\frac{1}{2}\sum_{i=1}^{g-1}(u_i-v_i)-z_3+\frac{1}{2}(z_2+z_4)+\frac{1}{2}\sum_{a}^{\{s_1\}}r_a
-\frac{1}{2}\sum_{a}^{\{s_2\}}r_a=\nonumber\\
&\hspace{0.7cm}=\sum_{a=1}^{\{s_1\}}r_a-\sum_{i=1}^{g-1}u_i+z_2-z_3-\Delta,\nonumber
\end{align}
\item $T_{-++-}$:
\begin{align}
&-\frac{1}{2}\sum_{i=1}^{g-1}(u_i-v_i)-z_3+\frac{1}{2}(z_2+z_4)-\frac{1}{2}\sum_{a}^{\{s_1\}}r_a
+\frac{1}{2}\sum_{a}^{\{s_2\}}r_a=\nonumber\\
&\hspace{0.7cm}=\sum_{a=1}^{\{s_2\}}r_a-\sum_{i=1}^{g-1}u_i+z_2-z_3-\Delta,\nonumber
\end{align}
\end{itemize}
where the gauge condition (\ref{gaugecond3}) was also used. With the further relation\footnote{The fact that 
the sum over all twists vanishes is a consequence of space-time
supersymmetry.} 
$h_4+h_5=0$, the correlation function takes the form
\begin{align}
&\mathcal{F}_g^{(\text{6d})}=\frac{\vartheta\!\!\left(\sum_{i=1}^{g-1}x_i+z_1-z_4-\Delta\right)
\vartheta\!\!\left(
\sum_{i=1}^{g-1}y_i-z_1+z_2-\Delta\right)}{ \prod_{a<b=1}^{4g-4}E(r_a,r_b)\prod_{a=1}^{4g-4}
\sigma^2(r_a)\prod_{j=1}^{g-1}E(x_i,z_4)E(y_i,z_1)E(u_i,z_2)E(v_i,z_3)}\cdot\nonumber\\
&\cdot\frac{\vartheta_{-h_4}\!\!\left(\sum_{a=1}^{\{s_1\}}r_a-
\sum_{i=1}^{g-1}u_i+z_2-z_3-\Delta\right)\vartheta_{-h_5}\!\!\left(\sum_{a=1}^{\{s_2\}}r_a-
\sum_{i=1}^{g-1}u_i+z_2-z_3-\Delta\right)}{E(z_1,z_2)E(z_1,z_4)E(z_2,z_3)E(z_3,z_4)}
\cdot\nonumber\\
&\cdot\prod_{i=1}^{g-1}\sigma(x_i)\sigma(y_i)\sigma(u_i)\sigma(v_i)\prod_{a<b}^{\{s_1\}}
E(r_a,r_b)\prod_{a<b}^{\{s_2\}}E(r_a,r_b)\prod_{i<j}^{g-1}E(x_i,x_j)E(y_i,y_j)E(u_i,u_j)
E(v_i,v_j)\cdot\nonumber\\
&\cdot\prod_{i=1}^{g-1}E(x_i,z_1)E(y_i,z_2)E(u_i,z_3)E(v_i,z_4)\prod_{a}^{\{s_1\}}
\partial X_4(r_a)\prod_{a}^{\{s_2\}}\partial X_5(r_a).\label{interampl}
\end{align}
Now we can in two occasions use the so-called bosonization formulas \cite{Verlinde:1986kw} to write
\begin{align}
\bullet\ &\frac{\vartheta\!\!\left(\sum_{i=1}^{g-1}x_i+z_1-z_4-\Delta\right)}
{\prod_{i=1}^{g-1}E(x_i,z_4)E(z_1,z_4)\sigma(z_4)}\prod_{i<j}^{g-1}E(x_i,x_j)
\prod_{i=1}^{g-1}E(x_i,z_1)\prod_{i=1}^{g-1}\sigma(x_i)\sigma(z_1)=\nonumber\\
&\hspace{0.5cm}=Z_1\text{det}\omega_i(x_1,x_2,\ldots,x_{g-1},z_1),\\
&\nonumber\\
\bullet\ &\frac{\vartheta\!\!\left(\sum_{i=1}^{g-1}y_i-z_1+z_2-\Delta\right)}{
\prod_{i=1}^{g-1}E(y_i,z_1)E(z_1,z_2)\sigma(z_1)}\prod_{i<j}^{g-1}E(y_i,y_j)
\prod_{i=1}^{g-1}E(y_i,z_2)\prod_{i=1}^{g-1}\sigma(y_i)\sigma(z_2)=\nonumber\\
&\hspace{0.5cm}=Z_1\text{det}\omega_i(y_1,y_2,\ldots,y_{g-1},z_2),
\end{align}
where $\omega$ are the abelian 1-differentials, of which there are $g$ on a compact 
Riemann surface of genus $g$. With the help of the gauge condition (\ref{gaugecond3}) and upon multiplying 
(\ref{interampl}) with
\begin{align}
1=\frac{\vartheta\!\!\left(\sum_{i=1}^{g-1}v_i+z_4-z_3-\Delta\right)}{\vartheta\!\!
\left(\sum_{a=1}^{4g-4}r_a-\sum_{i=1}^{g-1}u_i+z_2-z_3-3\Delta\right)},
\end{align}
also a third time the bosonization identity can be used
\begin{align}
&\frac{\vartheta\!\!\left(\sum_{i=1}^{g-1}v_i+z_4-z_3-\Delta\right)}
{\prod_{i=1}^{g-1}E(v_i,z_3)E(z_4,z_3)\sigma(z_3)}\prod_{i<j}^{g-1}E(v_i,v_j)
\prod_{i=1}^{g-1}E(v_i,z_4)\prod_{i=1}^{g-1}\sigma(v_i)\sigma(z_4)=\nonumber\\
&\hspace{0.5cm}=Z_1\text{det}\omega_i(v_1,v_2,\ldots,v_{g-1},z_4).
\end{align}
Taking into account all these identities and plugging them in (\ref{interampl}),
we obtain the result:
\begin{align}
&\mathcal{F}_g^{(\text{6d})}=\nonumber\\
&=\frac{\vartheta_{-h_4}\!\!\left(\sum_{a=1}^{\{s_1\}}r_a-\sum_{i=1}^{g-1}u_i+z_2-z_3-
\Delta\right)\vartheta_{-h_5}\!\!\left(\sum_{a=1}^{\{s_2\}}r_a-\sum_{i=1}^{g-1}u_i+z_2-z_3-
\Delta\right)}{\vartheta\!\!\left(\sum_{a=1}^{4g-4}r_a-\sum_{i=1}^{g-1}u_i+z_2-z_3-3
\Delta\right)\prod_{a<b=1}^{4g-4}E(r_a,r_b)\prod_{a=1}^{4g-4}\sigma^2(r_a)}\cdot\nonumber\\
&\cdot \frac{\prod_{i=1}^{g-1}\sigma(u_i)\prod_{a<b}^{\{s_1\}}E(r_a,r_b)\prod_{a<b}^{\{s_2\}}
E(r_a,r_b)\prod_{i<j}^{g-1}E(u_i,u_j)\sigma(z_3)}{\prod_{j=1}^{g-1}E(u_i,z_2)E(z_2,z_3)
\sigma(z_2)}\cdot\nonumber\\
&\cdot\prod_{i=1}^{g-1}E(u_i,z_3)\prod_{a}^{\{s_1\}}\partial X_4(r_a)\prod_{a}^{\{s_2\}}
\partial X_5(r_a)\cdot \text{det}\omega_i(x_i,z_1)\text{det}\omega_i(y_i,z_2)\text{det}
\omega_i(v_i,z_4).\nonumber
\end{align}
With the help of further identities of \cite{Verlinde:1986kw}, this can be rewritten in terms of correlation 
functions of the following kind
\begin{align}
\mathcal{F}_g^{(\text{6d})}=&\frac{\langle\prod_{a}^{\{s_1\}}\bar{\psi}_4(r_a)\bar{\psi}_4(z_2)\prod_{i=1}^{g-1}
\psi_4(u_i)\psi_4(z_3)\rangle\cdot\langle\prod_{a}^{\{s_2\}}\bar{\psi}_5(r_a)
\bar{\psi}_5(z_2)\prod_{i=1}^{g-1}\psi_5(u_i)\psi_5(z_3)\rangle}{\langle\prod_{a=1}^{4g-4}b(r_a)b(z_2)
\prod_{i=1}^{g-1}c(u_i)c(z_3)\rangle}\cdot\nonumber\\
&\cdot\prod_{a}^{\{s_1\}}\partial X_4(r_a)\prod_{a}^{\{s_2\}}\partial X_5(r_a)\ \text{det}
\omega_i(x_i,z_1)\text{det}\omega_i(y_i,z_2)\text{det}\omega_i(v_i,z_4),\label{resultcom2}
\end{align}
where the following relations were used
\begin{align}
&\frac{\vartheta_{-h_4}\!\!\left(\sum_{a}^{\{s_1\}}r_a-\sum_{i=1}^{g-1}u_i+z_2-z_3-\Delta\right)\prod_{a<b}^{\{s_1\}}E(r_a,r_b)\prod_{i<j}E(u_i,u_j)\prod_{a}^{\{s_1\}}E(r_a,z_2)}{\prod_{a,i}E(r_a,u_i)\prod_{a}^{\{s_1\}}E(r_a,z_3)\prod_{i=1}^{g-1}E(u_i,z_2)E(z_2,z_3)\prod_{i=1}^{g-1}\sigma(u_i)\sigma(z_3)}\cdot\nonumber\\*
&\hspace{0.5cm}\cdot \prod_{i=1}^{g-1}E(u_i,z_3)\prod_{a}^{\{s_1\}}\sigma(r_a)\sigma(z_2)=\langle\prod_{a}^{\{s_1\}}\bar{\psi}_4(r_a)\bar{\psi}_4(z_2)\prod_{i=1}^{g-1}\psi_4(u_i)\psi_4(z_3)\rangle,\nonumber
\end{align}
together with a similar definition for the correlator of $\psi_5$ and
\begin{align}
&\frac{\vartheta\!\!\left(\sum_{a=1}^{4g-4}r_a-\sum_{i=1}^{g-1}u_i+z_2-z_3-3\Delta\right)\prod_{a<b}^{4g-4}E(r_a,r_b)\prod_{i<j}E(u_i,u_j)\prod_{a=1}^{4g-4}E(r_a,z_2)}{\prod_{a,i}E(r_a,u_i)\prod_{a}^{4g-4}E(r_a,z_3)\prod_{i=1}^{g-1}E(u_i,z_2)E(z_2,z_3)\prod_{i=1}^{g-1}\sigma^3(u_i)\sigma^3(z_3)}\cdot\nonumber\\*
&\hspace{0.5cm}\cdot \prod_{i=1}^{g-1}E(u_i,z_3)\prod_{a}^{4g-4}\sigma^3(r_a)\sigma^3(z_2)=\langle\prod_{a=1}^{4g-4}b(r_a)b(z_2)\prod_{i=1}^{g-1}c(u_i)c(z_3)\rangle.\nonumber
\end{align}
Rewriting the expression (\ref{resultcom2}) using the operators of the $N=4$ superconformal algebra described in Section 
\ref{SectN4Top}, one finds 
\begin{align}
&\mathcal{F}_g^{(\text{6d})}=\nonumber\\
&=\frac{\langle\prod_{a=1}^{4g-4}G^-_{K3}(r_a)\prod_{i=1}^{g-1}J^{++}_{K3}(u_i)J^{--}_{K3}(z_2)
J^{++}_{K3}(z_3)\rangle}{\langle\prod_{a=1}^{4g-4}b(r_a)b(z_2)
\prod_{i=1}^{g-1}c(u_i)c(z_3)\rangle}
\cdot\text{det}\omega_i(x_i,z_1)\text{det}\omega_i(y_i,z_2)\text{det}\omega_i(v_i,z_4).
\nonumber
\end{align}

In the next step, using the fact that the gauge condition (\ref{gaugecond3}) still allows the freedom to move $g$
PCO points, we let the positions of $(g-1)$ of the PCO's collapse with the $u_i$ and another one with $z_3$, 
which with the help of the OPE's in Section \ref{SectN4Top} and Appendix \ref{AppendixSuperconformal} yields 
$\tilde{G}^+_{K3}$ 
(the fact that there is no singularity in this procedure owes to the $bc$ ghost system correlator 
in the denominator):
\begin{align}
&\mathcal{F}_g^{(\text{6d})}=\nonumber\\
&=\frac{\langle\prod_a^{3g-4}G^-_{K3}(r_a)\prod_{i=1}^{g-1}\tilde{G}^+_{K3}(u_i)J^{--}(z_2)
\tilde{G}^+_{K3}(z_3)\rangle}{\langle\prod_{a=1}^{3g-4}b(r_a)
b(z_2)\rangle}\cdot\text{det}\omega_i(x_i,z_1)\text{det}\omega_i(y_i,z_2)
\text{det}\omega_i(v_i,z_4).\label{resultcom21}
\end{align}
This amplitude is to be multiplied by the $(3g-3)$ Beltrami differentials folded with the $b$ ghosts 
to provide the correct measure for the integration over the genus $g$ moduli space. Note that
the ratio of the correlators appearing in (\ref{resultcom21}) is independent of the $(3g-3)$
positions $r_a$ and $z_2$. These positions can therefore be transmuted to the positions of
the operators $b$ that are folded with the Beltrami differentials; the correlation functions
of the latter then cancel with the denominator of (\ref{resultcom21}). Including now the
right-moving sector, we can integrate out the graviton and graviphoton insertion points giving finally
\begin{align}
\mathcal{F}_g^{(\text{6d})}=&\int_{\mathcal{M}_g}\langle|\prod_i^{3g-4}G^-_{K3}(\mu_i)
J^{--}_{K3}(\mu_{3g-3})|^2
\int\prod_{i=1}^{g}|\tilde{G}^+_{K3}|^2\rangle\cdot(\text{det}(\text{Im}\tau))^3,\label{resBerVafa}
\end{align}
where it is understood that the Beltrami differentials are totally anti-symmetrized and the absolute square indicates, that also the right moving contributions have been restored.
The additional factor of $(\text{det}(\text{Im}\tau))^3$ cancels exactly the zero-mode contribution 
of the space-time bosons and the remaining expression is topological and exactly the same as in 
\cite{Berkovits:1994vy} 
(equation (2.15)). Actually, by writing one of ${\tilde G}^+$'s as the contour integral 
$\tilde{G}^+_{K3}(z_3)=\oint \tilde{G}^+_{K3} J_{K3}(z_3)$ (see OPE's (\ref{N4K3}) and weights 
(\ref{twistweights})) and pulling off the contour integral, it can only encircle $J^{--}_{K3}$ which 
converts it to $G^-_{K3}$. Equation (\ref{resBerVafa}) can then be expressed in a more symmetric form:
\begin{align}
\mathcal{F}_g^{(\text{6d})}=&\int_{\mathcal{M}_g}\langle|\prod_i^{3g-3}G^-_{K3}(\mu_i)|^2
\int\prod_{i=1}^{g-1}|\tilde{G}^+_{K3}|^2\int |J_{K3}|^2\rangle_{int},
\label{result6ds}
\end{align}
where the subscript `$int$' means that the correlator is restricted to the internal ($K3$) part.
\subsubsection{Graviton Combination II}\label{GravComb2}
The other graviton combination different from the one considered previously in Section~\ref{GravComb1} is 
given by\\[10pt]
\begin{center}
\begin{tabular}{|c|c|c||c|c|c||c|c|}\hline
\textbf{insertion} & \textbf{number} & \textbf{position} & $\phi_1$ & $\phi_2$ & $\phi_3$ & 
$\phi_4$ & $\phi_5$\\\hline\hline
graviton & $1$ & $z_1$ & \parbox{0.7cm}{\vspace{0.2cm}$+1$\vspace{0.2cm}} & \parbox{0.7cm}
{\vspace{0.2cm}$+1$\vspace{0.2cm}} & 0 & 0 & 0 \\\hline 
 & $1$ & $z_2$ & \parbox{0.7cm}{\vspace{0.2cm}$+1$\vspace{0.2cm}} & \parbox{0.7cm}
{\vspace{0.2cm}$-1$\vspace{0.2cm}} & 0 & 0 & 0 \\\hline 
 & $1$ & $z_3$ & \parbox{0.7cm}{\vspace{0.2cm}$-1$\vspace{0.2cm}} & \parbox{0.7cm}
{\vspace{0.2cm}$+1$\vspace{0.2cm}} & 0 & 0 & 0 \\\hline 
 & $1$ & $z_4$ & \parbox{0.7cm}{\vspace{0.2cm}$-1$\vspace{0.2cm}} & \parbox{0.7cm}
{\vspace{0.2cm}$-1$\vspace{0.2cm}} & 0 & 0 & 0 \\\hline
\end{tabular}
\end{center}
${}$\\[10pt]
The amplitude can now be expressed as
\begin{align}
\mathcal{F}_g^{(\text{6d})}=\langle&\prod_{i=1}^{g-1}e^{-\frac{\varphi}{2}}e^{\frac{i}{2}
(\phi_1+\phi_2+\phi_3+\phi_4+\phi_5)}(x_i)\prod_{i=1}^{g-1}
e^{-\frac{\varphi}{2}}e^{\frac{i}{2}(\phi_1-\phi_2-\phi_3+\phi_4+\phi_5)}(y_i)
\cdot\nonumber\\
\cdot&\prod_{i=1}^{g-1}e^{-\frac{\varphi}{2}}
e^{\frac{i}{2}(-\phi_1+\phi_2-\phi_3+\phi_4+\phi_5)}(u_i)
\prod_{i=1}^{g-1}e^{-\frac{\varphi}{2}}
e^{\frac{i}{2}(-\phi_1-\phi_2+\phi_3+\phi_4+\phi_5)}(v_i)\cdot\nonumber\\
\cdot&e^{i(\phi_1+\phi_2)}(z_1)\cdot e^{i(\phi_1-\phi_2)}(z_2)
\cdot e^{i(-\phi_1+\phi_2)}(z_3)\cdot e^{-(i\phi_1+\phi_2)}(z_4)\cdot\nonumber\\
\cdot&\prod_a^{\{s_1\}}e^{\varphi}e^{-i\phi_4}\partial X_4(r_a)
\prod_a^{\{s_2\}}e^{\varphi}e^{-i\phi_5}\partial X_5(r_a)\rangle.
\label{harmLp}
\end{align} 
Going through similar steps as in the previous case, the result after performing the spin-structure sum is
\begin{align}
&\mathcal{F}_g^{(\text{6d})}=\nonumber\\
&=\frac{\vartheta_{-h_4}\!\!\left(\sum_{a=1}^{\{s_1\}}r_a-\sum_{i=1}^{g-1}u_i-z_3+z_2-
\Delta\right)\vartheta_{-h_5}\!\!\left(\sum_{a=1}^{\{s_2\}}r_a-\sum_{i=1}^{g-1}u_i-z_3+z_2-
\Delta\right)}{\vartheta\!\!\left(\sum_{a=1}^{4g-4}r_a-\sum_{i=1}^{g-1}u_i-z_3+z_2-3\Delta\right)
\prod_{a<b=1}^{4g-4}E(r_a,r_b)\prod_{a=1}^{4g-4}\sigma^2(r_a)}\cdot\nonumber\\
&\cdot \frac{\prod_{i=1}^{g-1}\sigma(u_i)\prod_{a<b}^{\{s_1\}}E(r_a,r_b)\prod_{a<b}^{\{s_2\}}
E(r_a,r_b)\prod_{i<j}^{g-1}E(u_i,u_j)\sigma(z_3)}{\prod_{j=1}^{g-1}E(u_i,z_2)E(z_2,z_3)
\sigma(z_2)}\cdot\nonumber\\
&\cdot\prod_{i=1}^{g-1}E(u_i,z_3)\prod_{a}^{\{s_1\}}\partial X_4(r_a)\prod_{a}^{\{s_2\}}
\partial X_5(r_a)\ \text{det}\omega_i(x_i,z_1)\text{det}\omega_i(y_i,z_2)\text{det}
\omega_i(v_i,z_4)\nonumber
\end{align}
Note, that this is exactly the same expression as (\ref{resultcom2}), such that the final answer can immediately 
be written down
\begin{align}
\mathcal{F}_g^{(\text{6d})}=&\int_{\mathcal{M}_g}\langle|\prod_i^{3g-4}G^-_{K3}(\mu_i)J^{--}_{K3}(\mu_{3g-3})|^2
\int\prod_{i=1}^{g}|\tilde{G}^+_{K3}|^2\rangle_{int}.
\label{result6d}
\end{align}
$ $\\[20pt]
Thus, we conclude that both generic helicity combinations of the superfield expression in 
(\ref{SuperExp}) lead to the same topological result, which is in perfect agreement with \cite{Berkovits:1994vy}.
\section{Type IIA on $K3 \times T^2$}\label{Section4dscattering}
After having established the results of \cite{Berkovits:1994vy} in the RNS formalism, the 
question arises whether similar topological amplitudes can be found for $N=4$ 
supersymmetry in 4 dimensions as well. Our basic strategy in approaching this 
question will be to toroidally compactify two additional dimensions and study 
again couplings on a compact genus $g$ Riemann surface. A major difference compared 
to the above cases is that in addition to the 4-dimensional gravity multiplet 
there are additional Kaluza-Klein fields which can take part in the scattering process.  

The easiest way to start out is to begin with the 6-dimensional graviton 
insertions of (\ref{fullRtensor}), compactify one additional plane and supplement 
the arising fields with $(2g-2)$ graviphotons from the 4-dimensional Weyl 
multiplet. Note, that now we consider a coupling proportional to 
$T_+^{2g-2}$ instead of $T_+^{4g-4}$. This is because of the cancellation of $U(1)$
charge, which is similar the the critical case of $N=2$ $F_g$'s, as explained in the 
Introduction. Moreover, if we started with a $T_+^{4g-4}$ 
coupling in 4 dimensions, there would be an additional factor of $\text{det}
(\text{Im}\tau)$ not being canceled by the space-time bosons, that 
would spoil the topological nature of the resulting amplitude. 
Considering only the case that the left-moving sector 
is symmetric to the right-moving one there are 3 apriori different cases to 
analyze:\footnote{For some remarks concerning asymmetric distributions of 
the left- and right-moving sector see Appendix \ref{AppendixLeftRightAsym}.}
\begin{itemize}
\item all of the 6-dimensional gravitons remain gravitons in the 4-dimensional amplitude
\item 2 out of 4 gravitons get converted into graviscalars in the process of compactification
\item all gravitons become graviscalars
\end{itemize}
Here by graviscalars we mean the moduli insertions corresponding to either the 
complex structure $U$ or the K\"ahler modulus $T$ of the torus $T^2$. 

In the following we will show that all three cases lead to the same 
topological expression. The result remains the same even if we replace graviscalar
insertions with NS KK graviphotons that we present in Appendix \ref{AppendixLeftRightAsym}.
A brief review of some basic tools which will be needed throughout the computation, namely the $N=4$ 
superconformal algebra for compactification on $K3\times T^2$ and its topological twist, is given in 
Section \ref{SectN4Top}.

\subsection{4 Graviton Case}\label{4gravT2}
We study the graviton setup of Section \ref{GravComb2} but regard the 3rd plane as the 
additionally compactified torus. 
Since we have now only $(2g-2)$ graviphoton insertions in the $\left(-1/2\right)$-ghost
picture, the number of PCO has also diminished to $(3g-3)$. The way to distribute the charges of 
the  graviphotons and PCO is
\begin{center}
\begin{tabular}{|c|c|c||c|c||c||c|c|}\hline
\textbf{insertion} & \textbf{number} & \textbf{position} & $\phi_1$ & $\phi_2$ & $\phi_3$ & 
$\phi_4$ & $\phi_5$\\\hline\hline
graviphoton & $g-1$ & $x_i$ & \parbox{0.7cm}{\vspace{0.2cm}$+\frac{1}{2}$\vspace{0.2cm}} & 
\parbox{0.7cm}{\vspace{0.2cm}$+\frac{1}{2}$\vspace{0.2cm}} & \parbox{0.7cm}{\vspace{0.2cm}$+
\frac{1}{2}$\vspace{0.2cm}} & \parbox{0.7cm}{\vspace{0.2cm}$+\frac{1}{2}$\vspace{0.2cm}} & 
\parbox{0.7cm}{\vspace{0.2cm}$+\frac{1}{2}$\vspace{0.2cm}} \\\hline 
 & $g-1$ & $y_i$ & \parbox{0.7cm}{\vspace{0.2cm}$-\frac{1}{2}$\vspace{0.2cm}} & \parbox{0.7cm}
{\vspace{0.2cm}$-\frac{1}{2}$\vspace{0.2cm}} & \parbox{0.7cm}{\vspace{0.2cm}$+\frac{1}{2}$
\vspace{0.2cm}} & \parbox{0.7cm}{\vspace{0.2cm}$+\frac{1}{2}$\vspace{0.2cm}} & \parbox{0.7cm}
{\vspace{0.2cm}$+\frac{1}{2}$\vspace{0.2cm}} \\\hline 
PCO & $g-1$ & $\{s_3\}$ & 0 & 0 &\parbox{0.7cm}{\vspace{0.2cm}$-1$\vspace{0.2cm}} & 0 & 0\\
\hline 
 & $g-1$ & $\{s_4\}$ & 0 & 0 & 0 &\parbox{0.7cm}{\vspace{0.2cm}$-1$\vspace{0.2cm}} & 0 \\
\hline 
 & $g-1$ & $\{s_5\}$ & 0 & 0 & 0 & 0 &\parbox{0.7cm}{\vspace{0.2cm}$-1$\vspace{0.2cm}} \\
\hline 
\end{tabular}
\end{center}
${}$\\[10pt]
The amplitude in this setup is given by\footnote{As before, only the left-moving part is 
considered. Furthermore, we introduce the superscript $(1)$ to distinguish this expression from others that will 
be computed in subsequent sections of the paper.}
\begin{align}
\mathcal{F}_g^{(1)}=\langle&\prod_{i=1}^{g-1}e^{-\frac{\varphi}{2}}e^{\frac{i}{2}(\phi_1+\phi_2+\phi_3+\phi_4+
\phi_5)}(x_i)\prod_{i=1}^{g-1}e^{-\frac{\varphi}{2}}e^{\frac{i}{2}(-\phi_1-\phi_2+\phi_3+\phi_4+
\phi_5)}(y_i)\cdot\nonumber\\
\cdot&e^{i(\phi_1+\phi_2)}(z_1)\cdot e^{i(\phi_1-\phi_2)}(z_2)\cdot e^{i(-\phi_1+\phi_2)}(z_3) 
\cdot e^{-i(\phi_1+\phi_2)}(z_4)\cdot\nonumber\\
\cdot&\prod_a^{\{s_3\}}e^{\varphi}e^{-i\phi_3}\partial X_3(r_a)\prod_a^{\{s_4\}}e^{\varphi}
e^{-i\phi_4}\partial X_4(r_a)\prod_a^{\{s_5\}}e^{\varphi}e^{-i\phi_5}\partial X_5(r_a)
\rangle,\nonumber
\end{align} 
which, after performing the contractions and the spin structure sum, together with some manipulations, 
using the gauge choice
\begin{align}
\sum_{a=1}^{3g-3}r_a=\sum_{i=1}^{g-1}y_i-z_1-z_2+z_3+z_4+2\Delta,
\label{gaugecond4}
\end{align} 
reads:
\begin{align}
\mathcal{F}_g^{(1)}&=\frac{\langle\prod_{a}^{\{s_3\}}\bar{\psi}_3\partial X_3(r_a)\bar{\psi}_3(z_2)\psi_3(\alpha)
\rangle\cdot\langle\prod_{a}^{\{s_4,s_5\}}G^-_{K3}(r_a)J^{--}_{K3}(z_2)J^{++}_{K3}(z_3)
\rangle}{\langle \prod_a^{3g-3}b(r_a)b(z_2)c(z_3)\rangle}\cdot\nonumber\\
&\hspace{0.5cm}\cdot \text{det}\omega_i(x_i,z_1)\text{det}\omega_i(y_i,z_4),\nonumber
\end{align}
where $\alpha$ is an arbitrary position on the Riemann surface and
\begin{align}
&\frac{\vartheta\!\!\left(\sum_{a}^{\{s_3\}}r_a+z_2-z_3-\Delta\right)\prod_{a<b}^{\{s_3\}}E(r_a,r_b)\prod_{a}^{\{s_3\}}E(r_a,z_2)\prod_{a}^{\{s_3\}}\sigma(r_a)\sigma(z_2)}{\prod_{a}^{\{s_3\}}E(r_a,z_3)E(z_2,z_3)\sigma(z_3)}=\nonumber\\
&\hspace{0.5cm}=\langle\prod_{a}^{\{s_3\}}\bar{\psi}_3(r_a)\bar{\psi}_3(z_2)\psi_3(\alpha)\rangle.\nonumber
\end{align} 

Now one may collapse one of the $r_a$ (for simplicity the last one $r_{3g-3}$ is chosen) with 
$z_3$. The resulting expression is:
\begin{align}
\mathcal{F}_g^{(1)}&=\frac{\langle\prod_{a}^{\{s_3\}}\bar{\psi}_3\partial X_3(r_a)\bar{\psi}_3(z_2)\psi_3(\alpha)
\rangle\cdot\langle\prod_{a=g}^{3g-4}G^-_{K3}(r_a)J^{--}_{K3}(z_2)\tilde{G}^+_{K3}(z_3)\rangle}
{\langle \prod_a^{3g-4}b(r_a)b(z_2)\rangle}\cdot\nonumber\\
&\hspace{0.5cm}\cdot \text{det}\omega_i(x_i,z_1)\text{det}\omega_i(y_i,z_4).\nonumber
\end{align}
This can also be written as (for simplicity we consider $\alpha=z_2$)
\begin{align}
\mathcal{F}_g^{(1)}&=\frac{\langle\prod_{a=1}^{3g-4}G^-(r_a)J^{--}_{K3}(z_2)\psi_3\bar{\psi}_3(z_2)\tilde{G}^+_{K3}
(z_3)\rangle}{\langle \prod_a^{3g-4}b(r_a)b(z_2)\rangle}\cdot \text{det}\omega_i(x_i,z_1)
\text{det}\omega_i(y_i,z_4)\nonumber
\end{align}
One can easily check that the first term in the above expression does not contain any poles with non-vanishing 
residue. One can therefore transport all $G^-$ and the $J^{--}_{K3}$ to the Beltrami differentials as before 
and including the right-moving sector and the integration over $x_i$, $y_i$, $z_1$ and $z_4$, one finds:  
\begin{align}
\mathcal{F}_g^{(1)}=\int_{\mathcal{M}_g}\langle|\prod_{a=1}^{3g-4}G^-(\mu_a)J^{--}_{K3}(\mu_{3g-3})|^2\int 
|\tilde{G}^+_{K3}(z_3)|^2\int|\psi_3\bar{\psi}_3(z_2)|^2 \rangle_{int}.
\label{result1}
\end{align}
where anti-symmetrization of all the $(3g-3)$ Beltrami differentials is understood.

Compared to the results of Section \ref{K3compamp}, we note the following formal\footnote{The comparison between 
(\ref{resBerVafa}) and (\ref{result1}) is only 
formal because the operators involved are from different superconformal algebras.} differences
\begin{itemize}
\item the $g$ insertions of $\int \tilde{G}^+_{K3}$ have been reduced to only one, 
which is due to the reduced number of graviphotons present in the new amplitude. Note that the $G^-$'s above 
are given by the sum of the corresponding operators in $K3$ ($G^-_{K3}$) and the ones in $T^2$ ($G^-_{T^2}$). 
The only non-zero contribution comes from terms 
where $(g-1)$ of them come from the $T^2$ part, which together with $J_{T^2}=\psi_3\bar\psi_3$ soak all the 
$g$ zero modes of 
the 1-differential $\bar{\psi}_3$ and one zero mode of the zero-differential $\psi_3$ in the twisted
theory. The remaining $(2g-4)$ $G^-$ come from $K3$ and together with $J^{--}_{K3}$ precisely account for
the $U(1)$ charge anomaly $(2g-2)$ of the twisted $K3$ SCFT.
\item the power of $\text{det}(\text{Im}\tau)$ coming from the integration of $x_i$, $y_i$, $z_1$ and $z_4$ 
has changed from 3 to 2, exactly canceling the zero mode contribution of the space-time bosons
which is now 4-dimensional as compared to 6-dimensional in Section \ref{K3compamp}.
\end{itemize}
By writing $\tilde{G}^+_{K3}(z_3)=\oint \tilde{G}^+_{K3} J_{K3}(z_3)$ and pulling off the 
contour integral which only converts $J^{--}_{K3}$ to $G^-_{K3}$, equation (\ref{result1}) can be
expressed in a more symmetric form:
\begin{align}
\mathcal{F}_g^{(1)}=\int_{\mathcal{M}_g}\langle|\prod_{a=1}^{3g-3}G^-(\mu_a)|^2\int 
|J_{K3}(z_3)|^2\int|J_{T^2}(z_2)|^2 \rangle_{int},
\label{result1a}
\end{align}
that can be compared to eq.(\ref{result6ds}) of the 6d case.

\subsection{2 Graviton - 2 Scalar Case}\label{2gr2sc}
We study the graviton setup of Section \ref{GravComb1} and again compactify the 3rd plane to 
a torus. Besides that, we keep the rest of the setup as in Section \ref{4gravT2}:
\begin{center}
\begin{tabular}{|c|c|c||c|c||c||c|c|}\hline
\textbf{insertion} & \textbf{number} & \textbf{position} & $\phi_1$ & $\phi_2$ & $\phi_3$ & 
$\phi_4$ & $\phi_5$\\\hline\hline
graviton & $1$ & $z_1$ & 0 & \parbox{0.7cm}{\vspace{0.2cm}$+1$\vspace{0.2cm}} & \parbox{0.7cm}
{\vspace{0.2cm}$+1$\vspace{0.2cm}} & 0 & 0 \\\hline 
 & $1$ & $z_2$ & \parbox{0.7cm}{\vspace{0.2cm}$+1$\vspace{0.2cm}} & \parbox{0.7cm}
{\vspace{0.2cm}$-1$\vspace{0.2cm}} & 0 & 0 & 0 \\\hline 
 & $1$ & $z_3$ & 0 & \parbox{0.7cm}{\vspace{0.2cm}$+1$\vspace{0.2cm}} & \parbox{0.7cm}
{\vspace{0.2cm}$-1$\vspace{0.2cm}} & 0 & 0 \\\hline 
 & $1$ & $z_4$ & \parbox{0.7cm}{\vspace{0.2cm}$-1$\vspace{0.2cm}} & \parbox{0.7cm}
{\vspace{0.2cm}$-1$\vspace{0.2cm}} & 0 & 0 & 0 \\\hline\hline
graviphoton & $g-1$ & $x_i$ & \parbox{0.7cm}{\vspace{0.2cm}$+\frac{1}{2}$\vspace{0.2cm}} & 
\parbox{0.7cm}{\vspace{0.2cm}$+\frac{1}{2}$\vspace{0.2cm}} & \parbox{0.7cm}
{\vspace{0.2cm}$+\frac{1}{2}$\vspace{0.2cm}} & \parbox{0.7cm}{\vspace{0.2cm}$+
\frac{1}{2}$\vspace{0.2cm}} & \parbox{0.7cm}{\vspace{0.2cm}$+\frac{1}{2}$\vspace{0.2cm}} \\\hline 
 & $g-1$ & $y_i$ & \parbox{0.7cm}{\vspace{0.2cm}$-\frac{1}{2}$\vspace{0.2cm}} & \parbox{0.7cm}
{\vspace{0.2cm}$-\frac{1}{2}$\vspace{0.2cm}} & \parbox{0.7cm}{\vspace{0.2cm}$+
\frac{1}{2}$\vspace{0.2cm}} & \parbox{0.7cm}{\vspace{0.2cm}$+\frac{1}{2}$\vspace{0.2cm}} & 
\parbox{0.7cm}{\vspace{0.2cm}$+\frac{1}{2}$\vspace{0.2cm}} \\\hline \hline
PCO & $g-1$ & $\{s_3\}$ & 0 & 0 &\parbox{0.7cm}{\vspace{0.2cm}$-1$\vspace{0.2cm}} & 0 & 0\\\hline 
 & $g-1$ & $\{s_4\}$ & 0 & 0 & 0 &\parbox{0.7cm}{\vspace{0.2cm}$-1$\vspace{0.2cm}} & 0 \\\hline 
 & $g-1$ & $\{s_5\}$ & 0 & 0 & 0 & 0 &\parbox{0.7cm}{\vspace{0.2cm}$-1$\vspace{0.2cm}} \\\hline 
\end{tabular}
\end{center}
${}$\\[10pt]
In this setup, the amplitude takes the form\footnote{We use the same symbol $\mathcal{F}_g^{(1)}$, since we 
find that this amplitude is actually identical to (\ref{result1a}).}
\begin{align}
\mathcal{F}_g^{(1)}=\langle&\prod_{i=1}^{g-1}e^{-\frac{\varphi}{2}}e^{\frac{i}{2}(\phi_1+\phi_2+\phi_3+\phi_4+
\phi_5)}(x_i)\prod_{i=1}^{g-1}e^{-\frac{\varphi}{2}}e^{\frac{i}{2}(-\phi_1-\phi_2+\phi_3+
\phi_4+\phi_5)}(y_i)\cdot\nonumber\\
\cdot&e^{i(\phi_2+\phi_3)}(z_1)\cdot e^{i(\phi_1-\phi_2)}(z_2)\cdot e^{i(\phi_2-\phi_3)}(z_3)
\cdot e^{-i(\phi_1+\phi_2)}(z_4)\cdot\nonumber\\
\cdot&\prod_a^{\{s_3\}}e^{\varphi}e^{-i\phi_3}\partial X_3(r_a)\prod_a^{\{s_4\}}e^{\varphi}
e^{-i\phi_4}\partial X_4(r_a)\prod_a^{\{s_5\}}e^{\varphi}e^{-i\phi_5}\partial X_5(r_a)\rangle.
\end{align} 
Since the calculations are quite similar to the case just discussed, we restrain from reporting the details of 
the computation, but content ourselves by writing down the final result:
\begin{align}
\mathcal{F}_g^{(1)}&=\int_{\mathcal{M}_g}\langle|\prod_{a=1}^{3g-4}G^-(\mu_a)J^{--}_{K3}(\mu_{3g-3})|^2\int 
|\tilde{G}^+_{K3}(z_3)|^2\int |J_{T^2}(z_2)|^2 \rangle_{int}\nonumber\\
&=\int_{\mathcal{M}_g}\langle|\prod_{a=1}^{3g-3}G^-(\mu_a)|^2\int 
|J_{K3}(z_3)|^2\int |J_{T^2}(z_2)|^2 \rangle_{int}.
\label{result2}
\end{align}
\subsection{4 Scalar Case}\label{4sc}
Finally, the 4 graviscalar case is obtained from the 4 graviton configuration of Section \ref{4gravT2}, by 
exchanging one of the space-time planes with the one of the internal tori for the configuration in 
\ref{GravComb2}. The setup of charges is hence as follows:
\begin{center}
\begin{tabular}{|c|c|c||c|c||c||c|c|}\hline
\textbf{insertion} & \textbf{number} & \textbf{position} & $\phi_1$ & $\phi_2$ & $\phi_3$ & 
$\phi_4$ & $\phi_5$\\\hline\hline
graviton & $1$ & $z_1$ & 0 & \parbox{0.7cm}{\vspace{0.2cm}$+1$\vspace{0.2cm}} & \parbox{0.7cm}
{\vspace{0.2cm}$+1$\vspace{0.2cm}}  & 0 & 0 \\\hline 
 & $1$ & $z_2$ & 0 & \parbox{0.7cm}{\vspace{0.2cm}$+1$\vspace{0.2cm}} & \parbox{0.7cm}
{\vspace{0.2cm}$-1$\vspace{0.2cm}} & 0 & 0  \\\hline 
 & $1$ & $z_3$ & 0 & \parbox{0.7cm}{\vspace{0.2cm}$-1$\vspace{0.2cm}} & \parbox{0.7cm}
{\vspace{0.2cm}$+1$\vspace{0.2cm}} & 0 & 0  \\\hline 
 & $1$ & $z_4$ & 0 & \parbox{0.7cm}{\vspace{0.2cm}$-1$\vspace{0.2cm}} & \parbox{0.7cm}
{\vspace{0.2cm}$-1$\vspace{0.2cm}} & 0 & 0  \\\hline\hline
graviphoton & $g-1$ & $x_i$ & \parbox{0.7cm}{\vspace{0.2cm}$+\frac{1}{2}$\vspace{0.2cm}} 
& \parbox{0.7cm}{\vspace{0.2cm}$+\frac{1}{2}$\vspace{0.2cm}} & \parbox{0.7cm}
{\vspace{0.2cm}$+\frac{1}{2}$\vspace{0.2cm}} & \parbox{0.7cm}{\vspace{0.2cm}$+\frac{1}{2}$
\vspace{0.2cm}} & \parbox{0.7cm}{\vspace{0.2cm}$+\frac{1}{2}$\vspace{0.2cm}} \\\hline 
 & $g-1$ & $y_i$ & \parbox{0.7cm}{\vspace{0.2cm}$-\frac{1}{2}$\vspace{0.2cm}} & \parbox{0.7cm}
{\vspace{0.2cm}$-\frac{1}{2}$\vspace{0.2cm}} & \parbox{0.7cm}{\vspace{0.2cm}$+\frac{1}{2}$
\vspace{0.2cm}} & \parbox{0.7cm}{\vspace{0.2cm}$+\frac{1}{2}$\vspace{0.2cm}} & \parbox{0.7cm}
{\vspace{0.2cm}$+\frac{1}{2}$\vspace{0.2cm}} \\\hline \hline
PCO & $g-1$ & $\{s_3\}$ & 0 & 0 &\parbox{0.7cm}{\vspace{0.2cm}$-1$\vspace{0.2cm}} & 0 & 0\\\hline 
 & $g-1$ & $\{s_4\}$ & 0 & 0 & 0 &\parbox{0.7cm}{\vspace{0.2cm}$-1$\vspace{0.2cm}} & 0 \\\hline 
 & $g-1$ & $\{s_5\}$ & 0 & 0 & 0 & 0 &\parbox{0.7cm}{\vspace{0.2cm}$-1$\vspace{0.2cm}} \\\hline 
\end{tabular}
\end{center}
${}$\\[10pt]
The corresponding amplitude reads 
\begin{align}
\mathcal{F}_g^{(1)}=\langle&\prod_{i=1}^{g-1}e^{-\frac{\varphi}{2}}e^{\frac{i}{2}(\phi_1+\phi_2+\phi_3+\phi_4+
\phi_5)}(x_i)\prod_{i=1}^{g-1}e^{-\frac{\varphi}{2}}e^{\frac{i}{2}(-\phi_1-\phi_2+\phi_3+
\phi_4+\phi_5)}(y_i)\cdot\nonumber\\
\cdot&e^{i(\phi_2+\phi_3)}(z_1)\cdot e^{i(\phi_2-\phi_3)}(z_2)\cdot e^{i(-\phi_2+\phi_3)}(z_3) 
\cdot e^{-i(\phi_2+\phi_3)}(z_4)\cdot\nonumber\\
\cdot&\prod_a^{\{s_3\}}e^{\varphi}e^{-i\phi_3}\partial X_3(r_a)\prod_a^{\{s_4\}}e^{\varphi}
e^{-i\phi_4}\partial X_4(r_a)\prod_a^{\{s_5\}}e^{\varphi}e^{-i\phi_5}\partial X_5(r_a)\rangle,
\nonumber
\end{align} 
which, by similar steps as before, is reduced to the same expression as in eq.(\ref{result2}).

\subsection{Comparison of the Results}
Comparing the final results of Sections \ref{4gravT2}, \ref{2gr2sc} and \ref{4sc}, 
one encounters that indeed all three are the same. 
This implies, that the topological amplitude does not depend 
on the precise details of the field content which is considered, but is only related to more 
general structures such as the compactification and the number of inserted fields. It 
therefore seems an interesting question, whether one can find different topological 
expressions if one varies some of these main properties of the amplitude. 
This question will be studied in the next section.
\section{Different Correlators}\label{ChangeAmpl}
The simplest way to find different correlators than the above ones is to consider different 
numbers of vertex insertions. The two feasible options are:
\begin{itemize}
\item changing the number of scalar and graviton insertions to 3: As we will demonstrate, 
although this will lead to a new topological expression one can show that it is 
identically vanishing.
\item changing the number of graviphotons: Since a change in the number of graviphoton 
insertions implies (via the balancing of the super-ghost charges) also a change in the 
number of PCO insertions, it is more convenient to consider just a different loop order, 
instead of altering the number of fields; we will be more precise below.
\end{itemize}
We will exploit both possibilities in this section.
\subsection{1 Graviton - 2 Scalar Amplitude}\label{3trivial}
Tampering with the number of scalar and graviton insertions raises the main problem of 
how to cancel the charges in the space-time part of the amplitude. One setup, which 
makes this possible is the following
\begin{center}
\begin{tabular}{|c|c|c||c|c||c||c|c|}\hline
\textbf{insertion} & \textbf{number} & \textbf{position} & $\phi_1$ & $\phi_2$ & $\phi_3$ 
& $\phi_4$ & $\phi_5$\\\hline\hline
graviton & $1$ & $z_1$ & \parbox{0.7cm}{\vspace{0.2cm}$+1$\vspace{0.2cm}} & \parbox{0.7cm}
{\vspace{0.2cm}$+1$\vspace{0.2cm}}  & 0 & 0 & 0 \\\hline 
graviscalar & $1$ & $z_2$ & \parbox{0.7cm}{\vspace{0.2cm}$-1$\vspace{0.2cm}} & 0 & 
\parbox{0.7cm}{\vspace{0.2cm}$+1$\vspace{0.2cm}} & 0 & 0  \\\hline 
 & $1$ & $z_3$ & 0 & \parbox{0.7cm}{\vspace{0.2cm}$-1$\vspace{0.2cm}} & \parbox{0.7cm}
{\vspace{0.2cm}$-1$\vspace{0.2cm}} & 0 & 0  \\\hline 
graviphoton & $g-1$ & $x_i$ & \parbox{0.7cm}{\vspace{0.2cm}$+\frac{1}{2}$\vspace{0.2cm}} 
& \parbox{0.7cm}{\vspace{0.2cm}$+\frac{1}{2}$\vspace{0.2cm}} & \parbox{0.7cm}{\vspace{0.2cm}
$+\frac{1}{2}$\vspace{0.2cm}} & \parbox{0.7cm}{\vspace{0.2cm}$+\frac{1}{2}$\vspace{0.2cm}} 
& \parbox{0.7cm}{\vspace{0.2cm}$+\frac{1}{2}$\vspace{0.2cm}} \\\hline 
 & $g-1$ & $y_i$ & \parbox{0.7cm}{\vspace{0.2cm}$-\frac{1}{2}$\vspace{0.2cm}} & 
\parbox{0.7cm}{\vspace{0.2cm}$-\frac{1}{2}$\vspace{0.2cm}} & \parbox{0.7cm}
{\vspace{0.2cm}$+\frac{1}{2}$\vspace{0.2cm}} & \parbox{0.7cm}{\vspace{0.2cm}$+
\frac{1}{2}$\vspace{0.2cm}} & \parbox{0.7cm}{\vspace{0.2cm}$+\frac{1}{2}$\vspace{0.2cm}} \\\hline 
PCO & $g-1$ & $\{s_3\}$ & 0 & 0 &\parbox{0.7cm}{\vspace{0.2cm}$-1$\vspace{0.2cm}} & 0 & 0\\\hline 
 & $g-1$ & $\{s_4\}$ & 0 & 0 & 0 &\parbox{0.7cm}{\vspace{0.2cm}$-1$\vspace{0.2cm}} & 0 \\\hline 
 & $g-1$ & $\{s_5\}$ & 0 & 0 & 0 & 0 &\parbox{0.7cm}{\vspace{0.2cm}$-1$\vspace{0.2cm}} \\\hline 
\end{tabular}
\end{center}
${}$\\[10pt]
Although all such amplitudes turn out to vanish, we still present here the computation for pedagogical reasons, showing in particular in a concrete example the gauge independence of gauge choice for the PCO positions. The amplitude written down in this setup is 
\begin{align}
\mathcal{F}_g^{(2)}=\langle&\prod_{i=1}^{g-1}e^{-\frac{\varphi}{2}}e^{\frac{i}{2}(\phi_1+\phi_2+\phi_3+
\phi_4+\phi_5)}(x_i)\prod_{i=1}^{g-1}e^{-\frac{\varphi}{2}}e^{\frac{i}{2}(-\phi_1-
\phi_2+\phi_3+\phi_4+\phi_5)}(y_i)\cdot\nonumber\\
\cdot&e^{i(\phi_1+\phi_2)}(z_1)\cdot e^{i(-\phi_1+\phi_3)}(z_2)\cdot e^{-i(\phi_2+
\phi_3)}(z_3)\cdot\nonumber\\
\cdot&\prod_a^{\{s_3\}}e^{\varphi}e^{-i\phi_3}\partial X_3(r_a)\prod_a^{\{s_4\}}
e^{\varphi}e^{-i\phi_4}\partial X_4(r_a)\prod_a^{\{s_5\}}e^{\varphi}e^{-i\phi_5}
\partial X_5(r_a)\rangle.
\end{align} 

This correlation function can be treated in quite the same way as the other ones considered above, 
however, there is a small subtlety concerning the gauge condition. Choosing the gauge 
\begin{align}
\sum_{a=1}^{3g-3}r_a=\sum_{i=1}^{g-1}y_i-z_1+z_2+2\Delta,\label{gaugecondit}
\end{align}
the result of the amplitude, following the same steps as before, is given by
\begin{align}
\mathcal{F}_g^{(2)}=&\int_{\mathcal{M}_{g}}\langle|\prod_{a=1}^{3g-3}G^-(\mu_{a})|^2
\int |J_{T^2}(z_3)|^2\rangle_{int}.\label{resultvan0}
\end{align}
However, choosing for example 
\begin{align}
\sum_{a=1}^{3g-3}r_a=\sum_{i=1}^{g-1}y_i-z_1+z_3+2\Delta,\label{gaugecondit2}
\end{align}
the spin structure sum gives a vanishing result, implying that the correlation function is zero. 
Therefore, the topological expression (\ref{resultvan0}) has to vanish as well, which can 
indeed be shown as follows.
In order to soak all the zero modes of the 3rd plane, the $G^-$ of (\ref{resultvan0}) have to be distributed 
in the following way among $T^2$ and $K3$
\begin{align}
\mathcal{F}_g^{(2)}=&\int_{\mathcal{M}_{g}}\langle|\prod_{a=1}^{g-1}G_{T^2}^-(\mu_{a})
\prod_{a=g}^{3g-3}G_{K3}^-(\mu_{a})|^2\int 
|J_{T^2}(z_3)|^2\rangle_{int}.
\end{align}
Using the OPE relations (\ref{N4K3}), one can express one of the $G^-_{K3}$'s as the contour integral
$G^-_{K3}=\oint\tilde{G}^+_{K3}J^{--}_{K3}$. Deforming the contour, the only possible operators that 
can be encircled are $G_{K3}^-$ yielding however zero residue. This entails
\begin{align}
\mathcal{F}_g^{(2)}=0,\label{zero}
\end{align}
consistently with the alternative vanishing through the different gauge choice (\ref{gaugecondit2}).
\subsection{Amplitudes at Genus $g+1$}\label{genusg+1}
As explained already before, instead of altering the number of graviphotons taking part 
in the scattering process it is equivalent to change the genus of the world-sheet. In other words, 
the same vertex insertions as in Section \ref{Section4dscattering} 
are considered, with the only difference, that an additional handle is attached to the surface.
Of course, the kinematics of all fields involved has to be altered, as well:
\begin{center}
\begin{tabular}{|c|c|c||c|c||c||c|c|}\hline
\textbf{insertion} & \textbf{number} & \textbf{position} & $\phi_1$ & $\phi_2$ & $\phi_3$ & 
$\phi_4$ & $\phi_5$\\\hline\hline
graviscalars & $1$ & $z_1$ & 0 & \parbox{0.7cm}{\vspace{0.2cm}$+1$\vspace{0.2cm}} & 
\parbox{0.7cm}{\vspace{0.2cm}$+1$\vspace{0.2cm}} & 0 & 0 \\\hline 
 & $1$ & $z_2$ & 0 & \parbox{0.7cm}{\vspace{0.2cm}$-1$\vspace{0.2cm}} & \parbox{0.7cm}
{\vspace{0.2cm}$+1$\vspace{0.2cm}} & 0 & 0  \\\hline 
gravitons & $1$ & $z_3$ & \parbox{0.7cm}{\vspace{0.2cm}$+1$\vspace{0.2cm}} & \parbox{0.7cm}
{\vspace{0.2cm}$+1$\vspace{0.2cm}} & 0 & 0 & 0  \\\hline 
 & $1$ & $z_4$ & \parbox{0.7cm}{\vspace{0.2cm}$-1$\vspace{0.2cm}} & \parbox{0.7cm}
{\vspace{0.2cm}$-1$\vspace{0.2cm}} & 0 & 0 & 0  \\\hline 
graviphoton & $g-1$ & $x_i$ & \parbox{0.7cm}{\vspace{0.2cm}$+\frac{1}{2}$\vspace{0.2cm}} 
& \parbox{0.7cm}{\vspace{0.2cm}$+\frac{1}{2}$\vspace{0.2cm}} & \parbox{0.7cm}
{\vspace{0.2cm}$+\frac{1}{2}$\vspace{0.2cm}} & \parbox{0.7cm}{\vspace{0.2cm}
$+\frac{1}{2}$\vspace{0.2cm}} & \parbox{0.7cm}{\vspace{0.2cm}$+\frac{1}{2}$\vspace{0.2cm}} \\
\hline 
 & $g-1$ & $y_i$ & \parbox{0.7cm}{\vspace{0.2cm}$-\frac{1}{2}$\vspace{0.2cm}} 
& \parbox{0.7cm}{\vspace{0.2cm}$-\frac{1}{2}$\vspace{0.2cm}} & \parbox{0.7cm}{\vspace{0.2cm}
$+\frac{1}{2}$\vspace{0.2cm}} & \parbox{0.7cm}{\vspace{0.2cm}$+\frac{1}{2}$\vspace{0.2cm}} 
& \parbox{0.7cm}{\vspace{0.2cm}$+\frac{1}{2}$\vspace{0.2cm}} \\\hline 
PCO & $g+1$ & $\{s_3\}$ & 0 & 0 &\parbox{0.7cm}{\vspace{0.2cm}$-1$\vspace{0.2cm}} & 0 & 0\\\hline 
 & $g-1$ & $\{s_4\}$ & 0 & 0 & 0 &\parbox{0.7cm}{\vspace{0.2cm}$-1$\vspace{0.2cm}} & 0 \\\hline 
 & $g-1$ & $\{s_5\}$ & 0 & 0 & 0 & 0 &\parbox{0.7cm}{\vspace{0.2cm}$-1$\vspace{0.2cm}} \\\hline 
\end{tabular}
\end{center}
${}$\\[10pt]
The two additional PCO are present because of the integration over the super-moduli of 
the new surface, and are used to balance the $\phi_3$ charge from the vertex insertions. 
Then, the amplitude can be written as
\begin{align}
\mathcal{F}_g^{(3)}=\langle&\prod_{i=1}^{g-1}e^{-\frac{\varphi}{2}}e^{\frac{i}{2}(\phi_1+\phi_2+\phi_3+
\phi_4+\phi_5)}(x_i)\prod_{i=1}^{g-1}e^{-\frac{\varphi}{2}}e^{\frac{i}{2}(-\phi_1-\phi_2+
\phi_3+\phi_4+\phi_5)}(y_i)\cdot\nonumber\\
\cdot&e^{i(\phi_2+\phi_3)}(z_1)\cdot e^{i(-\phi_2+\phi_3)}(z_2)\cdot e^{i(\phi_1+\phi_2)}
(z_3)\cdot e^{-i(\phi_1+\phi_2)}(z_4)\cdot\nonumber\\
\cdot&\prod_a^{\{s_3\}}e^{\varphi}e^{-i\phi_3}\partial X_3(r_a)\prod_a^{\{s_4\}}
e^{\varphi}e^{-i\phi_4}\partial X_4(r_a)\prod_a^{\{s_5\}}e^{\varphi}e^{-i\phi_5}
\partial X_5(r_a)\rangle.
\end{align} 

Since this amplitude is a bit non-standard because the number of graviphotons is $2g-2$ on a genus $g+1$
surface, we give some details of the calculation below.
By taking all the contractions we find 
\begin{align}
\mathcal{F}_g^{(3)}=&\frac{\vartheta_s\!\!\left(\frac{1}{2}\sum_{i=1}^{g-1}(x_i-y_i)+z_3-z_4\right)
\vartheta_s\!\!\left(\frac{1}{2}\sum_{i=1}^{g-1}(x_i-y_i)+z_1-z_2+z_3-z_4\right)}
{ \vartheta_s\!\!\left(\frac{1}{2}\sum_{i=1}^{g-1}(x_i+y_i)-\sum_{a=1}^{3g-1}r_a+
2\Delta\right)\prod_{a<b=1}^{3g-1}E(r_a,r_b)\prod_{a=1}^{3g-1}\sigma^2(r_a)}\cdot\nonumber\\
&\cdot\frac{\vartheta_s\!\!\left(\frac{1}{2}\sum_{i=1}^{g-1}(x_i+y_i)+z_1+z_2-
\sum_{a}^{\{s_3\}}r_a\right)\vartheta_{h_4,s}\!\!\left(\frac{1}{2}\sum_{i=1}^{g-1}
(x_i+y_i)-\sum_{a}^{\{s_4\}}r_a\right)}{\prod_{a}^{\{s_3\}}E(r_a,z_1)E(r_a,z_2)
\prod_{j=1}^{g-1}E(x_i,z_4)E(y_i,z_3)E^2(z_3,z_4)E(z_1,z_4)E(z_2,z_3)}\cdot\nonumber\\
&\cdot\vartheta_{h_5,s}\!\!\left(\frac{1}{2}\sum_{i=1}^{g-1}(x_i+y_i)-
\sum_{a}^{\{s_5\}}r_a\right)E(z_1,z_3)E(z_2,z_4)\prod_{i=1}^{g-1}\sigma(x_i)
\sigma(y_i)\cdot\nonumber\\
&\cdot\prod_{a<b}^{\{s_3\}}E(r_a,r_b)\prod_{a<b}^{\{s_4\}}E(r_a,r_b)
\prod_{a<b}^{\{s_5\}}E(r_a,r_b)\prod_{i<j}^{g-1}E(x_i,x_j)E(y_i,y_j)\cdot\nonumber\\
&\cdot\prod_{j=1}^{g-1}E(x_i,z_1)E(x_i,z_3)E(y_i,z_2)E(y_i,z_4)\prod_{a}^{\{s_3\}}
\partial X_3(r_a)\prod_{a}^{\{s_4\}}\partial X_4(r_a)\prod_{a}^{\{s_5\}}\partial X_5(r_a)\nonumber
\end{align}
In this case, the gauge condition takes the form
\begin{align}
&\frac{1}{2}\sum_{i=1}^{g-1}(x_i-y_i)+z_3-z_4=\frac{1}{2}\sum_{i=1}^{g-1}(x_i+y_i)-
\sum_{a=1}^{3g-1}r_a+2\Delta\nonumber\\
&\Rightarrow \sum_{a=1}^{3g-1}r_a=\sum_{i=1}^{g-1}y_i-z_3+z_4+2\Delta,\label{gaugecondg1}
\end{align}
which results in a cancellation of the first $\vartheta$-function in the numerator against 
the one in the denominator. After performing the spin structure sum and deploying 
bosonization identities, the result is given by ($\alpha$ is again an arbitrary position)
\begin{align}
\mathcal{F}_g^{(3)}=&\frac{\langle\prod_{a}^{\{s_3\}}\bar{\psi}_3\partial X_3(r_a)\psi_3(\alpha)\rangle\cdot
\langle\prod_a^{\{s_4\}}\bar{\psi}_4(r_a)\partial X_4\bar{\psi}_4(z_2)\rangle\cdot \langle
\prod_a^{\{s_5\}}\bar{\psi}_5\partial X_5(r_a)\bar{\psi}_5(z_2)\rangle}{\langle 
\prod_{a=1}^{3g-1}b(r_a)b(z_2)\rangle)}\cdot\nonumber\\
&\cdot\text{det}\omega_i(x_i,z_1,z_3)\text{det}\omega_i(y_i,z_2,z_4).
\end{align}
Using the expressions for the $N=4$ superconformal algebra, this may also be written as
\begin{align}
\mathcal{F}_g^{(3)}=&\frac{\langle\prod_{a}^{\{3g-1\}}G^-(r_a)\psi_3(\alpha)J^{--}(z_2)\rangle}{\langle 
\prod_{a=1}^{3g-1}b(r_a)b(z_2)\rangle)}\cdot\text{det}\omega_i(x_i,z_1,z_3)\text{det}
\omega_i(y_i,z_2,z_4)
\label{fgthree}
\end{align}
One immediately sees, that there are no poles or zeros if one of the $r_a$ approaches 
$z_2$, which means, that the above expression is independent of $z_2$. This entails, 
that one can transport the $(3g-1)$ $G_-$ and the $J^{--}$ to the $3g$ Beltrami 
differentials. The integral over $x_i,y_i,z_1,z_2,z_3,z_4$ can be performed without 
obstacles (including the right-moving sector) yielding a $(\text{det}(\text{Im}\tau))^2$ which is needed to cancel 
the bosonic space-time part. The remaining internal part can then be written as 
(a complete antisymmetrization of the Beltrami differentials is understood):
\begin{align}
\mathcal{F}_g^{(3)}=&\int_{\mathcal{M}_{g+1}}\langle|\prod_{a=1}^{3g-1}G^-(\mu_{a})
J^{--}_{K3}(\mu_{3g})\psi_3(\alpha)|^2\rangle_{int}.\label{ampg1ext}
\end{align}
This amplitude seems to be the `minimal' topological amplitude, in the sense, 
that it contains the minimum number of insertions of additional SCFT-operators, 
while still being non-trivial and topological. As we will see in the next section, 
this amplitude comprises even more interesting and pleasant features.
\section{Duality Mapping}\label{sect:dual}
After having established a number of topological amplitudes on the type IIA side at arbitrary loop order, 
the question is to which order they are mapped on the 
heterotic side if the appropriate duality mapping is applied. Indeed, we will find that most 
amplitudes, especially the 6d couplings of \cite{Berkovits:1994vy} are mapped to 
higher loop orders, whose computation seems to be out of reach.

For simplicity, we will consider in this chapter also for the 4d amplitudes 
the decompactification limit ($T^2\to \mathbb{C}$), so that the duality map stays in 6 dimensions. 
This strategy has the further advantage, that we 
do not have to deal with moduli insertions, but only with gravitons and graviphotons.
According to \cite{Kiritsis:2000zi} the only relevant information in this case is the behavior 
of the 6d string coupling constant $g_s$ and the metric $G_{\mu\nu}$, which are given by
\begin{align}
g_s^{\text{IIA}}&= \frac{1}{g_s^{\text{HET}}},\\
(G^{\text{IIA}})_{\mu\nu}&= \frac{(G^{\text{HET}})_{\mu\nu}}{(g_s^{\text{HET}})^2}.
\end{align}
For further convenience, we then compute the following building blocks of the amplitudes.

First, every amplitude contains the corresponding integration measure
\begin{align}
\sqrt{\text{det}G^{\text{IIA}}}= (g_s^{\text{HET}})^{-6}\sqrt{\text{det}G^{\text{HET}}},\nonumber
\end{align}
since the determinant of the metric in 6 dimensions 
behaves like the 6th power of $G_{\mu\nu}$. Furthermore, the behavior of 
powers of the Riemann tensor is given by
\begin{align}
&R^4_{\big|\text{IIA}}= (g_s^{\text{HET}})^{8}R^4_{\big|\text{HET}},\nonumber\\
&R^3_{\big|\text{IIA}}= (g_s^{\text{HET}})^{6}R^3_{\big|\text{HET}},\nonumber
\end{align}
since the Riemann tensor scales in the same way as the metric 
and there are 8 (6) inverse metrics necessary to contract all indices of four (three) Riemann 
tensors. Similarly, the behavior of the graviphoton field strength reads:
\begin{align}
T^2_{\big|\text{IIA}}= (g_s^{\text{HET}})^2T^2_{\big|\text{HET}}.\nonumber
\end{align}

Using the above relations, the mapping of the amplitudes of the previous chapters on the heterotic side is given by:
\begin{itemize}
\item $(g_s^{\text{IIA}})^{2g-2}\int\sqrt{\text{det} G^{\text{IIA}}} 
(R^4T^{4g-4})_{\big|\text{IIA}} = (g_s^{\text{HET}})^{2g}\int\sqrt{\text{det} 
G^{\text{HET}}} (R^4T^{4g-4})_{\big|\text{HET}}$\\
This means, that the type II $g$-loop amplitude $\mathcal{F}_g^{(\text{6d})}$ (which we studied in Section 
\ref{SectTopAmpK3}) is mapped to a $(g+1)$-loop amplitude on the heterotic side.
\item $(g_s^{\text{IIA}})^{2g-2}\int\sqrt{\text{det} G^{\text{IIA}}} 
(R^4T^{2g-2})_{\big|\text{IIA}} = (g_s^{\text{HET}})^2\int\sqrt{\text{det} 
G^{\text{HET}}} (R^4T^{2g-2})_{\big|\text{HET}}$\\
This entails that the type II amplitude $\mathcal{F}_g^{(1)}$ of Section 
\ref{Section4dscattering}  is mapped to a 2-loop amplitude on the heterotic side.
\item $(g_s^{\text{IIA}})^{2g-2}\int\sqrt{\text{det} G^{\text{IIA}}} 
(R^3T^{2g-2})_{\big|\text{IIA}} = \int\sqrt{\text{det} G^{\text{HET}}} 
(R^3T^{2g-2})_{\big|\text{HET}}$\\
In other words, the trivially vanishing $\mathcal{F}_g^{(2)}$ found in Section \ref{3trivial} 
get mapped to a 1-loop computation in the heterotic theory.
\item $(g_s^{\text{IIA}})^{2g}\int\sqrt{\text{det} G^{\text{IIA}}} 
(R^4T^{2g-2})_{\big|\text{IIA}} = \int\sqrt{\text{det} G^{\text{HET}}} 
(R^4T^{2g-2})_{\big|\text{HET}}$\\
Finally, the $\mathcal{F}_g^{(3)}$ of Section \ref{genusg+1} get mapped to 1-loop 
in the heterotic side.
\end{itemize}
Since higher order corrections than 1-loop calculations do not seem feasible and the 
amplitude with $R^3$ was already found to be trivially vanishing in the type IIA 
theory, we will restrict ourselves to the heterotic computation of the $\mathcal{F}_g^{(3)}$, 
which will be performed in the next section.
\section{Heterotic on $T^6$}\label{sect:hetT6}
The aim of this section is - as advertised - to compute the heterotic dual $\mathcal{F}_g^{(3,\text{HET})}$ 
of the amplitude $\mathcal{F}_g^{(3)}$ (\ref{ampg1ext}). To this end the right-moving sector has also to be 
taken into 
account, which we have disregarded up to now because we considered it to behave in the same way as the left 
moving one.\footnote{This is only up to sign changes of the internal theory because of the chiralities in 
the type IIA theory.}
The computation on the type II side revealed, that the amplitude is topological for 
the left and right sectors separately. This entails, that there are various 
possibilities of how the left-moving computation can be completed by adding right-moving components. 
For concreteness, let us complete the amplitudes in such a way, that the 
scattering occurs between two gravitons and two graviscalars (from a 4-dimensional 
point of view). Still there are two different possibilities corresponding to 
type IIA and type IIB superstring theory. Since in 6 dimensions, type IIB is not dual to heterotic, 
in the following we will concentrate only on type IIA.

\subsection{Type IIA Setting}\label{Completion1}
\begin{center}
\begin{tabular}{|c|c|c||c|c||c||c|c||c|c||c||c|c|}\hline
\textbf{field} & \textbf{nr.} & \textbf{pos.} & \parbox{0.4cm}{\vspace{0.2cm}$\phi_1$\vspace{0.2cm}} & 
\parbox{0.4cm}{\vspace{0.2cm}$\phi_2$\vspace{0.2cm}} & \parbox{0.4cm}{\vspace{0.2cm}$\phi_3$\vspace{0.2cm}} & 
\parbox{0.4cm}{\vspace{0.2cm}$\phi_4$\vspace{0.2cm}} 
& \parbox{0.4cm}{\vspace{0.2cm}$\phi_5$\vspace{0.2cm}} & \parbox{0.4cm}{\vspace{0.2cm}
$\tilde{\phi_1}$\vspace{0.2cm}} & \parbox{0.4cm}{\vspace{0.2cm}$\tilde{\phi_2}$\vspace{0.2cm}} & 
\parbox{0.4cm}{\vspace{0.2cm}$\tilde{\phi_3}$\vspace{0.2cm}} & 
\parbox{0.4cm}{\vspace{0.2cm}$\tilde{\phi_4}$\vspace{0.2cm}} & \parbox{0.4cm}{\vspace{0.2cm}$\tilde{\phi_5}
$\vspace{0.2cm}}\\\hline\hline
scalar & $1$ & $z_1$ & 0 & \parbox{0.7cm}{\vspace{0.2cm}$+1$\vspace{0.2cm}} & 
\parbox{0.7cm}{\vspace{0.2cm}$+1$\vspace{0.2cm}} & 0 & 0 & 0 & \parbox{0.7cm}{\vspace{0.2cm}$+1$\vspace{0.2cm}} & 
\parbox{0.7cm}{\vspace{0.2cm}$-1$\vspace{0.2cm}} 
& 0 & 0 \\\hline 
 & $1$ & $z_2$ & 0 & \parbox{0.7cm}{\vspace{0.2cm}$-1$\vspace{0.2cm}} & \parbox{0.7cm}{\vspace{0.2cm}$+1$
\vspace{0.2cm}} & 0 & 0  & 0 & \parbox{0.7cm}{\vspace{0.2cm}$-1$
\vspace{0.2cm}} & \parbox{0.7cm}{\vspace{0.2cm}$-1$\vspace{0.2cm}} & 0 & 0  \\\hline 
graviton & $1$ & $z_3$ & \parbox{0.7cm}{\vspace{0.2cm}$+1$\vspace{0.2cm}} & \parbox{0.7cm}{\vspace{0.2cm}
$+1$\vspace{0.2cm}} & 0 & 0 & 0 & \parbox{0.7cm}{\vspace{0.2cm}$+1$\vspace{0.2cm}} & \parbox{0.7cm}
{\vspace{0.2cm}$+1$\vspace{0.2cm}} & 0 & 0 & 0 \\\hline 
& $1$ & $z_4$ & \parbox{0.7cm}{\vspace{0.2cm}$-1$\vspace{0.2cm}} & \parbox{0.7cm}{\vspace{0.2cm}$-1$
\vspace{0.2cm}} & 0 & 0 & 0 & \parbox{0.7cm}{\vspace{0.2cm}$-1$
\vspace{0.2cm}} & \parbox{0.7cm}{\vspace{0.2cm}$-1$\vspace{0.2cm}} & 0 & 0 & 0 \\\hline 
graviph. & $g-1$ & $x_i$ & \parbox{0.7cm}{\vspace{0.2cm}$+\frac{1}{2}$\vspace{0.2cm}} & 
\parbox{0.7cm}{\vspace{0.2cm}$+\frac{1}{2}$\vspace{0.2cm}} & \parbox{0.7cm}{\vspace{0.2cm}$+\frac{1}{2}$
\vspace{0.2cm}} & \parbox{0.7cm}{\vspace{0.2cm}$+
\frac{1}{2}$\vspace{0.2cm}} & \parbox{0.7cm}{\vspace{0.2cm}$+\frac{1}{2}$\vspace{0.2cm}}  
& \parbox{0.7cm}{\vspace{0.2cm}$+\frac{1}{2}$\vspace{0.2cm}} & \parbox{0.7cm}
{\vspace{0.2cm}$+\frac{1}{2}$\vspace{0.2cm}} & \parbox{0.7cm}{\vspace{0.2cm}$-\frac{1}{2}$\vspace{0.2cm}} & 
\parbox{0.7cm}{\vspace{0.2cm}$-\frac{1}{2}$
\vspace{0.2cm}} & \parbox{0.7cm}{\vspace{0.2cm}$-\frac{1}{2}$\vspace{0.2cm}}\\\hline 
 & $g-1$ & $y_i$ & \parbox{0.7cm}{\vspace{0.2cm}$-\frac{1}{2}$\vspace{0.2cm}} & 
\parbox{0.7cm}{\vspace{0.2cm}$-\frac{1}{2}$\vspace{0.2cm}} & \parbox{0.7cm}
{\vspace{0.2cm}$+\frac{1}{2}$\vspace{0.2cm}} & \parbox{0.7cm}{\vspace{0.2cm}$+
\frac{1}{2}$\vspace{0.2cm}} & \parbox{0.7cm}{\vspace{0.2cm}$+\frac{1}{2}$\vspace{0.2cm}}& 
\parbox{0.7cm}{\vspace{0.2cm}$-\frac{1}{2}$\vspace{0.2cm}} & \parbox{0.7cm}
{\vspace{0.2cm}$-\frac{1}{2}$\vspace{0.2cm}} & \parbox{0.7cm}{\vspace{0.2cm}$-
\frac{1}{2}$\vspace{0.2cm}} & \parbox{0.7cm}{\vspace{0.2cm}$-\frac{1}{2}$
\vspace{0.2cm}} & \parbox{0.7cm}{\vspace{0.2cm}$-\frac{1}{2}$\vspace{0.2cm}} \\\hline 
PCO & $g+1$ & $\{s_3\}$ & 0 & 0 &\parbox{0.7cm}{\vspace{0.2cm}$-1$
\vspace{0.2cm}} & 0 & 0 & 0 & 0 &\parbox{0.7cm}{\vspace{0.2cm}$+1$
\vspace{0.2cm}} & 0 & 0\\\hline 
 & $g-1$ & $\{s_4\}$ & 0 & 0 & 0 &\parbox{0.7cm}{\vspace{0.2cm}$-1$
\vspace{0.2cm}} & 0 & 0 & 0 & 0 &\parbox{0.7cm}{\vspace{0.2cm}$+1$
\vspace{0.2cm}} & 0 \\\hline 
 & $g-1$ & $\{s_5\}$ & 0 & 0 & 0 & 0 &\parbox{0.7cm}
{\vspace{0.2cm}$-1$\vspace{0.2cm}} & 0 & 0 & 0 & 0 &\parbox{0.7cm}{\vspace{0.2cm}$+1$\vspace{0.2cm}}\\\hline 
\end{tabular}
\end{center}
${}$\\[10pt]
In order to figure out the vertex combination for the heterotic side, especially the 
scalar insertions have to be precisely identified and the duality map has to be applied:
\begin{itemize}
\item scalars and gravitons\\
In terms of $0$-ghost picture vertex operators of the type II theory, the scalar and graviton 
insertions read 
\begin{align}
&(\partial X^3-ip_2\psi^2\psi^3)(\bar{\partial} \bar{X}^3-ip_2\psi^2\bar{\psi}^3)
e^{ip_2X^2}(z_1),\nonumber\\
&(\partial X^3-i\bar{p}_2\bar{\psi}^2\psi^3)(\bar{\partial} \bar{X}^3-i\bar{p}_2
\bar{\psi}^2\bar{\psi}^3)e^{i\bar{p}_2\bar{X}^2}(z_2),\nonumber\\
&(\partial X^2-ip_1\psi^1\psi^2)(\bar{\partial} X^2-ip_1\psi^1\psi^2)e^{ip_1X^1}(z_3),\nonumber\\
&(\partial \bar{X}^2-i\bar{p}_2\bar{\psi}^2\bar{\psi}^1)(\bar{\partial} \bar{X}^2-i
\bar{p}_2\bar{\psi}^2\bar{\psi}^1)e^{i\bar{p}_2\bar{X}^2}(z_4).\nonumber
\end{align}
The following corresponding momenta and helicities can then be extracted:
\begin{center}
\begin{tabular}{|c|c|c||c|c|c||c|c|}\hline
$\phi_1$ & $\phi_2$& $\phi_3$ & \parbox{0.4cm}{\vspace{0.1cm}$\tilde{\phi}_1$\vspace{0.1cm}} & \parbox{0.4cm}
{\vspace{0.1cm}$\tilde{\phi}_2$\vspace{0.1cm}} & 
\parbox{0.4cm}{\vspace{0.1cm}$\tilde{\phi}_3$\vspace{0.1cm}} & \textbf{momentum} & 
\textbf{helicity} \\\hline\hline
$0$ & $+$ & $+$ & $0$ & $+$ & $-$ & $p_2$ & $h_{3,\bar{3}}$ \\\hline
$0$ & $-$ & $+$ & $0$ & $-$ & $-$ & $\bar{p}_2$ & $h_{3,\bar{3}}$ \\\hline
$+$ & $+$ & $0$ & $+$ & $+$ & $0$ & $p_1$ & $h_{2,2}$ \\\hline
$-$ & $-$ & $0$ & $-$ & $-$ & $0$ & $\bar{p}_2$ & $h_{\bar{1},\bar{1}}$ \\\hline
\end{tabular}
\end{center}
The last two vertices are self-dual gravitons of the 4d supergravity 
multiplet, which is still true after the duality map. The first two insertions 
are however moduli of the compact torus. Since the first part of these vertices without 
momenta reads $\partial X^3\bar{\partial}\bar{X}^3$, which corresponds to 
$(1,1)$ form, it is clear that these graviscalars are the $T^2$ K\"ahler 
moduli. Upon Type IIA -- Heterotic duality, they are mapped to 
the dilaton on the heterotic side \cite{Kiritsis:2000zi}. The corresponding heterotic vertices
therefore read
\begin{center}
\begin{tabular}{|c|c||c|c||c|c|}\hline
\textbf{field} & \textbf{position} & \textbf{mom.} & \textbf{helicity} & 
\textbf{vertex} \\\hline\hline
dilaton & $z_1$ & $p_2$ & \parbox{0.4cm}{\vspace{0.2cm}$-$\vspace{0.2cm}} & 
$(\partial Z^\mu-ip_2\psi^2\chi^\mu)\bar{\partial}Z_\mu e^{ip_2 X^2}$ \\\hline
& $z_2$ & $\bar{p}_2$ & \parbox{0.4cm}{\vspace{0.2cm}$-$\vspace{0.2cm}} & 
$(\partial Z^\nu-i\bar{p}_2\bar{\psi}^2\chi^\nu)\bar{\partial}Z_\nu e^{i\bar{p}_2 \bar{X}^2}$ \\\hline
graviton & $z_3$ & $p_1$ & \parbox{0.7cm}{\vspace{0.2cm}$h_{2,2}$\vspace{0.2cm}} & 
$(\partial X^2-ip_1\psi^1\psi^2)\bar{\partial}X^2 e^{ip_1 X^1}$ \\\hline
 & $z_4$ & $\bar{p}_2$ & \parbox{0.7cm}{\vspace{0.2cm}$h_{\bar{1},\bar{1}}$\vspace{0.2cm}} & 
$(\partial \bar{X}^1-i\bar{p}_2\bar{\psi}^2\bar{\psi}^1)
\bar{\partial}\bar{X}^1 e^{i\bar{p}_2 \bar{X}^2}$ \\\hline
\end{tabular}
\end{center}
\item RR graviphotons
\begin{center}
\begin{tabular}{|c|c||c|c||c||c|c||c|c||c||c|c|}\hline
label & number & $\phi_1$ & $\phi_2$ & $\phi_3$ & $\phi_4$ & $\phi_5$ & 
\parbox{0.4cm}{\vspace{0.1cm}$\tilde{\phi}_1$\vspace{0.1cm}} & \parbox{0.4cm}
{\vspace{0.1cm}$\tilde{\phi}_2$\vspace{0.1cm}} & \parbox{0.4cm}{\vspace{0.1cm}
$\tilde{\phi}_3$\vspace{0.1cm}} & \parbox{0.4cm}{\vspace{0.2cm}$\tilde{\phi}_4$
\vspace{-0.2cm}} & \parbox{0.4cm}{\vspace{0.1cm}$\tilde{\phi}_5$\vspace{0.1cm}} \\ \hline\hline
$x_i$ & $g-1$ & $+$ & $+$ & $+$ & $+$ & $+$ & $+$ & $+$ & $-$ & $-$ & $-$\\ \hline
$y_i$ & $g-1$ & $-$ & $-$ & $+$ & $+$ & $+$ & $-$ & $-$ & $-$ & $-$ & $-$\\ \hline
\end{tabular}
\end{center}
The relevant part of the graviphoton vertex operator in 4 dimensions is given by
\begin{align}
:e^{-\frac{\varphi+\tilde{\varphi}}{2}}p_\mu\epsilon_\nu\left[S^a 
{(\sigma^{\mu\nu})_a}^b\tilde{S}^b\mathcal{S}\tilde{\mathcal{S}}+S_{\dot{a}}
{(\bar{\sigma}^{\mu\nu})^{\dot{a}}}_{\dot{b}}\tilde{S}^{\dot{b}}
\bar{\mathcal{S}}\bar{\tilde{\mathcal{S}}}\right]e^{ip\cdot Z}:.
\end{align}
In terms of the helicity combinations, the spin fields $S_a$ are given by
\begin{align}
&S_a=\left(\begin{array}{c}e^{\frac{i}{2}(\phi_1+\phi_2+\phi_3)} \\ 
e^{\frac{i}{2}(\phi_1-\phi_2-\phi_3)} \\ e^{\frac{i}{2}(-\phi_1+\phi_2-\phi_3)} \\ 
e^{\frac{i}{2}(-\phi_1-\phi_2+\phi_3)}\end{array}\right),\\
&\tilde{S}_a=\left(\begin{array}{c}e^{-\frac{i}{2}(\phi_1+\phi_2+\phi_3)} \\ 
e^{\frac{i}{2}(-\phi_1+\phi_2+\phi_3)} \\ e^{\frac{i}{2}(\phi_1-\phi_2+\phi_3)} \\ 
e^{\frac{i}{2}(\phi_1+\phi_2-\phi_3)}\end{array}\right).
\end{align}
One can now analyze the two different vertices entering in the setting of the table:
\begin{itemize}
\item vertex 1:\\
The matrix $p_\mu \epsilon_\nu{(\sigma^{\mu\nu})_a}^b$ has to read
\begin{align}
p_\mu \epsilon_\nu{(\sigma^{\mu\nu})_a}^b\sim\left(
\begin{array}{cccc} 0 & 0 & 0 & 1 \\ 0 & 0 & 0 & 0 \\ 0 & 0 & 0 & 0 \\ 0 & 0 & 0 & 0
\end{array}\right).
\end{align}
The relevant (reduced) Lorentz generators are those which have a 
non-vanishing $(1,4)$ entry, namely:
\begin{align}
&\sigma^{02}=\frac{1}{2}\left(\begin{array}{cccc} 0 & 0 & 0 & -1 \\ 0 & 0 & 1 & 0 \\ 
0 & 1 & 0 & 0 \\ -1 & 0 & 0 & 0\end{array}\right), &&\sigma^{03}=\frac{i}{2}
\left(\begin{array}{cccc} 0 & 0 & 0 & 1 \\ 0 & 0 & 1 & 0 \\ 0 & -1 & 0 & 0 \\ 
-1 & 0 & 0 & 0\end{array}\right),\\
&\sigma^{12}=\frac{i}{2}\left(\begin{array}{cccc} 0 & 0 & 0 & 1 \\ 0 & 0 & -1 & 0 \\ 
0 & 1 & 0 & 0 \\ -1 & 0 & 0 & 0\end{array}\right), &&\sigma^{13}=\frac{1}{2}
\left(\begin{array}{cccc} 0 & 0 & 0 & 1 \\ 0 & 0 & 1 & 0 \\ 0 & 1 & 0 & 0 \\ 
1 & 0 & 0 & 0\end{array}\right).
\end{align}
We thus have the linear equation
\begin{align}
a\sigma^{02}+b\sigma^{03}+c\sigma^{12}+d\sigma^{02}=\left(\begin{array}{cccc} 0 & 0 & 0 & 1 \\ 
0 & 0 & 0 & 0 \\ 0 & 0 & 0 & 0 \\ 0 & 0 & 0 & 0\end{array}\right).\label{detequ}
\end{align}
Solving for the coefficients $(a,b,c,d)$, one finds:
\begin{align}
p_\mu \epsilon_\nu{(\sigma^{\mu\nu})_a}^b&\sim -\frac{1}{2}\left[
(\sigma^{02}+i\sigma^{03})+i(\sigma^{12}+i\sigma^{13})\right]\nonumber\\
&=\frac{1}{2}\left[(\sigma^{20}+i\sigma^{30})+i(\sigma^{21}+i\sigma^{31})\right],\nonumber
\end{align}
which, upon switching to the complex basis given in (\ref{compbas1})-(\ref{compbas3}), 
entails the two possibilities
\begin{align}
&(p_A,\epsilon_B)=\left\{\begin{array}{c} (\bar{p}_1,\bar{\epsilon}_2) \\ 
(\bar{p}_2,\bar{\epsilon}_1)\end{array}\right.
\end{align}
\item vertex 2:\\
Here the matrix $p_\mu \epsilon_\nu{(\sigma^{\mu\nu})_a}^b$ has to read
\begin{align}
p_\mu \epsilon_\nu{(\sigma^{\mu\nu})_a}^b\sim
\left(\begin{array}{cccc} 0 & 0 & 0 & 0 \\ 0 & 0 & 0 & 0 \\ 0 & 0 & 0 & 0 \\ 
1 & 0 & 0 & 0\end{array}\right).
\end{align}
The relevant (reduced) Lorentz generators are the same as in the case above, 
but the determining equation (\ref{detequ}) changes to:
\begin{align}
a\sigma^{02}+b\sigma^{03}+c\sigma^{12}+d\sigma^{02}\left(\begin{array}{cccc} 0 & 0 & 0 & 0 \\ 0 & 0 & 0 & 0 \\ 
0 & 0 & 0 & 0 \\ 
1 & 0 & 0 & 0\end{array}\right).
\end{align}
Solving now for the coefficients $(a,b,c,d)$ reveals
\begin{align}
p_\mu \epsilon_\nu{(\sigma^{\mu\nu})_a}^b&\sim -\frac{1}{2}
\left[(\sigma^{02}-i\sigma^{03})-i(\sigma^{12}-i\sigma^{13})\right]\nonumber\\
&=\frac{1}{2}\left[(\sigma^{20}-i\sigma^{30})-i(\sigma^{21}-i\sigma^{31})\right],\nonumber
\end{align}
which upon the same complexifications as above entails
\begin{align}
&(p_A,\epsilon_B)=\left\{\begin{array}{c} (p_1,\epsilon_2) \\ (p_2,\epsilon_1)\end{array}
\right..
\end{align}
\end{itemize}
For concreteness we then choose the following momenta and helicities  
\begin{align}
\left(\begin{array}{c}(p^1_A,\epsilon^1_B) \\ (p^2_A,\epsilon^2_B)
\end{array}\right)=\left(\begin{array}{c} (\bar{p}_1,\bar{\epsilon}_2) \\ (p_2,\epsilon_1)
\end{array}\right).\label{finhel}
\end{align}

Finally, under the R-symmetry $SO(4) \sim SU(2)_L \times SU(2)_R$ (from the 6d point of view
which we are using here to determine the map from IIA to heterotic), the graviphotons transform
in the $(2,2)$ representation. The specific vertices we have chosen correspond to graviphotons carrying
charge $(+1/2,-1/2)$ with respect to the Cartan generators of the two $SU(2)$'s. On the heterotic side, 
the R-symmetry group $SO(4)$ acts on the tangent space of $T^4$. Let us choose (real) coordinates 
$Z^\alpha$ on $T^4$ with $\alpha=6,7,8,9$ normalized so that their two point functions 
\begin{align}
\langle\partial Z^\alpha(z) \partial Z^\beta(w)\rangle = -\frac{\delta^{\alpha\beta}}{(z-w)^2},
\end{align}
then the above choice of graviphotons entails 
choosing a complex direction say $X^4= \frac{1}{\sqrt{2}}(Z^6-iZ^7)$ (see also (\ref{compbas4})). The 
fermionic partner of $X^A$ ($A$ takes now the values $4,\bar{4},5,\bar{5}$) will be denoted as before by 
$\psi^A$ in the following.

Plugging the momenta and helicities of (\ref{finhel}) into the $0$-ghost picture vertex operator
\begin{align}
V^{(0),A}=\epsilon_B\left(\partial X^A-ip\cdot\chi \psi^A\right)\bar{\partial}X^Be^{ip\cdot Z},
\end{align}
the two different vertices appearing in the amplitude take the following form (see also \cite{Antoniadis:1995zn})
\begin{align}
&V^A_F=(\partial X^A-i\bar{p}_2\bar{\psi}^2\psi^A)\bar{\partial}\bar{X}^1
e^{i\bar{p}_2\bar{X}^2},\label{gravphothetvert1}\\
&V^A_F=(\partial X^A-ip_1\psi^1\psi^A)\bar{\partial}X^2e^{ip_1X^1}.\label{gravphothetvert2}
\end{align}
\end{itemize}
\subsection{Contractions}
We summarize the relevant vertex operators on the heterotic side:
\begin{align}
&V_{\varphi}(p_2)=(\partial Z^\mu-ip_2\psi^2\chi^\mu)\bar{\partial}Z_\mu e^{ip_2 X^2},\nonumber\\
&V_{\varphi}(\bar{p}_2)=(\partial Z^\nu-i\bar{p}_2\bar{\psi}^2\chi^\nu)
\bar{\partial}Z_\nu e^{i\bar{p}_2 \bar{X}^2},\nonumber\\
&V_h(p_1)=(\partial X^2-ip_1\psi^1\psi^2)\bar{\partial}X^2 e^{ip_1 X^1},\nonumber\\
&V_h(\bar{p}_2)=(\partial \bar{X}^1-i\bar{p}_2\bar{\psi}^2\bar{\psi}^1)
\bar{\partial}\bar{X}^1 e^{i\bar{p}_2 \bar{X}^2},\nonumber\\
&V^A_F(p_1)=(\partial X^A-ip_1\psi^1\psi^A)\bar{\partial}X^2e^{ip_1X^1},\nonumber\\
&V^A_F(\bar{p}_2)=(\partial X^A-i\bar{p}_2\bar{\psi}^2\psi^A)\bar{\partial}\bar{X}^1
e^{i\bar{p}_2\bar{X}^2}.\nonumber
\end{align}
The next question is which parts of the vertex operators have to be considered. 
This will be determined below for each type of vertex separately:
\begin{itemize}
\item gravitons: since we are considering couplings of the Riemann tensor, there have 
to be two derivatives from every vertex, implying that each graviton has to contribute 
two momenta. Thus, only the following terms of the vertices do contribute to the amplitude
\begin{align}
V_h(p_1)=&\left(\partial X^2-\ \fbox{\parbox{1.4cm}{$ip_1\psi^1\psi^2$}}\ \right)\ 
\fbox{\parbox{0.8cm}{$\bar{\partial}X^2$}}\  \left(1+\fbox{\parbox{1cm}{$ip_1X^1$}}+
\ldots\right),\nonumber\\
V_h(\bar{p}_2)=&\left(\partial \bar{X}^1-\ \fbox{\parbox{1.4cm}{$i\bar{p}_2
\bar{\psi}^2\bar{\psi}^1$}}\ \right)\ \fbox{\parbox{0.8cm}{$\bar{\partial}\bar{X}^1$}}\  
\left(1+\fbox{\parbox{1cm}{$i\bar{p}_2\bar{X}^2$}}+\ldots\right);\nonumber
\end{align}
\item graviphotons: Since the graviphoton vertex holds only one momentum, and $\psi$'s cannot contract among 
themselves, it is clear 
that only the following terms can contribute 
\begin{align}
V^A_F(p_1)=&\left(\ \fbox{\parbox{0.85cm}{$\partial X^A$}}\ -ip_1\psi^1
\psi^A\right) \ 
\fbox{\parbox{0.8cm}{$\bar{\partial}X^2$}}\  \left(1+\fbox{\parbox{1cm}{$ip_1X^1$}}+
\ldots\right),\nonumber\\
V^A_F(\bar{p}_2)=&\left(\ \fbox{\parbox{0.85cm}{$\partial X^A$}}\ -i\bar{p}_2\bar{\psi}^2
\psi^A\right)\ \fbox{\parbox{0.8cm}{$\bar{\partial}\bar{X}^1$}}\  \left(1+
\fbox{\parbox{1cm}{$i\bar{p}_2\bar{X}^2$}}+\ldots\right);\nonumber
\end{align}
\item dilaton: Comparing with the computation on the type II side, it is obvious 
that there have to be 2 momenta for each vertex. Furthermore, it is necessary that 
the fermionic part contributes so that the spin structure sum gives a non-vanishing result:  
\begin{align}
V_{\varphi}(p_2)=&\left(\partial Z^\mu-\ \fbox{\parbox{1.4cm}{$ip_2\psi^2\chi^\mu$}}\ 
\right)\ \fbox{\parbox{0.8cm}{$\bar{\partial}Z_\mu$}}\  \left(1+\fbox{\parbox{1cm}
{$ip_2X^2$}}+\ldots\right),\nonumber\\
V_{\varphi}(\bar{p}_2)=&\left(\partial Z^\nu-\ \fbox{\parbox{1.4cm}{$i\bar{p}_2
\bar{\psi}^2\chi^\nu$}}\ \right)\ \fbox{\parbox{0.8cm}{$\bar{\partial}Z_\nu$}}\  
\left(1+\fbox{\parbox{1cm}{$i\bar{p}_2\bar{X}^2$}}+\ldots\right).\nonumber
\end{align}
Furthermore, from the trace part only the following term leads to non-vanishing contractions
\begin{align}
V_{\varphi}(p_2)=&\left(\partial \bar{X}^2-\ \fbox{\parbox{1.4cm}{$ip_2\psi^2\bar{\psi}^1$}}\ 
\right)\ \fbox{\parbox{0.8cm}{$\bar{\partial}X^1$}}\  \left(1+\fbox{\parbox{1cm}{$ip_2X^2$}}+
\ldots\right),\nonumber\\
V_{\varphi}(\bar{p}_2)=&\left(\partial X^2-\ \fbox{\parbox{1.4cm}
{$i\bar{p}_2\bar{\psi}^2\psi^1$}}\ \right)\ \fbox{\parbox{0.8cm}{$\bar{\partial}\bar{X}^1$}}\  
\left(1+\fbox{\parbox{1cm}{$i\bar{p}_2\bar{X}^2$}}+\ldots\right).\nonumber
\end{align}
\end{itemize}
Note that if $X^1$ and ${\bar X}^1$ are replaced by $X^2$ and ${\bar X}^2$, respectively,
in the two scalar vertices above, the contribution of right movers is reduced to a total
derivative ($\bar\partial (X^2)^2$ and $\bar\partial ({\bar X}^2)^2$) which yields zero
result. Thus, the above terms have been chosen in such a way that all contractions are apriori non-vanishing.
\subsection{Computation of the Space-Time Correlator}
Taking into account the momentum structure of the above vertices, one finds
\begin{align}
A_g&=\langle V_h(p_1)V_h(\bar{p}_2)V_{\varphi}(p_2)V_{\varphi}(\bar{p}_2)\prod_{i=1}^{g-1}
V_F(p_1^{(i)})\prod_{j=1}^{g-1}V_F(\bar{p}_2^{(j)})\rangle=\nonumber\\
&\equiv (p_1)^4(\bar{p}_2)^4\prod_{i,j=1}^{g-1}p_1^{(i)}\bar{p}_2^{(j)}((g+1)!)^2\mathcal{F}_g^{(3,\text{HET})}.
\label{spacetimecor}
\end{align}
The contractions of the above vertices lead to the following (rather formal) expression 
for the amplitude\footnote{Similar as in \cite{Antoniadis:1995zn}, one can show that the 
contributions of even and odd spin structures are the same.}
\begin{eqnarray}
\mathcal{F}_g^{(3,\text{HET})}&=&\frac{1}{((g+1)!)^2}\int \frac{d^2\tau}{\tau_2^3}
\frac{1}{\bar{\eta}^{24}}\langle\prod_{i=1}^{g+1}\int d^2x_iX^1\bar{\partial} X^2(x_i)
\prod_{j=1}^{g+1}\int d^2y_j\bar{X}^2\bar{\partial}\bar{X}^1(y_j)\rangle\cdot\nonumber\\
&~&\hspace{1cm}\cdot\sum_{(P_L,P_R) \in \Gamma^{(6,22)}}\left(P^A_L\right)^{2g-2}
q^{\frac{1}{2}P_L^2}\bar{q}^{\frac{1}{2}P_R^2},
\end{eqnarray}
where $\tau=\tau_1+i\tau_2$ is the Teichm\"uller parameter of the world-sheet torus with $q=e^{2\pi i\tau}$, 
and $P_L,P_R$ are the left and right momenta of the $\Gamma^{(6,22)}$ compactification lattice. Possible 
additional constant prefactors were dropped since they are not relevant for our analysis. Note that $X^A$'s 
being the same complex field 
(eg. $\frac{1}{\sqrt{2}}(Z^6-iZ^7)$) do not have any singular OPE among themselves and therefore they 
contribute only through their zero modes. 
In this expression, still the (normalized) correlator of the space-time fields $X^{1,2}$ has 
to be computed. To this end, the following generating functional can be introduced:
\begin{align}
G(\lambda,\tau,\bar{\tau})&=\sum_{h=0}^\infty\frac{1}{(h!)^2}\left(\frac{\lambda}{\tau_2}\right)^{2h}
\langle\prod_{i=1}^h\int d^2x_iX^1\bar{\partial} X^2(x_i)\prod_{j=1}^{h}
\int d^2y_j\bar{X}^2\bar{\partial}\bar{X}^1(y_j)\cdot\nonumber\\
&\hspace{1cm}\cdot \int d^2w_1X^2
\bar{\partial}X^2(w_1)\int d^2w_2\bar{X}^2\bar{\partial}\bar{X}^2(w_2)\rangle=\nonumber\\
&=\sum_{h=1}^\infty\lambda^{2h}G_{h}(\tau,\bar{\tau})
\end{align}
In \cite{Antoniadis:1995zn}, this generating functional was calculated and shown to be
\begin{equation}
G(\lambda,\tau,\bar{\tau})=\left(\frac{2\pi i\lambda\bar{\eta}^3}{\bar{\Theta}_1
(\lambda,\bar{\tau})}\right)^2
\text{exp}\left(-\frac{\pi\lambda^2}{\tau_2}\right),
\end{equation}
where $\Theta_1$ is the usual odd theta-function defined by
\begin{align}
\Theta_1(z,\tau)=\sum_{n\in\mathbb{Z}}q^{\frac{1}{2}\left(n-\frac{1}{2}\right)^2}e^{2\pi i 
\left(z-\frac{1}{2}\right)\left(n-\frac{1}{2}\right)}.
\end{align}
It can be obtained from (\ref{thetadef}) by the choice $(\alpha_1,\alpha_2)=\left(-\frac{1}{2},-\frac{1}{2}\right)$. 
We recall here some properties satisfied by $G_g (\tau, \bar{\tau})$:
\begin{eqnarray}
&& G_g(\frac{a\tau+b}{c\tau+d}, \frac{a\bar{\tau}+b}{c\bar{\tau}+d}) = 
\bar{\tau}^{2g} G_g(\tau,\bar{\tau}), \nonumber\\
&&\frac{\partial}{\partial \tau} G_g= -\frac{i\pi}{2\tau_2^2} G_{g-1}.
\label{Ggeq}
\end{eqnarray}
With this space-time correlator, the final expression for the amplitude 
takes the form
\begin{equation}
\mathcal{F}_g^{(3,\text{HET})}=\int \frac{d^2\tau}{\tau_2^3}\frac{1}{\bar{\eta}^{24}} \tau_2^{2g+2} G_{g+1} 
\sum_{(P_L,P_R) \in \Gamma^{(6,22)}}
\left(P^A_L\right)^{2g-2}
q^{\frac{1}{2}P_L^2}\bar{q}^{\frac{1}{2}P_R^2}.\label{hetampres}
\end{equation}

\section{Harmonicity Relation for Type II Compactified on $K3$}\label{harmoIIK3}
\subsection{Holomorphicity for Type II Compactified on $CY_3$}\label{sectholom}
After having established different topological amplitudes in various
configurations, we will now look for the analog of the harmonicity
relation discussed in \cite{Berkovits:1994vy, Ooguri:1995cp}, that
generalizes holomorphicity of $N=2$ F-terms involving vector
multiplets. Before plunging into the computation for the new 
$N=4$ topological amplitudes on $K3\times T^2$, we first review the
known results for $N=2$ compactifications on a 
Calabi-Yau threefold $CY_3$ and for $N=4$ on $K3$.

Type II theory compactified on a $CY_3$ exhibits $N=2$ supersymmetry and the relevant multiplets are the 4d, $N=2$ supergravity Weyl multiplet $W^{(\text{4d})}_{\mu\nu}$ and the (chiral) Yang-Mills multiplet $V$. The former contains as bosonic components the graviton and the graviphoton (see \cite{deRoo:1980mm,Bergshoeff:1980is})\footnote{In our superfield expressions, numerical factors are dropped.},
\begin{align}
W^{(\text{4d})}_{\mu\nu}=T_{\mu\nu}+(\theta_L\sigma^{\rho\tau}\theta_R)R_{\mu\nu\rho\tau}+\text{fermions},
\end{align}
while in a chiral basis, the later (which has a complex scalar and a vector-field as bosonic components) can be written as the following ultrashort superfield
\begin{align}
V=\phi(x)+\theta_L\sigma^{\mu\nu}\theta_R F_{\mu\nu}+\text{fermions}.\label{vectCY}
\end{align}
Note that $V$ depends only on half of the Grassmann coordinates, namely only the chiral ones $\theta_L,\theta_R$ but not on the anti-chiral ones $\bar{\theta}_L,\bar{\theta}_R$.

As shown in \cite{Antoniadis:1993ze}, topological amplitudes at the $g$-loop level correspond to the coupling
\begin{align}
\int d^4x\int d^2\theta_Ld^2\theta_R F_g(V)((W^{(\text{4d})})^{2})^g=\int d^4x R^2T^{2g-2}F_g(\phi(x))+\ldots
\label{holoeq}
\end{align}
where $F_g$ is captured by the partition function of the $N=2$ topological string and the dots stand for additional terms arising from the superspace integral.
The right hand side exhibits the fact, that $F_g$ depends only on the scalars of the Yang-Mills multiplets $\phi$, but not on their complex conjugates $\bar{\phi}$. Na\"ively, one therefore can write down the following holomorphicity equation
\begin{align}
\partial_{\bar{\phi}}F_g=0\, .
\label{holomorphicity}
\end{align} 
However in \cite{Bershadsky:1993cx}, it was shown that this equation is spoiled by the so-called holomorphic anomaly, which stems from additional boundary contributions when taking the derivative in (\ref{holomorphicity}). From the effective field theory point of view, the anomaly is due to the propagation of massless states in the loops \cite{Antoniadis:1993ze}.
\subsection{6d $N=2$ Harmonicity Equation}\label{N2der}
Recalling the topological amplitudes of Section \ref{SectTopAmpK3} for type II string theory compactified on $K3$, they correspond to the following $N=2$ supersymmetric term of the six-dimensional effective action:
\begin{align}
\int d^6x\int d^4\theta^1_L\int d^4\theta^1_R\mathcal{F}_g^{(\text{6d})}(q^{ij})({W_{a_1}}^{b_1}{W_{a_2}}^{b_2}{W_{a_3}}^{b_3}{W_{a_4}}^{b_4}\epsilon^{a_1a_2a_3a_4}\epsilon_{b_1b_2b_3b_4})^g +\ldots
\label{superintBerVafa}
\end{align}
where the dots stand for possible additional terms needed for complete supersymmetrization, 
following from integration over harmonic superspace 
(see discussion below and \cite{Galperin:1984av}).
Here ${W_a}^b$ is the 6d $N=2$ Weyl multiplet already defined in (\ref{Wdef}) with the $SU(2)_L\times SU(2)_R$ indices fixed to $i=1,j=1$ \cite{Berkovits:1998ex}, and we included a (apriori arbitrary) function of the 6d $N=2$ matter multiplet $q^{ij}$ with the following component expansion
\begin{align}
q^{ij}=\lambda^{ij}+\theta_L^i\chi_R^j+\theta^{j}_R\chi^i_L+\bar{\theta}_L^i\bar{\chi}_R^j+\bar{\theta}^{j}_R\bar{\chi}^i_L+(\theta_L^i\sigma^{\mu\nu}\theta^j_R)F^+_{\mu\nu}+(\bar{\theta}_L^i\bar{\sigma}^{\mu\nu}\bar{\theta}^j_R)F^-_{\mu\nu}+\ldots,\label{6dmatter}
\end{align}
with the pair of indices $(i,j)$ transforming under $SU(2)_L\times SU(2)_R$ and $F^+$ ($F^-$) denoting the (anti)-self-dual part of the field strength. The scalars $\lambda^{ij}$ satisfy the hermiticity relation 
\begin{align}
\bar{\lambda}_{ij}=\epsilon_{ik}\epsilon_{jl}\lambda^{kl}.
\label{complexconj}
\end{align}
In passing from the 4d case discussed in the previous section to the 6d one it is necessary to introduce harmonic superspace in order to obtain a short version of the multiplet $q^{ij}$~\cite{Andrianopoli:1999vr}. Moreover, one has to restore the R-symmetry indices and rewrite (\ref{superintBerVafa}) in a covariant way. Since the R-symmetry group is 
\begin{align}
SU(2)\times SU(2),
\end{align}
the harmonic coordinates should parameterize the coset manifold
\begin{align}
\frac{SU(2)_{L}}{U(1)_{L}}\times \frac{SU(2)_{R}}{U(1)_{R}}.
\end{align}

The BPS character of the (covariantized) effective action term (\ref{superintBerVafa}) should then restrict the 
moduli dependence $q^{ij}$ of $\mathcal{F}_g^{(\text{6d})}$'s, generalizing the holomorphicity equation 
(\ref{holomorphicity}). Such an equation was found in Ref.~\cite{Berkovits:1994vy} by covariantizing the 
corresponding string amplitudes (\ref{harmL}) and (\ref{harmLp}). This can be done by introducing two constant 
doublets $u^L$ and $u^R$, under $SU(2)_L$ and $SU(2)_R$ respectively, satisfying $|u^L|^2=|u^R|^2=1$, 
\begin{align}
u^{L,i}\equiv\epsilon^{ij}u_j^{L}=\overline{u_i^{L}},
\label{updown}
\end{align}
and similar for $u^R_i$, and
replace the Weyl superfields ${W_a}^b\equiv({W_a}^b)^{11}$ in eq.~(\ref{superintBerVafa}) by 
$({W_a}^b)^{ij}u^L_iu^R_j$. The moduli dependent coefficient $\mathcal{F}_g^{(\text{6d})}$ is then promoted 
to a function of $u^L$ and $u^R$, as well. The moduli $\lambda_{ij}$ couple to the string world
sheet action as:
\begin{align}
S\to S&+\int \left(\bar{\lambda}^\alpha_{22}\oint G^+\oint \bar{G}^+\bar{\phi}_\alpha+\bar{\lambda}^\alpha_{12}\oint \tilde{G}^+\oint \bar{\tilde{G}}^+\bar{\phi}_\alpha+\right.\nonumber\\
&+\left.\bar{\lambda}^\alpha_{21}\oint G^+\oint \bar{\tilde{G}}^+\bar{\phi}_\alpha+\bar{\lambda}^\alpha_{11}\oint \tilde{G}^+\oint \bar{G}^+\bar{\phi}_\alpha\right).
\end{align}
In the twisted theory $G^+$, $\tilde{G}^+$ (and their right moving counterparts) are dimension one operators. By
deforming the contours of $G^+$ and $\tilde{G}^+$ and using $N=4$ superconformal algebra,
it was then shown~\cite{Berkovits:1994vy} that 
$\mathcal{F}_g^{(\text{6d})}$  satisfy the following equation modulo possible contact terms and contributions 
from the boundaries
of the world-sheet moduli space:
\begin{align}
\frac{\partial}{\partial u_i^{L}}\frac{D}{D\lambda^{ij}}\mathcal{F}_g^{(\text{6d})}=0,
\label{BerVafaHarm}
\end{align}
for fixed $j$, which by the hermiticity relation satisfied by $\lambda$ can also be written as:
\begin{align}
\epsilon_{ik}\frac{\partial}{\partial u_i^{L}}\frac{D}{D\bar{\lambda}_{kj}}\mathcal{F}_g^{(\text{6d})}=0.
\label{BerVafaHarmmod}
\end{align}
Note that since the index $j$ is free there are actually two equations which transform as a doublet of
$SU(2)_R$ (there is actually another doublet of equation transfoming under $SU(2)_L$ obtained by 
exchanging $u_L$ and $u_R$). 
It turned out however, as realized in \cite{Ooguri:1995cp}, that besides boundary anomalous contributions, there are contact terms that spoil each of the
two equations for given $j$, and only a certain linear combination of the two which is
$SU(2)$ invariant is free from both boundary and contact terms. 
The correct equation was found to be:  
\begin{align}
\sum_{i,j,k=1}^2u^R_k\epsilon_{ij}\frac{d}{du^L_i}\frac{D}{D\bar{\lambda}_{jk}^\alpha}
\mathcal{F}_g^{(\text{6d})}=0.
\label{trueharm}
\end{align}
There is also another equation obtained by exchanging $u_L$ and $u_R$:
\begin{align}
\sum_{i,j,k=1}^2u^L_k\epsilon_{ij}\frac{d}{du^R_i}\frac{D}{D\bar{\lambda}_{kj}^\alpha}
\mathcal{F}_g^{(\text{6d})}=0.
\label{trueharmR}
\end{align}
\section{Harmonicity Relation for Type II Compactified on $K3\times T^2$}\label{sect:harmrelIIA}
\subsection{Supergravity Considerations}\label{sugracons}
After having reviewed in the previous section the known computations and results, we now turn to our main 
objective, which is to establish a similar equation to (\ref{BerVafaHarm}) or (\ref{trueharm}) for the 
compactification on $K3\times T^2$. The vector multiplets in the resulting 4-dimensional $N=4$ theory 
are described by the field strength multiplet $Y^{ij}_A$ where $i,j$ transform in the fundamental
representation of the R-symmetry group $SU(4)$ and index $A$ labels the vector multiplet (in the context of string theory $A=1,..,22$). $Y^{ij}_A$ is antisymmetric and therefore transforms in the antisymmetric 6-dimensional representation of $SU(4)$ and satisfies the reality condition
\begin{equation}
Y^{ij}_A=\frac{1}{2}\epsilon^{ijkl}Y_{kl,A},~~~~Y_{k\ell,A}=\bar{Y}^{kl}_A.
\end{equation}
As in the 6d $N=2$ case discussed before, one has to introduce harmonic superspace variables in order to 
shorten in particular the $Y^{ij}_A$ multiplet. In this case, 
the harmonic coordinates should parametrize generically the coset manifold~\cite{Ivanov:1984ut}:
\begin{align}
\frac{G}{H} \equiv \frac{SU(4)}{U(1)\times U(1)\times U(1)}.
\end{align}
Here we have chosen $H$ to be the maximal abelian subgroup of $G$ as in~\cite{Andrianopoli:1999vr}. Hence, we have a set of four harmonic 
coordinates $u_i^I$, $I=1,..,4$ (together with their conjugates $u_I^i$), 
each transforming as a 
\textbf{4} (and $(\bar{\bf{4}})$) of $SU(4)$, which differ by their charges with respect to the three $U(1)$'s:
\begin{align}
&u_i^1\equiv u_i^{(1,0,1)}, &&u_i^2\equiv u_i^{(-1,0,1)}, &&u_i^3\equiv u_i^{(0,1,-1)}, &&u_i^4\equiv 
u_i^{(0,-1,-1)}.
\end{align}
They satisfy the conditions
\begin{equation}
u_i^I u_I^j = \delta_i^j,~~~~~~u_I^i u_i^J= \delta_I^J,
~~~~~~~~\epsilon^{i_1..i_4}u^1_{i_1}..u^4_{i_4}=1,
\label{harmonicconstraint}
\end{equation}
where an implicit summation over repeated indices is understood.
The derivatives $D^I_J$ which preserve this condition are given by
\begin{equation}
D^I_J= u^I_i \frac{\partial}{\partial u^J_i} - u_J^i \frac{\partial}{\partial u_I^i},~~~~~\sum_I D^I_I=0
\label{derivativeconstraint}
\end{equation}
These derivatives satisfy an $SU(4)$ algebra. The diagonal ones, namely for $I=J$, modulo the trace are just the charge 
operators of the subgroup $H$. The remaining generators for $I<J$ and their conjugates for $I>J$ define
the subalgebras $L^+$ and $L^-$ (corresponding to positive and negative roots respectively) 
in the Cartan decomposition of $G=H+L^+ + L^-$. The highest weight in a given irreducible representation 
is defined by the condition (called H-analyticity condition) that it is annihilated by $L^+$. One 
must further impose G-analyticity condition which ensures that the superfield depends only on half
of the Grassman variables and hence gives a short multiplet. The
details can be found in \cite{Andrianopoli:1999vr}. The resulting superfield is\footnote{We thank E. Sokatchev for showing us the precise form of the superfields $Y$ and $K$.}
\begin{equation}
Y^{12}_A=u^1_iu^2_j\phi^{ij}_A+\theta_3 \sigma^{\mu\nu}\theta_4 F_{+,\mu\nu,A}+
\bar{\theta}^1\bar{\sigma}^{\mu\nu}\bar{\theta}^2F_{-,\mu\nu,A}+\ldots
\end{equation}
where $\theta_I$ and $\bar{\theta}^I$ 
are harmonic projected Grassman coordinates $\theta_I= u^i_I \theta_i$ and $\bar{\theta}^I=u^I_i \bar{\theta}^i$, $F_{\pm,\mu\nu}$ are self-dual and anti-selfdual parts of the gauge field strength
and dots refer to  terms involving fermions and derivatives of bosonic
fields. By choosing different Cartan decompositions (i.e. different choices of positive roots) we can get
different superfields $Y^{IJ}$ but here we will focus on $(IJ)=(12)$.

In the present case, we also have the gravitational multiplet which comes with graviphotons that transform
in the 6-dimensional representation of $SU(4)$. Unfortunately the shortening of the gravitational multiplet
using harmonic variables is not known at present. However, as we will see below, we will be able to
reproduce the string amplitudes constructed in this paper by postulating the superfield
\begin{eqnarray}
K_{\mu\nu}^{12}&=&u^1_i u^2_j T_{+,\mu\nu}^{ij}+(\theta_3\sigma^{\rho\tau}\theta_4)R_{+,\mu\nu\rho\tau}+
\bar{\theta}^{1,\dot{a}}\bar{\theta}^{2,\dot{b}}(\sigma_{\mu\nu})^{ab}\partial_{a\dot{a}}
\partial_{b\dot{b}}
\Phi_++\ldots, \nonumber\\
\tilde{K}_{\mu\nu}^{12}&=&u^i_3 u^j_4\epsilon_{ijkl} T_{-,\mu\nu}^{kl}+(\bar{\theta}^1\bar{\sigma}^{\rho\tau}\bar{\theta}^2)
R_{-,\mu\nu\rho\tau}+
\theta_3^a \theta_4^b(\bar{\sigma}_{\mu\nu})^{\dot{a}\dot{b}}\partial_{a\dot{a}}\partial_{b\dot{b}}
\Phi_- +\ldots\
\label{Kijsuperfield}
\end{eqnarray}
where
$T^{ij}_{\pm\mu \nu}$ are the self-dual and anti-self-dual parts of
the graviphoton
field strengths. More generally there should be superfields $K_{\mu\nu}^{IJ}$ (and $\tilde{K}_{\mu\nu}^{IJ}$)
obtained by different
harmonic projections whose lowest component will be given by $u^I_i u^J_j T^{ij}_{+,\mu\nu}$
(and $u^I_i u^J_j T^{ij}_{-,\mu\nu}$)
but here we will focus on the effective action terms involving the fields with $(IJ)=(12)$,  
in a way similar to the 6d case~\cite{Berkovits:1994vy}. In the absence of a rigorous $N=4$ supergravity 
construction, eq.(\ref{Kijsuperfield}) should be thought of as a convenient way of summarizing the results of string amplitudes constructed in this paper. Actually, the above expressions can  be obtained by taking two supercovariant derivatives of the 4d $N=4$ Weyl multiplet 
$W_{N=4}$~\cite{Bergshoeff:1980is}: 
$K^{IJ}_{\mu\nu}\simeq {\cal D}^I\sigma_{\mu\nu}{\cal D}^J W_{N=4}$ and
$\tilde{K}_{\mu\nu}^{IJ}=\epsilon^{IJMN}{\bar K}_{\mu\nu,MN}$.

Using these superfields we can write down the following couplings:
\begin{eqnarray}
S_1&=&\int d^4x \int d^2\theta_3\int d^2\theta_4\int d^2\bar{\theta}^1\int d^2\bar{\theta}^2\, 
\mathcal{F}_g^{(1)}(K_{\mu\nu}^{12}K^{\mu\nu,12})^g(\tilde{K}_{\mu\nu}^{12}\tilde{K}^{\mu\nu,12}),
\label{amplitudesuper1}\\
S_3&=&\int d^4x \int d^2\theta_3\int d^2\theta_4\int d^2\bar{\theta}^1\int d^2\bar{\theta}^2\, 
\mathcal{F}_g^{(3)}(K_{\mu\nu}^{12}K^{\mu\nu,12})^{g+1},\label{amplitudesuper2}
\end{eqnarray}
where $\mathcal{F}_g^{(1)}$ and $\mathcal{F}_g^{(3)}$ depend on the vector multiplets and harmonic
variables\footnote{Here by a slight abuse of notation we are using the same symbols as the ones
appearing in the topological string amplitudes, however we will see shortly the relation $\mathcal{F}_g^{(1)}$ and $\mathcal{F}_g^{(3)}$ appearing in the above equations to the corresponding 
quantities appearing in the string amplitudes.}. Whether such terms give rise to consistent
supergravity couplings is not clear at present and needs to be understood.
By performing now the integration over the Grassmann coordinates $\theta^i$, one finds that the two expressions 
(\ref{amplitudesuper1}) and (\ref{amplitudesuper2}) give rise to the following effective action terms
\begin{eqnarray}
S_1\!\!\!\! &=&\!\!\!\! \int d^4 x \mathcal{F}_g^{(1)}\left[R_+^2R_-^2T_+^{2g-2}+
(\partial\partial \Phi_+)^2(\partial\partial \Phi_-)^2T_+^{2g-2}+
(\partial\partial \Phi_+)(\partial\partial \Phi_-)R_+R_-T_+^{2g-2}\right]+\ldots\nonumber\\
S_3\!\!\!\! &=&\!\!\!\! \int d^4 x
\mathcal{F}_g^{(3)}R_+^2(\partial\partial \Phi_+)^2T_+^{2g-2}+\ldots
\end{eqnarray}
It follows that the two couplings in (\ref{amplitudesuper1}) and (\ref{amplitudesuper2}) correspond to the 
two classes of topological amplitudes $\mathcal{F}_g^{(1)}$ and $\mathcal{F}_g^{(3)}$ discussed in sections 
\ref{Section4dscattering} and \ref{ChangeAmpl}, respectively. They reproduce in particular the corresponding 
amplitudes $R_+^2R_-^2T_+^{2g-2}$ at genus $g$ and $R_+^2(dd\Phi_+)^2T_+^{2g-2}$ at genus $g+1$ found above.

In order to find the corresponding 
generalized harmonicity equation, we will follow the same procedure as before by going back to string theory 
and covariantize the respective string amplitudes under $SU(4)$. As mentioned earlier there are several different projections labelled by $(IJ)$. In order to pick out $(IJ)=(12)$ we can choose the 
graviphoton field strength  $T_{+,\mu\nu}^{ij}$ to be proportional to $u_1^{[i} u_2^{j]}$ times some self dual field strength. Due to the orthogonality relations (\ref{harmonicconstraint}) this will precisely pick out the lowest component of $K_{\mu\nu}^{12}$ and the resulting amplitudes will be 
given by $\mathcal{F}_g^{(1)}$ and $\mathcal{F}_g^{(3)}$ appearing in (\ref{amplitudesuper1}) 
and (\ref{amplitudesuper2}) respectively. On the other hand choosing such graviphoton field strength
means that the vertex operators for the graviphoton field strengths in the string computations  should be folded with $u_1^{[i} u_2^{j]}$. This will render the topological string amplitudes a
dependence on the harmonic variables and it is those quantities that compute $\mathcal{F}_g^{(1)}$ and $\mathcal{F}_g^{(3)}$ appearing in (\ref{amplitudesuper1}) 
and (\ref{amplitudesuper2}). In the next section we will show, by going to 
the heterotic duals of these amplitudes, that $\mathcal{F}_g^{(3)}$ satisfies the equation:
\begin{align}
\epsilon^{ijkl} D^1_i D^2_j\frac{D}{D\phi^{kl}}\mathcal{F}^{(3)}_g=0,
\label{harmrelN4}
\end{align}
up to an anomaly boundary term, very much in analogy to the $N=2$ $F_g$'s. Here the derivatives $D^I_i$
are defined as:
\begin{equation}
D^I_i\equiv u_i^J D^I_J = u_i^J u_k^I\frac{\partial}{\partial u_k^J}-\frac{\partial}{\partial u_I^i}
\label{DIi}
\end{equation}
and we have relaxed the determinant condition and the related trace condition in eqs.(\ref{harmonicconstraint}) and (\ref{derivativeconstraint}). This can always be done by
introducing another variable say $\lambda$ and let $u_i^I \rightarrow u_i^I \lambda$ and $u^i_I\rightarrow u^i_I /\lambda$. One can check that as a result $[D^I_i, D^J_j]=0$.  Note that acting
on functions only of $u_1^i$ and $u_2^i$, as will be the case in the following, only the second term on the right hand side of eq.(\ref{DIi}) will be relevant.

Equation (\ref{harmrelN4}) is
$SU(4)$ invariant analogous to the `correct' equations (\ref{trueharm})
and (\ref{trueharmR}) in the 6-dimensional case, which were singlets
under the corresponding R-symmetry group $SU(2)_L \times SU(2)_R$. 
It turns out that there are stronger harmonicity equations with only one 
harmonic derivative ($D^1_i$ or $D^2_i$), that transform
covariantly under $SU(4)$ and are similar to the six dimensional ones 
discussed in the previous section.
However, here, we will restrict ourselves to the weaker version (\ref{harmrelN4}).\footnote{We thank 
E. Sokatchev for pointing this out to us. The stronger version of the harmonicity equation will
appear elsewhere~\cite{prep}.}
Moreover, by considering the decompactification limit in six dimensions, $\mathcal{F}_g^{(3)}$ is mapped to a `semi-topological' quantity similar to $\mathcal{F}_g^{(6d)}$, that satisfies the same (6d) harmonicity equations as (\ref{trueharm}) and (\ref{trueharmR}). This supports the claim that all such equations (\ref{trueharm}), (\ref{trueharmR}) and (\ref{harmrelN4}) should be a consequence of the BPS character of these couplings generalizing the $N=2$ holomorphicity to extended supersymmetries.
 
\subsection{Further Investigation}
The next logical step would be to test this relation using directly the $\mathcal{F}_g$'s computed in 
Sections \ref{Section4dscattering} and \ref{ChangeAmpl} for type II theory and by this procedure determine a 
possible harmonicity equation. However in doing so, one encounters several problems: 
\begin{itemize}
\item The topological expression cannot be used, because some of the operator insertions necessary for testing 
(\ref{harmrelN4}) are 
from the RR sector\footnote{Recall that in type II compactified on $K3\times T^2$ only $SU(2)_L\times SU(2)_R
\times U(1)$ subgroup of the R-symmetry group $SU(4)$ is manifest where $U(1)$ is the rotation group acting on the
tangent space of $T^2$. In particular the full $SU(4)$ group mixes NS-NS and R-R sectors. Since eq.(\ref{harmrelN4}) is $SU(4)$ invariant it involves deforming by both NS-NS and R-R moduli.}, for which no representation in terms of the superconformal algebra exists. 
One could therefore try to check the equation only perturbatively in RR field insertions.
\item The corresponding correlation functions with additional RR insertions lead to rather difficult expressions 
which are hard to compute.
\end{itemize}
For these reasons, we change our strategy and switch to the heterotic side. As we have shown, the 
$\mathcal{F}_g^{(3)}$ can be computed upon duality mapping by a one-loop torus amplitude on the heterotic side, 
which is easy to obtain. In this way, the harmonicity relation can be tested and we can also verify whether it 
contains any anomalous terms.
\section{Harmonicity Relation for Heterotic on $T^6$}\label{sect:harmHet}
\subsection{Derivation of the Harmonicity Relation}
To discuss the harmonicity relation on the heterotic side we first need to express
the $\mathcal{F}^{(3,\text{HET})}_g$ in an $SO(6) \sim SU(4)$ covariant way. In computing $\mathcal{F}^{(3,\text{HET})}_g$ 
we used a particular combination of graviphotons which was labeled say by $G$, 
with $\partial X^G$ appearing in the vertex being 
$(\partial Z^6 - i\partial Z^7)/\sqrt{2}$.
Since $Z^4, Z^5,...,Z^9$ transform as vector of 
$SO(6)$, this particular choice of the graviphoton is not $SO(6)$ invariant. 
Using the fact that the vector of $SO(6)$ can be expressed as antisymmetric tensor product of two $SU(4)$ 
fundamentals, we can define
\begin{align}
X_{ij} = Z^\alpha (C\sigma_\alpha)_{ij},
\end{align}
where $\sigma_\alpha$ are the $(4\times 4)$ part of the $SO(6)$ gamma 
matrices acting on the chiral spinor (i.e. $SU(4)$ fundamental representation {\bf 4}) and $C$ is the 
charge conjugation matrix such that $(C\sigma_\alpha)^T = -C\sigma_\alpha$. 
It is easy to see that $X_{ij}$ are
complex and satisfy the reality condition $\overline{X_{ij}} = \frac{1}{2}\epsilon^{ijkl} X_{kl}$. Furthermore
we normalize them so that their 2-point function is 
\begin{align}
\langle \partial X_{ij}(z) \partial X_{kl}(w)\rangle = - \frac{2\epsilon_{ijkl}}{(z-w)^2}. 
\end{align}
In real notation, this essentially means that $X^{ij} = Z^\alpha + iZ^\beta$ for 
some $\alpha$ and $\beta$ different
from each other, where $Z^\alpha$ are canonically normalized so that their kinetic term comes with the
factor $1/2$. As a result the zero mode part of the $\partial X_{ij}$ is just given by
$2\pi P^L_{ij}$ and the lattice part of the left-moving partition function is $q^{\frac{1}{2} P_L^2}$ where
$(P^L)^2 = \frac{1}{8} \epsilon^{ijkl} P^L_{ij} P^L_{kl}$. 

As explained at the end of section 8, in order to pick $K^{12}_{\mu\nu}$ we must choose the graviphoton 
field strength to be proportional to $u_1^{[i} u_2^{j]}$. This means that the graviphoton vertex must be
folded with $u_1^{[i} u_2^{j]}$:
\begin{equation}
V_F(u_1,u_2) =  \frac{1}{2\pi}\int d^2 z u_1^{i} u_2^{j}(\partial X_{ij} -ip.\chi \psi_{ij})\bar{\partial} Z^{\mu} e^{ip.Z}(z,\bar{z}),
\end{equation} 
and define $\mathcal{F}^{(3,\text{HET})}_g(u_1,u_2)$ as in (\ref{spacetimecor}) with all the $2g-2$ $V_F$ replaced by 
$V_F(u_1,u_2)$. 
An important point to notice is that the singular term in the OPE is
\begin{equation}
\langle u_1^{i} u_2^{j} \partial X_{ij}(z) ~u_1^k u_2^l \partial X_{kl}(w)\rangle = -\frac{2u_1^i u_2^j u_1^k u_2^l \epsilon_{ijkl}}{(z-w)^2} = 0.
\end{equation}
This implies that $\partial X_{ij}$ are replaced by their lattice momenta $2\pi P^L_{ij}$ inside
the correlation function. The final result can be read off from eq.(\ref{hetampres})
\begin{equation}
\mathcal{F}^{(3,\text{HET})}_g(u_1,u_2)=\int \frac{d^2\tau}
{\tau_2^3}\frac{1}{\bar{\eta}^{24}} \tau_2^{2g+2} G_{g+1} \sum_{(P^L,P^R) \in \Gamma^{(6,22)}}
\left(u_1^i u_2^j P_{ij}^L\right)^{2g-2}
q^{\frac{1}{2}(P^L)^2}\bar{q}^{\frac{1}{2}(P^R)^2}\, .
\label{hetampresuv}
\end{equation}

The marginal deformations are given by the $(1,1)$ operators $V_{ij,A}=-\frac{1}{2\pi}
\partial X_{ij} \bar{J}_A$, where $\bar{J}_A$ are the right-moving currents for $A=1,...,22$ normalized so that 
the 2-point function satisfies
\begin{align}
\langle\bar{J}_A(\bar{z}) \bar{J}_B(\bar{w})\rangle = \frac{\delta_{AB}}{(\bar{z}-\bar{w})^2}.
\end{align} 
Let us denote by $D_{ij, A}$ the covariant derivative with respect to moduli which corresponds 
to inserting the marginal operator $V_{ij,A}$ in various correlation functions. It is easy to see that the 
resulting
variation in the lattice momenta are given by   
\begin{equation}
D_{ij,A} P^L_{kl} =  \epsilon_{ijkl} P^R_A,  ~~~~~  
D_{ij,A} P^R_B = \delta_{AB} P^L_{ij}.
\end{equation}
This definition of the covariant derivative naturally follows from
the insertion of the marginal operators as we will see below. However
what this covariant derivative precisely means geometrically in $N=4$
supergravity needs to be understood.

The resulting variation in the topological amplitude $\mathcal{F}^{(3,\text{HET})}_g(u,v) $ is given by
\begin{align}
\epsilon^{ijkl}D_{kl,A} &\mathcal{F}^{(3,\text{HET})}_g(u_1,u_2) = 2\int \frac{d^2\tau}{\tau_2^3}\frac{1}{\bar{\eta}^{24}} 
\tau_2^{2g+2} G_{g+1} \sum_{(P^L,P^R) \in \Gamma^{(6,22)}}
\left(u_1 P^L u_2\right)^{2g-3}\cdot \nonumber\\ 
&\cdot\big[(2g-2)(u_1^i u_2^j- u_1^j u_2^i)+\frac{\pi i}{2}(\tau-\bar{\tau})\epsilon^{ijkl} (P^L)_{kl}\left(u_1P^L u_2\right)\big]P^R_A~
q^{\frac{1}{2}(P^L)^2}\bar{q}^{\frac{1}{2}(P^R)^2},
\label{hetanomaly1}
\end{align}
where $(u_1 P^L u_2) \equiv u_1^k u_2^l P^L_{kl}$. This equation can also be obtained directly by inserting 
the marginal operator $V_{ij,M}$ in the definition of $\mathcal{F}^{(3,\text{HET})}_g(u,v)$ and using the 
following
formula for the correlation function:
\begin{eqnarray}
\label{corr}
\langle (\partial X_{kl}\bar{J}_A)(w,\bar{w})&&\hspace{-1cm}\prod_{a=1}^{2g-2} (u_1 \partial X(z_a) u_2)
\rangle ~=~(2\pi)^{2g} P^L_{kl}P^R_I (u_1 P^L u_2)^{2g-2} \\ 
&+&\hspace{-0.2cm} 2\sum_{a=1}^{2g-2}(\partial_w^2 \log \theta_1 (w-z_a) ) 
\epsilon_{klmn} u_1^m u_2^n (2\pi)^{2g-2}(u_1 P^L u_2)^{2g-3} P^R_A.\nonumber
\end{eqnarray}
Now we integrate $w$ in the second term by partial integration using the formula
\begin{equation}
\int d^2 w \partial_w^2 \log \theta_1 (w-z_a) = \int d^2 w [\partial_w 
\bigl{(}\partial_w \log \theta_1 (w-z_a) + 2\pi i\frac{\text{Im}(w-z_a)}{\tau_2} 
\bigr{)} - \frac{\pi}{\tau_2} ]
= -\pi,
\label{corrint}
\end{equation}
where in the second equality we have used the fact that 
$(\partial_w \log \theta_1 (w-z_a) + 2\pi i\frac{\text{Im}(w-z_a)}{\tau_2})$ is periodic on the torus.  
Using eqs. (\ref{corrint}) and (\ref{corr}) we obtain eq.(\ref{hetanomaly1}). 

Now we can apply the derivative operators $D^1_i$ and $D^2_j$ defined in eq.(\ref{DIi}) with the result
{\allowdisplaybreaks
\begin{eqnarray}
D^2_j [\left(u_1P^L u_2\right)^{2g-3}( u_1^i u_2^j-u_1^j u_2^i)] &=&  -2g u_1^i  \left(u_1P^L u_2\right)^{2g-3},
\nonumber \\
&&\nonumber\\
D^1_i D^2_j [\left(u_1 P^L u_2\right)^{2g-3} ( u_1^i u_2^j-u_1^j u_2^i)] &=& 2g(2g+1) 
\left(u_1 P^L u_2\right)^{2g-3}, \nonumber \\
&\nonumber\\
D^2_j[\epsilon^{ijkl}P^L_{kl}\left(u_1 P^L u_2\right)^{2g-2}] &=& -(2g-2) \epsilon^{ijkl}P^L_{kl}
\left(u_1 P^L u_2\right)^{2g-3} (u_1^m P^L_{mj}),\nonumber \\
&\nonumber\\
D^1_i D^2_j [\epsilon^{ijkl}P^L_{kl}\left(u_1 P^L u_2\right)^{2g-2}] &=& (2g-2) [ 8 P_L^2  
\left(u_1 P^L u_2\right)^{2g-3} + \nonumber\\ 
&&+(2g-3)  \left(u_1 P^L u_2\right)^{2g-4} \epsilon^{ijkl} P^L_{kl} (u_1^m P^L_{mj}) (P^L_{in} u_2^n) ] \nonumber \\
&=& 2(2g-2)(2g+1)  \left(u_1 P^L u_2\right)^{2g-3} (P^L)^2,\label{uvder}
\end{eqnarray}�where in the last equality we have used the relation 
\begin{align}
\epsilon^{ijkl} P^L_{kl} (u_1^m P^L_{mj}) 
(P^L_{in} u_2^n) = 2  \left(u_1 P^L u_2\right) P_L^2,
\end{align}
which can be proven by comparing the coefficients of
$u_1^m u_2^n$ on both sides. Using eqs. (\ref{uvder}) in (\ref{hetanomaly1}) we obtain
\begin{align}
\epsilon^{ijkl}D^1_i D^2_j D_{kl,A} \mathcal{F}^{(3,\text{HET})}_g(u_1,u_2) &= 4i(2g-2)(2g+1)\int d^2\tau
\frac{1}{\bar{\eta}^{24}}  G_{g+1}\cdot \nonumber\\
&\cdot\frac{\partial}{\partial \tau} \tau_2^{2g} 
\sum_{(P^L,P^R) \in \Gamma^{(6,22)}} \left(u_1 P^L u_2\right)^{2g-3} P^R_A~
q^{\frac{1}{2}(P^L)^2}\bar{q}^{\frac{1}{2}(P^R)^2}.
\label{hetanomaly2}
\end{align}
Note that apart from the factor $G_{g+1}$, the total derivative with respect to $\tau$
can also be understood from the requirement of world-sheet modular invariance.
We can now carry out a partial integration with respect to $\tau$. Since the boundary term 
is modular invariant as follows from the second equation of (\ref{Ggeq}) and in the infrared limit $\tau_2 
\rightarrow \infty$ there is exponential suppression due to the presence of 
$P_L$ (for $g > 1$) we conclude that boundary terms vanish. The only contribution therefore
comes when the $\tau$-derivative acts on $G_{g+1}$. Using the second equation of (\ref{Ggeq})
we get: 
\begin{align}
\epsilon^{ijkl}D^1_i D^2_j D_{kl,A} \mathcal{F}^{(3,\text{HET})}_g(u_1,u_2) &= -2 \pi (2g-2)(2g+1)
\int d^2\tau
\frac{1}{\bar{\eta}^{24}}  G_{g}  \tau_2^{2g-2}\cdot\nonumber\\
&\cdot\sum_{(P^L,P^R) \in \Gamma^{(6,22)}} \left(u_1 P^L u_2\right)^{2g-3} P^R_A~ 
q^{\frac{1}{2}(P^L)^2}\bar{q}^{\frac{1}{2}(P^R)^2}.
\end{align}
By using  eq.(\ref{hetanomaly1}) we can rewrite the right hand side of the above equation to obtain the
following recursion relation:
\begin{equation}
\epsilon^{ijkl}D^1_i D^2_j D_{kl,A} \mathcal{F}^{(3,\text{HET})}_g(u_1,u_2) = 
(2g-2)(2g+1) u_1^i u_2^j D_{ij,A} \mathcal{F}^{(3,\text{HET})}_{g-1}(u_1,u_2).
\label{harmohet}
\end{equation}

This equation is satisfied also for $g=1$ but in a trivial way since by construction
$\mathcal{F}^{(3,\text{HET})}_1$ does not depend on $u$ and $v$. As a result, the left hand 
side vanishes and similarly the right hand side is zero due to the 
presence of the $(2g-2)$ factor. However, there may be a non-trivial equation satisfied 
by $\mathcal{F}^{(3,\text{HET})}_1$ analogous to the $t\bar{t}$ equation in the 
$N=2$ case, but we will not attempt to analyze it here.
  
The fact that the right hand side involves $\mathcal{F}^{(3,\text{HET})}_g$ with a lower value of $g$ suggests 
that on the type II side
there must be boundary contributions coming from degeneration limits of the Riemann surface, as in 
the holomorphicity equation (\ref{holomorphicity}) of the $N=2$ $F_g$'s.
It will be interesting to study the harmonicity condition in the type II side and check whether
that relation is mapped to the heterotic relation proven above under the duality map in
the appropriate weak coupling limit. However, as mentioned above, studying such relation in the
type II side involves introducing RR moduli fields which complete the $SO(6)$ representation of the $K3$ moduli. 
Even though turning on RR moduli in the Polyakov action is difficult, one may study 
this relation in the first order perturbation with respect to RR moduli.   

\subsection{Decompactification to $T^4$ and Connection to the Harmonicity Relation for Type II on $K3$}
Focusing on the heterotic 1-loop expression on $T^6$
\begin{align}
\mathcal{F}^{(3,\text{HET})}_g=\int \frac{d^2\tau}{\tau_2^3}\frac{\tau_2^{2g+2}}{\bar{\eta}^{24}}G_{g+1}
(\tau,\bar{\tau})\left(u_1P^Lu_2\right)^{2g-2}q^{\frac{1}{2}(P^L)^2}\bar{q}^{\frac{1}{2}(P^R)^2},\label{het1loop}
\end{align}
we want to study the behavior of the harmonicity equation (\ref{harmohet}) in the decompactification limit for one of the $T^2$. If harmonicity equations are a consequence of supersymmetry, in this limit, 
it is expected that (\ref{trueharm}) and (\ref{trueharmR}) are recovered.
More precisely, by heterotic - type II duality studied in Section \ref{sect:dual}, $\mathcal{F}^{(3,\text{HET})}_g$ is mapped to the $T^2$ decompactification limit of $\mathcal{F}^{(3)}_g$. From its explicit form computed in Section \ref{genusg+1}, the resulting expression is not anymore topological on K3, since there are leftover $\det({\rm Im}\tau)$ factors from the non-compact space-time coordinates.
Indeed, from eq.~(\ref{fgthree}), in the limit where the torus coordinate $X_3$ is non compact, the 
$\partial X_3$ from the left-moving sector and $\bar\partial{\bar X}_3$ from the right-movers give rise to contact terms $\langle\partial X_3\bar\partial{\bar X}_3\rangle\sim \omega_a({\rm Im}\tau)^{-1}_{ab} \bar{\omega}_b$, where $a,b=1,...,g+1$ where $g+1$ is the genus of the surface and $\omega_a$ are
the normalized holomorphic Abelian differentials. This leads to:
\begin{eqnarray}
\mathcal{F}^{(3)}_g&\sim&\int_{\mathcal{M}_{g+1}}{1\over \det({\rm Im}\tau)} 
\epsilon^{a_1..a_{g+1}}\epsilon^{d_1..d_{g+1} }\prod_{i=1}^{g+1}\int\mu_i~ \omega_{a_i} \omega_{b_i}({\rm Im}\tau)^{-1}_{b_i c_i} \int \bar{\mu}_i~\bar{\omega}_{c_i} \bar{\omega}_{d_i}\nonumber\\
&~&~~~~~~~\langle |\prod_{a=g+2}^{3g-1} G^-_{K3}(\mu_a)J^{--}_{K3}(\mu_{3g})|^2\rangle\, ,
\end{eqnarray}
where a complete antisymmetrization of the Beltrami's is understood. Despite its non topological nature, this quantity still satisfies the 6d harmonicity equations (\ref{trueharm}) and (\ref{trueharmR}), as we show below by considering the decompactification limit of the heterotic expression (\ref{het1loop}).

In (\ref{het1loop}), $P^L$ are the lattice vectors of a $\Gamma^{(6,22)}$. If we split $T^6$ into 
$T^4\times T^2$, this lattice gets broken to
\begin{align}
\Gamma^{(6,22)}\to \Gamma^{(4,20)}\otimes \Gamma^{(2,2)}.
\end{align}
Let us further denote by $P_{13}$ and its complex conjugate $P_{24}$ of $\Gamma^{(6,22)}$ to be entirely in $\Gamma^{(2,2)}$
and the remaining four $P_{12}$, $P_{14}$ and their complex conjugates $P_{34}$, $P_{23}$ to be
in $\Gamma^{(4,20)}$. $SU(4)$ now splits into $SU(2)_L\times SU(2)_R$ where $SU(2)_L$ acts on the
indices $(1,3)$ and $SU(2)_R$ on the indices $(2,4)$, respectively. In the decompactification limit
$P_{13}$ and $P_{24}$ decouple, so the relevant part of the $SU(4)$ harmonics $u_I^i$ can be assembled
into the harmonics of $SU(2)_L\times SU(2)_R$ with the identification $u_1^i, u_3^i \rightarrow
v_1^{a}, v_2^{a}$ and $u_2^i, u_4^i \rightarrow
v_{\dot{1}}^{\dot{a}}, v_{\dot{2}}^{\dot{a}}$, where we have used dotted and undotted indices $a,\dot{a}=1,2$ to indicate that they are respectively $SU(2)_R$ and $SU(2)_L$ harmonics. In this notation, 
$\Gamma^{(4,20)}$ lattice vectors are denoted by:
\begin{align}
P^{(4,20),L}_{a\dot{b}}\ ,\ P^{(4,20),R}_A\hspace{2cm}\text{with}\ a,\dot{b}=1,2
\end{align}
and now the vector multiplet label $A=1,\ldots,20$. Moreover
\begin{align}
(P^{(4,20),L})^2\equiv \frac{1}{2}P^{(4,20),L}_{a_1\dot{b}_1}P^{(4,20),L}_{a_2\dot{b}_2}\epsilon^{a_1a_2}\epsilon^{\dot{b}_1\dot{b}_2}\, .
\end{align}
The $SU(2)_L\times SU(2)_R$ harmonics satisfy:
\begin{align}
&v_1^a=\epsilon^{ab}v_{1,b}=\overline{v_a^1}=v^{2a}, &&v_{\dot{1}}^{\dot{a}}=\epsilon^{\dot{a}\dot{b}}v_{\dot{1},\dot{b}}=\overline{v_{\dot{a}}^{\dot{1}}}
=v^{\dot{2}\dot{a}},
\end{align} 
as well as $|v_1|^2=|v_2|^2=|v_{\dot{1}}|^2= |v_{\dot{2}}|^2=1$, where $|v_1|^2= v_1^a v_a^1$ etc.

Expression (\ref{het1loop}) in the decompactification limit of $T^2$ then goes into\footnote{The $\Gamma^{(2,2)}$ contribution leads to a $\tau_2^{-1}$ factor, together with the $T^2$ volume which is dropped.}
\begin{align}
\mathcal{F}^{(3,\text{HET})}_g\sim\int \frac{d^2\tau}{\tau_2^4}\frac{\tau_2^{2g+2}}{\bar{\eta}^{24}} 
G_{g+1}(\tau,\bar{\tau})\sum_{(P_L,P_R)\in \Gamma^{(4,20)}}\left(v_1^aP^{L}_{a\dot{b}}v_{\dot{1}}^{\dot{b}}\right)^{2g-2}
q^{\frac{1}{2}P_{L}^2}\bar{q}^{\frac{1}{2}P_R^2}\, .
\end{align}
Introducing the differential operator similar to (\ref{DIi})
\begin{align}
D_a^I=\sum_{J=1}^2v_a^JD_J^I\equiv\sum_{J=1}^2v_a^J\left(v_b^I\frac{\partial}{\partial v_b^J}-u_J^b\frac{\partial}{\partial v_I^b}\right)=\sum_{J=1}^2v_a^Jv_b^I\frac{\partial}{\partial v_b^J}-\frac{\partial}{\partial v_I^a},
\end{align}
where we have used
\begin{align}
\sum_{I=1}^2v_a^Iv_I^b=\delta_a^b\, .
\end{align}
Note that since $\mathcal{F}^{(3,\text{HET})}_g$ involves only $v_1^a$ and $v_{\dot{1}}^{\dot{a}}$,
$D_a^1$ reduces essentially to the second term on the right hand side namely $-\frac{\partial}{\partial v_I^a}$. Now we can apply the operator
\begin{align}
\epsilon^{ab}D^1_a\frac{D}{D\lambda_A^{b\dot{a}}}
\end{align}
onto $\mathcal{F}^{(3,\text{HET})}_g$, with the following result:
\begin{align}
&\epsilon^{ab}D^1_a\frac{D}{D\lambda_A^{b\dot{a}}}\mathcal{F}^{(3,\text{HET})}_g\sim \nonumber\\
&\sim (2g-2)\epsilon_{\dot{a}\dot{b}}v_{\dot{1}}^{\dot{b}}\int \frac{d^2\tau}{\tau_2^4} \frac{\tau_2^{2g+2}}{\bar{\eta}^{24}}G_{g+1}(\tau,\bar{\tau})\quad\cdot\nonumber\\
&\qquad\cdot\quad \sum_{(P_L,P_R)\in \Gamma^{(4,20)}}
[(2g-1) -2\pi \tau_2 P_{L}^2] P^R_A
\left(v_1^cP^{L}_{c\dot{c}}v_{\dot{1}}^{\dot{c}}\right)^{2g-3}q^{\frac{1}{2}P_{L}^2} 
\bar{q}^{\frac{1}{2}P_R^2}\nonumber\\
&=(2g-2)\epsilon_{\dot{a}\dot{b}}v_{\dot{1}}^{\dot{b}}\int d^2\tau G_{g+1}(\tau,\bar{\tau})\frac{\partial}{\partial \tau} \frac{\tau_2^{2g-1}}{\bar{\eta}^{24}} \sum_{(P_L,P_R)\in \Gamma^{(4,20)}}P^R_A
\left(v_1^cP^{L}_{c\dot{c}}v_{\dot{1}}^{\dot{c}}\right)^{2g-3}q^{\frac{1}{2}P_{L}^2}
\bar{q}^{\frac{1}{2}P_R^2}\nonumber\\
&\sim (2g-2)\epsilon_{\dot{a}\dot{b}}v_{\dot{1}}^{\dot{b}}(v_1^c v_{\dot{1}}^{\dot{c}} \frac{D}{D\lambda^{c\dot{c}}})
\mathcal{F}^{(3,\text{HET})}_{g-1}\, .
\end{align}
Note that the last expression is again an anomalous term coming from a lower genus contribution. Multiplying now by $v_{\dot{1}}^{\dot{a}}$ and summing over $\dot{a}$ we see that the right hand side vanishes. This equation resembles exactly the harmonicity relation (\ref{trueharm}) of \cite{Ooguri:1995cp}, without any anomaly.
Moreover, by exchanging $v_1$ with $v_{\dot{1}}$, one finds also a second equation which has precisely the same form as (\ref{trueharmR}).

\section{Conclusions}\label{sect:conclusion}

In summary, we have studied $N=4$ topological amplitudes in string theory. Unlike the $N=2$ case of the 
topological Calabi-Yau $\sigma$-model, the $N=4$ topological partition function vanishes trivially and one 
has to consider multiple correlation functions. In the `critical' case of $K3$ $\sigma$-model, the definition 
is essentially unique and the number of additional vertex insertions increases with the genus $g$ of the 
Riemann surface \cite{Berkovits:1994vy}. Furthermore, it computes a physical string amplitude in the 6d 
effective action with four gravitons and $(4g-4)$ graviphotons, of the type $R^4T^{4g-4}$.
However, we found that contrary to the $N=2$ case, on the heterotic side these amplitudes are not simplified 
because they start receiving contributions at higher loops $(g+1)$. On the other hand, it turns out that on 
$K3\times T^2$ one has a lot of freedom.
In this work, we made a systematic study of possible definitions with the following properties: 
(1) they are `economic', involving a minimum number of additional vertices;
(2) they compute some (gravitational) couplings in the low energy 4d string effective action; (3) they have 
interesting heterotic duals that allow their study by explicit computations.

Indeed, we found two such non-trivial definitions. The first uses two additional integrated vertices of the 
$U(1)$ currents of $K3$ and $T^2$, $J_{K3}$ and $J_{T^2}$ (in left and right movers separately) and computes 
the physical coupling of the 4d string effective action term $R_+^2R_-^2T_+^{2g-2}$ (and its supersymmetric 
completion), where the $+/-$ indices refer to self-dual/anti-self-dual parts of the corresponding field-strength. 
Unfortunately, on the heterotic side, they start receiving contributions at two loops. In the second definition, 
the $T^2$ integrated current $J_{T^2}=\psi_3\bar\psi_3$ is replaced by just the fermionic coordinate $\psi_3$ 
of dimension zero in the topological theory. Taking into account the right-moving sector, this amounts to 
differentiation with respect to the $T^2$ K\"ahler modulus, which is needed for obtaining a non-vanishing result. 
It turns out that this is the `minimal' definition that computes the physical coupling of the term 
$R_+^2(dd\Phi_+)^2T_+^{2g-2}$, with $\Phi_+$ a KK graviscalar corresponding to $T^2$ K\"ahler modulus. 
In the heterotic theory compactified on $T^6$, $\Phi_+$ is mapped to the dilaton and this amplitude is mapped 
at one loop in the appropriate limit, and thus, it can be studied as the $N=2$ $F_g$'s.

We also studied the dependence of the above couplings, that we call $N=4$ $\mathcal{F}_g$, on the 
compactification moduli which belong to 4d $N=4$ supermultiplets. Being BPS-type, they satisfy a harmonicity 
equation upon introducing appropriate harmonic superspace variables, that generalizes holomorphicity of 
$N=2$ $F_g$'s and the harmonicity of the 6d case found in \cite{Berkovits:1994vy, Berkovits:1998ex,Ooguri:1995cp}. 
We derived this equation on the heterotic side, where the full $SU(4)$ R-symmetry is manifest, and we 
uncovered an anomaly due to boundary contributions, analog to the holomorphic anomaly of $N=2$. 
It will be interesting to further study this equation on the type II side, compute explicit examples of 
the minimal $N=4$ $\mathcal{F}_g^{(3)}$'s and work out possible applications of the $N=4$ topological amplitudes, 
such as in the entropy of $N=4$ black holes.

Harmonicity equation in the topological theory side is essentially a statement of decoupling
of the BRS exact states although in the $N=4$ theory it appears in a more complicated way. In the
untwisted theory this BRS operator is translated to certain space-time supersymmetry generators
(this statement can be made precise at least in $N=2$ theories). Therefore
decoupling of BRS exact states should imply that the corresponding terms in the effective action
must be BPS operators. Therefore this equation must be a consequence of supersymmetry and the
shortening of the multiplets.
An interesting open question is to understand what this harmonicity equation precisely means in the context
of $N=4$ supergravity. 

\section*{Acknowledgements}
We thank S.~Ferrara, W.~Lerche, T.~Maillard, M.~Mari$\tilde{\text{n}}$o, M. Weiss and especially E. Sokatchev for enlightening discussions. 
This work was supported in part by the European Commission under the RTN contract MRTN-CT-2004-503369 and in part
by the INTAS contract 03-51-6346. The work of S.H. was supported by the Austrian Bundesministerium f\"ur Bildung, 
Wissenschaft und Kultur.

\def\thesubsection{\Alph{section}.\arabic{subsection}}

\appendix
\section{Gamma Matrices and Lorentz Generators}\label{AppendixGamma}
In any number of space-time dimensions, the Lorentz generators are defined as commutators of 
the $\Gamma$-matrices 
\begin{align}
{(\Sigma^{\mu\nu})_{\underline{a}}}^{\underline{b}}=-\frac{i}{4}
{[\Gamma^\mu,\Gamma^\nu]_{\underline{a}}}^{\underline{b}}\label{deflor}
\end{align} 
Upon choosing a suitable basis, the Lorentz generators can be brought into diagonal 
form, comprising the $\sigma^{\mu\nu}$ used in Section \ref{SectTopAmpK3}
\begin{align}
{(\Sigma^{\mu\nu})_{\underline{a}}}^{\underline{b}}=\left(\begin{array}{cc}
{(\sigma^{\mu\nu})_a}^b & 0 \\ 0 & {(\bar{\sigma}^{\mu\nu})^{\dot{a}}}_{\dot{b}} 
\end{array}\right).
\end{align}
The superspace integrals in Section \ref{SectTopAmpK3} involve traces over the 
`reduced' $\sigma^{\mu\nu}$, which are related to traces over the full 
$\Sigma^{\mu\nu}$ using the following identities
\begin{align}
&\text{tr}\ \Sigma^{\mu\nu}=\text{tr}\ \sigma^{\mu\nu}=0,\\
&\text{tr}\ (\Sigma^{\mu\nu}\Sigma^{\rho\sigma})=2\text{tr}\ (\sigma^{\mu\nu}
\sigma^{\rho\sigma})=\frac{d_{\text{spin}}}{4}(\eta^{\mu\rho}\eta^{\nu\sigma}-
\eta^{\mu\sigma}\eta^{\nu\rho}),\label{sigsig2}\\
&\text{tr}(\Sigma^{\mu_1\nu_1}\Sigma^{\mu_2\nu_2}\Sigma^{\mu_3\nu_3}\Sigma^{\mu_4\nu_4})
=\text{tr}(\sigma^{\mu_1\nu_1}\sigma^{\mu_2\nu_2}\sigma^{\mu_3\nu_3}\sigma^{\mu_4\nu_4})+
\text{tr}(\sigma^{\mu_4\nu_4}\sigma^{\mu_3\nu_3}\sigma^{\mu_2\nu_2}\sigma^{\mu_1\nu_1}),
\label{sigsig4}
\end{align}
with\footnote{Note, that the transition to Minkowski space is easily achieved by replacing 
$\Sigma^{0\mu}\to i\Sigma^{0\mu}$.}
\begin{align}
\eta^{\mu\nu}=\text{diag}(1,\ldots,1),
\end{align}
and $d_{\text{spin}}$ the spinor dimension, which in the case of 6 space-time dimensions is $8$.

For completeness we also list all the $\sigma^{\mu\nu}$ 
\small
{\allowdisplaybreaks
\begin{align}
&\sigma^{01}=\frac{1}{2}\left(
\begin{array}{cccc}
 1 & 0 & 0 & 0  \\
 0 & 1 & 0 & 0  \\
 0 & 0 & -1 & 0 \\
 0 & 0 & 0 & -1
\end{array}
\right), 
&&\sigma^{02}=\frac{1}{2}\left(
\begin{array}{cccc}
 0 & 0 & 0 & -1 \\
 0 & 0 & 1 & 0  \\
 0 & 1 & 0 & 0  \\
 -1 & 0 & 0 & 0
\end{array},
\right)
&&\sigma^{03}=\frac{1}{2}\left(
\begin{array}{cccc}
 0 & 0 & 0 & i  \\
 0 & 0 & i & 0  \\
 0 & -i & 0 & 0 \\
 -i & 0 & 0 & 0
\end{array}
\right),\nonumber\\
&\nonumber\\
&\sigma^{04}=\frac{1}{2}\left(
\begin{array}{cccccccc}
 0 & 0 & 1 & 0 \\
 0 & 0 & 0 & 1 \\
 1 & 0 & 0 & 0 \\
 0 & 1 & 0 & 0 
\end{array}
\right),
&&\sigma^{05}=\frac{1}{2}\left(
\begin{array}{cccc}
 0 & 0 & -i & 0 \\
 0 & 0 & 0 & i  \\
 i & 0 & 0 & 0  \\
 0 & -i & 0 & 0
\end{array}
\right),
&&\sigma^{12}=\frac{1}{2}\left(
\begin{array}{cccc}
 0 & 0 & 0 & i \\
 0 & 0 & -i & 0  \\
 0 & i & 0 & 0 \\
 -i & 0 & 0 & 0 
\end{array}
\right), \nonumber\\
&\nonumber\\
&\sigma^{13}=\frac{1}{2}\left(
\begin{array}{cccc}
 0 & 0 & 0 & 1 \\
 0 & 0 & 1 & 0 \\
 0 & 1 & 0 & 0 \\
 1 & 0 & 0 & 0 
\end{array}
\right), 
&&\sigma^{14}=\frac{1}{2}\left(
\begin{array}{cccc}
 0 & 0 & -i & 0 \\
 0 & 0 & 0 & -i \\
 i & 0 & 0 & 0 \\
 0 & i & 0 & 0 
\end{array}
\right),
&&\sigma^{15}=\frac{1}{2}\left(
\begin{array}{cccc}
 0 & 0 & -1 & 0 \\
 0 & 0 & 0 & 1  \\
 -1 & 0 & 0 & 0 \\
 0 & 1 & 0 & 0 
\end{array}
\right), \nonumber\\
&\nonumber\\
&\sigma^{23}=\frac{1}{2}\left(
\begin{array}{cccc}
 1 & 0 & 0 & 0 \\
 0 & -1 & 0 & 0 \\
 0 & 0 & 1 & 0 \\
 0 & 0 & 0 & -1
\end{array}
\right),
&&\sigma^{24}=\frac{1}{2}\left(
\begin{array}{cccc}
 0 & i & 0 & 0 \\
 -i & 0 & 0 & 0  \\
 0 & 0 & 0 & -i  \\
 0 & 0 & i & 0
\end{array}
\right), 
&&\sigma^{25}=\frac{1}{2}\left(
\begin{array}{cccc}
 0 & 1 & 0 & 0 \\
 1 & 0 & 0 & 0 \\
 0 & 0 & 0 & 1 \\
 0 & 0 & 1 & 0 
\end{array}
\right),\nonumber\\
&\nonumber\\
&\sigma^{34}=\frac{1}{2}\left(
\begin{array}{cccc}
 0 & 1 & 0 & 0 \\
 1 & 0 & 0 & 0 \\
 0 & 0 & 0 & -1 \\
 0 & 0 & -1 & 0
\end{array}
\right), 
&&\sigma^{35}=\frac{1}{2}\left(
\begin{array}{cccc}
 0 & -i & 0 & 0 \\
 i & 0 & 0 & 0 \\
 0 & 0 & 0 & -i \\
 0 & 0 & i & 0
\end{array}
\right),
&&\sigma^{45}=\frac{1}{2}\left(
\begin{array}{cccc}
 1 & 0 & 0 & 0 \\
 0 & -1 & 0 & 0 \\
 0 & 0 & -1 & 0 \\
 0 & 0 & 0 & 1
\end{array}
\right).\nonumber
\end{align}}
\normalsize
\section{Generalized Self-Duality in 6 Dimensions}\label{AppendixSelfDual}
The standard notion of (anti-)self-duality 
\begin{align}
\check{F}=F,\hspace{1cm}\text{or}\hspace{1cm}\check{F}=-F,
\end{align}
does not make sense in 6 dimensions. However, there is another possibility to generalize 
the notion of self-duality. Considering for example the vertex of a RR field in the 
$\left(-1/2\right)$-ghost picture (\ref{vertexgraviphoton})
\begin{align}
V_{T}^{\left(-1/2\right)}(p,\epsilon)=:e^{-\frac{1}{2}(\varphi+\tilde{\varphi})}
p_\nu\epsilon_{\mu}\bigg[\underbrace{S^a{(\sigma^{\mu\nu})_a}^b\tilde{S}_b 
\mathcal{S}\tilde{\mathcal{S}}(z,
\bar{z})}_{\text{self-dual}}+\underbrace{S_{\dot{a}}{(\bar{\sigma}^{\mu\nu})^{\dot{a}}}_{\dot{b}}
\tilde{S}^{\dot{b}}\bar{\mathcal{S}}
\bar{\tilde{\mathcal{S}}}(z,\bar{z})}_{\text{anti-self-dual}}\bigg]e^{ip\cdot Z}:,
\end{align}
the two terms corresponding by definition to self-dual or anti-self-dual terms of the field strength 
tensor. This entails that it is possible to project to (anti-)self-dual parts of a 2-index antisymmetric 
tensor by just contracting with ${(\sigma^{\mu\nu})_a}^b$ (${(\bar{\sigma}^{\mu\nu})^{\dot{a}}}_{\dot{b}}$):
\begin{align}
{{F^+}_a}^b\equiv {(\sigma^{\mu\nu})_a}^b F_{\mu\nu},\\
{{F^-}^{\dot{a}}}_{\dot{b}}\equiv {(\bar{\sigma}^{\mu\nu})^{\dot{a}}}_{\dot{b}} F_{\mu\nu}.
\end{align}
Since the Lorentz-generators can be defined in any number of dimensions, it is clear, that 
this notion of self-duality applies also in 6 dimensions. Note
however that although the 4-dimensional ${(\sigma^{\mu\nu})_a}^b$ 
and ${(\bar{\sigma}^{\mu\nu})^{\dot{a}}}_{\dot{b}}$ project to linearly independent 
sub-spaces, this is no longer true in 6 dimensions.
\section{Theta Functions and Riemann Identity}\label{AppendixRiemanntot}
\subsection{Theta Functions, Spin Structures and Prime Forms}
In this appendix, we basically review chapter 3 of \cite{Verlinde:1986kw}. Consider a compact 
Riemann surface of genus $g$ and choose a canonical basis of homology cycles $a_i,b_i$ 
according to Figure \ref{fig:cycles}.
\begin{figure}[htb]
\begin{center}
\epsfig{file=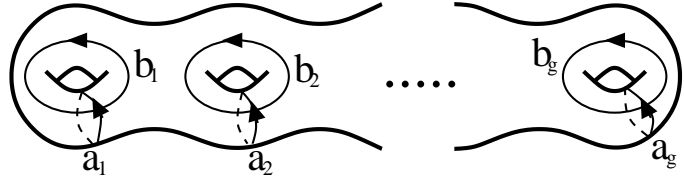, width=9cm}
\caption{A basis of homology cycles on a genus $g$ compact Riemann surface $\Sigma$.}
\label{fig:cycles}
\end{center}
\end{figure} 
$\ $\\[10pt]
Furthermore, there is a normalized basis of holomorphic 1-forms $\omega_i$, with 
$i=1,\ldots g$, whose integrals along the homology cycles read
\begin{align}
\oint_{a_i}\omega_j=\delta_{ij}, &&\oint_{b_i}\omega_j=\tau_{ij}, 
\end{align}
where $\tau_{ij}$ is a complex, symmetric $g\times g$ matrix which is called period matrix.

One can then associate `coordinates' with points on the Riemann surface. 
For this, we choose a base point $P_0$ and define for each 
point $P$ the Jacobi map\footnote{In computations, for notational simplicity, we drop the distinction 
between the point on the Riemann surface and the corresponding Jacobi map.}
\begin{align}
\mathcal{I}:\ P\to z_i(P)=\int_{P_0}^P\omega_i.
\end{align}
In this respect, $z$ is an element of the complex torus
\begin{align}
J(\mathcal{M}_g)=\mathbb{C}^g/(\mathbb{Z}^g+\tau \mathbb{Z}^g).
\end{align}
One can now define the Riemann theta function \cite{Fay} for $z\in J(\mathcal{M}_g)$
\begin{align}
\vartheta (z,\tau)=\sum_{n\in \mathbb{Z}^g}e^{i\pi n_i\tau_{ij}n_j+2\pi i n_i z_i}.
\end{align}
Shifting $z$ by a vector of the lattice $\mathbb{Z}^g+\tau \mathbb{Z}^g$, the Riemann 
theta function transforms as
\begin{align}
\vartheta(z+\tau n+m,\tau)=e^{-i\pi n\tau n-2\pi imz}\vartheta(z,\tau).
\end{align}
We also use the Riemann vanishing theorem: There exists a divisor class 
$\Delta$ of degree $g-1$ such that $\vartheta(z,\tau)=0$ if and only if there are $g-1$ 
points $p_1,\ldots p_{g-1}$ on $\mathcal{M}_g$ such that
\begin{align}
z=\Delta-\sum_{i=1}^{g-1}p_i
\end{align}
Furthermore, to every spin structure $\alpha$, one can associate a theta function with 
characteristics $(\alpha_1,\alpha_2)\in\left(\frac{1}{2}\mathbb{Z}^g/\mathbb{Z}^g\right)$ defined by
\begin{align}
\vartheta_\alpha(z,\tau)=e^{i\pi \alpha_1\tau\alpha_1+2\pi i\alpha_1(z+\alpha_2)}
\vartheta(z+\tau\alpha_1+\alpha_2,\tau).\label{thetadef}
\end{align}

Besides $\vartheta$-functions, the prime forms $E(x,y)$ enter also 
in our computations. One can view the prime form as a generalization of the 
holomorphic function $x-y$  on the Riemann sphere. The precise definition is
\begin{align}
E(x,y)=\frac{f_\alpha(x,y)}{h_\alpha(x)h_\alpha(y)}
\end{align}
where $h_\alpha$ is a holomorphic $\frac{1}{2}$-differential and $f_\alpha$ is given by
\begin{align}
f_\alpha(x,y)=\vartheta_\alpha(\int_x^y\omega).
\end{align}
One can show, that $E$ is independent of $\alpha$. Moreover, it is antisymmetric in $x,y$ exchange and has 
a simple root at $x=y$:
\begin{align}
&E(x,y)=-E(y,x),&&\text{and} &&\lim_{x\to y}E(x,y)=\mathcal{O}(x-y).
\end{align}
\subsection{Riemann Addition Theorem}\label{AppendixRiemann}
The sum over all different spin structures for the product of four $\vartheta$-functions gives:
\begin{align}
\sum_s&\vartheta_s\!(a)\vartheta_s\!(b)\vartheta_s\!(c)\vartheta_s\!(d)=\nonumber\\
&=\vartheta\!\bigg(\underbrace{\frac{a+b+c+d}{2}}_{=T_{++++}}\bigg)\vartheta\!
\bigg(\underbrace{\frac{-a-b+c+d}{2}}_{=T_{--++}}\bigg)\vartheta\!\bigg(\underbrace
{\frac{-a+b-c+d}{2}}_{=T_{-+-+}}\bigg)\vartheta\!
\bigg(\underbrace{\frac{-a+b+c-d}{2}}_{=T_{-++-}}\bigg)
\end{align}
where we omitted overall phase factors which are irrelevant for our computations.
\section{The $N=4$ Superconformal Algebra}\label{AppendixSuperconformal}
In this appendix, we review the $N=4$ superconformal algebra and explain our notation. 
We examine both four and six dimensional compactifications, or equivalently operators 
on $K3$ and $K3\times T^2$. We remind the reader however, that for the sake of simplicity of our calculations, 
we treat $K3$ only at its orbifold (free field) realization.

\subsection{Superconformal Algebra for Compactifications on $K3$}\label{AppendixSuperconformal6d}
The starting point is the spin-field of the internal theory which after bosonization of the fermionic 
coordinates $\psi_A=e^{i\phi_A}$, can be written with the help of two free 2d scalar fields
\begin{align}
\mathcal{S}=e^{\frac{i}{2}(\phi_4+\phi_5)},\hspace{1.5cm}\mathcal{S}^\dagger e^{-\frac{i}{2}(\phi_4+\phi_5)}.
\end{align}
According to \cite{Banks:1988yz} the current algebra (conformal dimension $1$ operators) of 
the SCFT can be uncovered by considering OPE of the spin fields among themselves:
\begin{align}
&\mathcal{S}(z)\mathcal{S}^\dagger(w)\sim\frac{1}{(z-w)^{\frac{3}{4}}}+
(z-w)^{\frac{1}{4}}J_{K3}+\ldots,\label{algconf1}\\
&\mathcal{S}(z)\mathcal{S}(w)\sim \frac{J^{++}_{K3}(w)}{(z-w)^{\frac{1}{4}}}+
\ldots,\label{algconf2}\\
&\mathcal{S}^\dagger(z)\mathcal{S}^\dagger(w)\sim \frac{J^{--}_{K3}(w)}
{(z-w)^{\frac{1}{4}}}+\ldots.\label{algconf3}
\end{align}
where the currents are given by
\begin{align}
&J_{K3}=\psi_4\bar{\psi}_4+\psi_5\bar{\psi}_5, && J^{++}_{K3}=\psi_4\psi_5, &&J^{--}_{K3}
=\bar{\psi}_4\bar{\psi}_5
\label{N4curs}
\end{align}
and satisfy an $SU(2)$ current algebra
\begin{align}
&J^{--}_{K3}J^{++}_{K3}\sim J_{K3}, && J_{K3}J^{++}_{K3}\sim 2J^{++}_{K3}, &&J_{K3}J^{--}_{K3}
=-2J_{K3}^{--}.\label{K3SU2curalg}
\end{align}

On the other hand, the $N=1$ world-sheet supercurrent is given by
\begin{align}
T_{F,K3}=\psi_4\partial\bar{X}_4+\psi_5\partial\bar{X}_5+
\bar{\psi}_4\partial X_4+\bar{\psi}_5\partial X_5
=G^+_{K3}+G^-_{K3}\nonumber
\end{align}
where $G^\pm_{K3}$ are the two (complex conjugate) supercurrents of the internal $N=2$ SCFT:
\begin{align}
&G^+_{K3}=\psi_4\partial \bar{X}_4+\psi_5\partial\bar{X}_5,\label{K3scurrent1}\\
&G^{-}_{K3}=\bar{\psi}_4\partial X_4+\bar{\psi}_5\partial X_5.
\label{K3scurrent2}
\end{align}
Indeed, $G^\pm_{K3}$ and the $U(1)$ current $J_{K3}$ form an $N=2$ superconformal algebra with the energy 
momentum tensor
\begin{align}
T_{B,K3}=\partial X_4\partial{\bar X}_4+\partial X_5\partial{\bar X}_5+
{1\over 2}(\psi_4{\leftrightarrow\atop\displaystyle{\partial\atop~}}{\bar\psi}_4+
\psi_5{\leftrightarrow\atop\displaystyle{\partial\atop~}}{\bar\psi}_5).\nonumber
\end{align}

In order to generate the $N=4$ superconformal algebra, one uses the $SU(2)$ 
currents (\ref{K3SU2curalg}) acting into (\ref{K3scurrent1}) and (\ref{K3scurrent2}) to find two additional 
supercurrents:
\begin{align}
&\tilde{G}^+_{K3}=-\psi_5\partial X_4+\psi_4\partial X_5,\\
&\tilde{G}^-_{K3}=-\bar{\psi}_5\partial \bar{X}_4+\bar{\psi}_4\partial \bar{X}_5.
\end{align}
Note that the four supercurrents form two $SU(2)$
doublets $(G^+_{K3},\tilde{G}^-_{K3})$ and $(\tilde{G}^+_{K3},G^-_{K3})$, while the superscripts denote $\pm 1$ 
unit of $U(1)$ charge. Schematically, 
one can move within one doublet with the help of the $SU(2)$ currents $J^{\pm\pm}_{K3}$
\begin{center}
\begin{tabular}{ccc}
$G^{+}_{K3}$ & \hspace{2cm} & $G^{-}_{K3}$\\
&&\\
$J^{++}_{K3}$\huge $\uparrow\ \downarrow$\normalsize $J^{--}_{K3}$ & & $J^{--}_{K3}$\huge $\uparrow\ 
\downarrow$\normalsize $J^{++}_{K3}$  \\
&&\\
$\tilde{G}^{-}_{K3}$ & \hspace{2cm} & $\tilde{G}^{+}_{K3}$
\end{tabular}
\end{center}
The OPE among the supercurrents read
 \begin{center}
\begin{tabular}{ccc}
$G^{+}_{K3}$ & \huge$\leftarrow$\normalsize \fbox{\parbox{2.8cm}{$\frac{J_{K3}}{z^2}-
\frac{T^B_{K3}-{1\over 2}\partial J_{K3}}{z}$}}\huge$\rightarrow$\normalsize & $G^{-}_{K3}$\\
&&\\
\parbox{0.6cm}{\huge$\uparrow$\normalsize\\\fbox{\parbox{0.2cm}{0}}\\ \huge$\downarrow$
\normalsize} &
\begin{tabular}{ccc}
\huge$\nwarrow$\normalsize&&\huge$\nearrow$\normalsize\\
&\fbox{\parbox{2.4cm}{$\mp\frac{2J_{K3}^{\pm\pm}}{z^2}\mp\frac{\partial J_{K3}^{\pm\pm}}{z}$}}&\\
\huge$\swarrow$\normalsize&&\huge$\searrow$\normalsize
\end{tabular}
 & \parbox{0.6cm}{\huge$\uparrow$\normalsize\\\fbox{\parbox{0.2cm}{0}}\\ \huge$\downarrow$
\normalsize} \\
&&\\
$\tilde{G}^{-}_{K3}$ & \huge$\leftarrow$\normalsize \fbox{\parbox{3.1cm}{$-\frac{J_{K3}}{z^2}+
\frac{T^B_{K3}-{1\over 2}\partial J_{K3}}{z}$}}\huge$\rightarrow$\normalsize & $\tilde{G}^{+}_{K3}$
\end{tabular}
\end{center}
Summarizing, the full $N=4$ superconformal algebra is given by the following OPE's:
\begin{align}
&J^{--}_{K3}(z)J^{++}_{K3}(0)\sim\frac{J_{K3}(0)}{z},\nonumber
\end{align}
\begin{align}
&J^{--}_{K3}(z)G^+(0)\sim\frac{\tilde{G}^-_{K3}(0)}{z}, &&J_{K3}^{++}(z)\tilde{G}_{K3}^-(0)
\sim-\frac{G^+_{K3}(0)}{z},\nonumber\\
&J_{K3}^{++}(z)G^-_{K3}(0)\sim \frac{\tilde{G}^+_{K3}(0)}{z}, &&J_{K3}^{--}(z)
\tilde{G}^+_{K3}(0)\sim-\frac{G^-_{K3}(0)}{z},\nonumber\\
&J_{K3}^{--}(z)G^-_{K3}(0)\sim 0, && J_{K3}^{++}(z)G^+_{K3}(0)\sim0,\nonumber\\
&J_{K3}^{--}(z)\tilde{G}^-_{K3}(0)\sim 0, && J_{K3}^{++}(z)\tilde{G}^+_{K3}(0)\sim0,
\nonumber\\
&G_{K3}^+(z)\tilde{G}^-_{K3}(0)\sim 0,&& G_{K3}^-(z)\tilde{G}^+_{K3}(0)\sim 0,\nonumber
\end{align}
\begin{align}
&G^+_{K3}(z)G^-_{K3}(0)\sim\frac{J_{K3}(0)}{z^2}-\frac{T^B_{K3}(0)-{1\over 2}\partial J_{K3}(0)}{z},
\nonumber\\
&\tilde{G}^+_{K3}(z)\tilde{G}^-_{K3}(0)\sim\frac{J_{K3}(0)}{z^2}-\frac{T^B_{K3}(0)-{1\over 2}
\partial J_{K3}(0)}{z},\nonumber\\
&\tilde{G}^+_{K3}(z)G^+_{K3}(0)\sim\frac{2J^{++}_{K3}(0)}{z^2}+\frac{\partial J^{++}_{K3}(0)}{z},
\nonumber\\
&\tilde{G}^-_{K3}(z)G^-_{K3}(0)\sim\frac{2J^{--}_{K3}(0)}{z^2}+\frac{\partial J^{--}_{K3}(0)}{z},
\nonumber
\end{align}
while for any operator $O^q_{K3}$ of $U(1)$ charge $q$, one has:
\begin{align}
J_{K3}(z)O^q_{K3}(0)\sim q\frac{O^q_{K3}(0)}{z}.
\nonumber
\end{align}
\subsection{Superconformal Algebra for Compactifications on $K3\times T^2$}
\label{AppendixSuperconformal4d}
Following the procedure of the $N=4$ case in 6 dimensions 
(\ref{algconf1})-(\ref{algconf3}), the starting point is given by 
the internal spin-fields. In the case of $K3\times T^2$, 
the number of spin-fields is doubled
\begin{align}
&\mathcal{S}_1=e^{\frac{i}{2}(\phi_3+\phi_4+\phi_5)},&&\mathcal{S}_1^\dagger
=e^{-\frac{i}{2}(\phi_3+\phi_4+\phi_5)},\\
&\mathcal{S}_2=e^{\frac{i}{2}(\phi_3-\phi_4-\phi_5)},&&\mathcal{S}_2^\dagger
=e^{-\frac{i}{2}(\phi_3-\phi_4-\phi_5)}.
\end{align} 
The conformal dimension $1$ generators are again obtained according to 
\cite{Banks:1988yz} by studying the OPE of the spin fields
\begin{align}
&\mathcal{S}_I(z)\mathcal{S}_J^\dagger(w)\sim\frac{\delta_{IJ}}{(z-w)^{\frac{3}{4}}}+
(z-w)^{\frac{1}{4}}J^{IJ}+\ldots,\\
&\mathcal{S}_I(z)\mathcal{S}_J(w)\sim\frac{\Psi^{IJ}(w)}{(z-w)^{\frac{1}{4}}}+\ldots,
\label{12objectdef}\\
&\mathcal{S}_I^\dagger(z)\mathcal{S}_J^\dagger(w)\sim\frac{\Psi^{\dagger IJ}(w)}
{(z-w)^{\frac{1}{4}}}+\ldots,
\end{align}
which can be done in a straight-forward way with the result:
\begin{align}
&\mathcal{S}_1(z)\mathcal{S}_1^\dagger(w)=\frac{1}{(z-w)^{\frac{3}{4}}}-\frac{1}{2}
\left(\psi_3\bar{\psi}_3+\psi_4\bar{\psi}_4+\psi_5\bar{\psi_5}\right)(z-w)^{\frac{1}{4}}+
\ldots\nonumber\\
&\mathcal{S}_2(z)\mathcal{S}_2^\dagger(w)=\frac{1}{(z-w)^{\frac{3}{4}}}-\frac{1}{2}
\left(\psi_3\bar{\psi}_3-\psi_4\bar{\psi}_4-\psi_5\bar{\psi_5}\right)(z-w)^{\frac{1}{4}}+
\ldots\nonumber\\
&\mathcal{S}_1(z)\mathcal{S}_2^\dagger(w)=(z-w)^{\frac{1}{4}}\psi_4\psi_5+\ldots\nonumber\\
&\mathcal{S}_2(z)\mathcal{S}_1^\dagger(w)=(z-w)^{\frac{1}{4}}\bar{\psi}_4\bar{\psi}_5+\ldots\nonumber\\
&\mathcal{S}_1(z)\mathcal{S}_1(w)=\psi_3\psi_4\psi_5(z-w)^{\frac{3}{4}}+\ldots\nonumber\\
&\mathcal{S}_2(z)\mathcal{S}_2(w)=\bar{\psi}_3\bar{\psi}_4\bar{\psi}_5(z-w)^{\frac{3}{4}}+\ldots\nonumber\\
&\mathcal{S}_1(z)\mathcal{S}_2(w)=\frac{\psi_3}{(z-w)^{\frac{1}{4}}}+\ldots\nonumber\\
&\mathcal{S}_1^\dagger(z)\mathcal{S}^\dagger_2(w)=\frac{\bar{\psi}_3}{(z-w)^{\frac{1}{4}}}+\ldots\nonumber
\end{align}
Upon taking suitable linear combinations of these expressions, one finds besides the $SU(2)$ currents 
(\ref{N4curs}), the following operators:
\begin{align}
&J_{T^2}=\psi_3\bar{\psi}_3,\\
&\Psi^{12}=\psi_3,\\
&\Psi^{12\dagger}=\bar{\psi}_3,\\
&\Psi^{11}=\Psi^{22}=0,
\end{align}
where 
$J_{T^2}$ is an additional $U(1)$ current $J_{T^2}$ associated to $T^2$. Thus, we now have an $SU(2)\times U(1)$ 
R-symmetry group.

The construction of the dimension-3/2 supercurrents is not as simple as in the previous case, 
because we have first to decouple the $U(1)$ from $SU(2)\times U(1)$. We start again with the $N=1$ internal 
supercurrent energy momentum tensor for the case of 6 compact  dimensions
\begin{align}
T_F=\psi_3\partial \bar{X}_3+\psi_4\partial \bar{X}_4+\psi_5
\partial \bar{X}_5+\bar{\psi}_3\partial X_3+\bar{\psi}_4\partial X_4+\bar{\psi}_5
\partial X_5.\nonumber
\end{align} 
Instead of splitting its expression directly into two parts as above, we follow \cite{Banks:1988yz} 
and make the following ansatz
\begin{align}
T_F=\sum_{q_3,q_4,q_5}e^{iq_3\phi_3}e^{iq_4\phi_4}e^{iq_5\phi_5}
\tilde{T}_F^{(q_3,q_4,q_5)},
\end{align}
where only the following combinations of charges are allowed
\begin{align}
\left(\begin{array}{c}q_3 \\ q_4 \\ q_5\end{array}\right)\in\left\{\left(
\begin{array}{c}\pm1 \\ 0 \\ 0\end{array}\right), \left(\begin{array}{c}0 \\ 
\pm1 \\ 0 \end{array}\right),\left(\begin{array}{c}0 \\ 0 \\ \pm 1\end{array}\right)\right\}.
\end{align}
It is now easy to see that
\begin{align}
&\tilde{T}_F^{(1,0,0)}=\partial \bar{X}_3, &&\tilde{T}_F^{(-1,0,0)}=\partial X_3,\\
&\tilde{T}_F^{(0,1,0)}=\partial \bar{X}_4, &&\tilde{T}_F^{(0,-1,0)}=\partial X_4,\\
&\tilde{T}_F^{(0,0,1)}=\partial \bar{X}_5, &&\tilde{T}_F^{(0,0,-1)}=\partial X_5.
\end{align}
At this point, it is possible, to separate the $U(1)$ part by setting
\begin{align}
&G^-_{T^2}=\bar{\psi}_3\partial X_3, &&G^{+}_{T^2}=\psi_3\partial\bar{X}_3,&&J_{T^2}\psi_3\bar{\psi}_3,
\end{align}
while the $K3$ operators are the same as before.
For further convenience, we also introduce the complex supercurrent of the $N=2$ subalgebra
\begin{align}
G^-=G^-_{T^2}+G^-_{K3}=\bar{\psi}_3\partial X_3+\bar{\psi}_4\partial X_4+\bar{\psi}_5
\partial X_5,
\end{align}
which is used in several computations in the text.
\section{Amplitudes with NS-NS Graviphotons}\label{AppendixLeftRightAsym}
In this appendix, we present two possibilities for insertions of graviphotons from the NS 
sector in the type II amplitudes discussed in Section \ref{Section4dscattering}.
For instance, the 6d helicity combination (\ref{exception1}) for the $R^4$ gives rise to the following setup 
of left and right-moving components:
\begin{center}
\begin{tabular}{|c|c|c||c|c||c||c|c||c|c||c||c|c|}\hline
\textbf{ins.} & \textbf{nr.} & \textbf{pos.} & \parbox{0.4cm}{\vspace{0.3cm}$\phi_1$
\vspace{-0.2cm}} & \parbox{0.4cm}{\vspace{0.2cm}$\phi_2$\vspace{0.2cm}} & \parbox{0.4cm}
{\vspace{0.2cm}$\phi_3$\vspace{0.2cm}} & \parbox{0.4cm}{\vspace{0.2cm}$\phi_4$\vspace{0.2cm}} 
& \parbox{0.4cm}{\vspace{0.2cm}$\phi_5$\vspace{0.2cm}} & \parbox{0.4cm}{\vspace{0.2cm}$
\tilde{\phi_1}$\vspace{0.2cm}} & \parbox{0.4cm}{\vspace{0.2cm}$\tilde{\phi_2}$\vspace{0.2cm}} 
& \parbox{0.4cm}{\vspace{0.2cm}$\tilde{\phi_3}$\vspace{0.2cm}} & \parbox{0.4cm}{\vspace{0.2cm}$
\tilde{\phi_4}$\vspace{0.2cm}} & \parbox{0.4cm}{\vspace{0.2cm}$\tilde{\phi_5}$\vspace{0.2cm}}\\
\hline\hline
graviph. & $1$ & $z_1$ & 0 & \parbox{0.7cm}{\vspace{0.2cm}$+1$\vspace{0.2cm}} & \parbox{0.7cm}
{\vspace{0.2cm}$+1$\vspace{0.2cm}} & 0 & 0 & \parbox{0.7cm}{\vspace{0.2cm}$+1$\vspace{0.2cm}} & 
\parbox{0.7cm}{\vspace{0.2cm}$+1$\vspace{0.2cm}} & 0 & 0 & 0 \\\hline 
 & $1$ & $z_2$ & 0 & \parbox{0.7cm}{\vspace{0.2cm}$+1$\vspace{0.2cm}} & \parbox{0.7cm}
{\vspace{0.2cm}$-1$\vspace{0.2cm}} & 0 & 0 & \parbox{0.7cm}{\vspace{0.2cm}$-1$\vspace{0.2cm}} & 
\parbox{0.7cm}{\vspace{0.2cm}$+1$\vspace{0.2cm}} & 0 & 0 & 0 \\\hline 
graviton & $1$ & $z_3$ & \parbox{0.7cm}{\vspace{0.2cm}$+1$\vspace{0.2cm}} & \parbox{0.7cm}
{\vspace{0.2cm}$-1$\vspace{0.2cm}} & 0 & 0  & 0 & \parbox{0.7cm}{\vspace{0.2cm}$+1$
\vspace{0.2cm}} & \parbox{0.7cm}{\vspace{0.2cm}$-1$\vspace{0.2cm}} & 0 & 0 & 0  \\\hline 
 & $1$ & $z_4$ & \parbox{0.7cm}{\vspace{0.2cm}$-1$\vspace{0.2cm}} & \parbox{0.7cm}
{\vspace{0.2cm}$-1$\vspace{0.2cm}} & 0 & 0 & \parbox{0.7cm}{\vspace{0.2cm}$-1$\vspace{0.2cm}} 
& \parbox{0.7cm}{\vspace{0.2cm}$-1$\vspace{0.2cm}} & 0 & 0 & 0 & 0 \\\hline 
graviph. & $g-1$ & $x_i$ & \parbox{0.7cm}{\vspace{0.2cm}$+\frac{1}{2}$\vspace{0.2cm}} & 
\parbox{0.7cm}{\vspace{0.2cm}$+\frac{1}{2}$\vspace{0.2cm}} & \parbox{0.7cm}{\vspace{0.2cm}
$+\frac{1}{2}$\vspace{0.2cm}} & \parbox{0.7cm}{\vspace{0.2cm}$+\frac{1}{2}$\vspace{0.2cm}} 
& \parbox{0.7cm}{\vspace{0.2cm}$+\frac{1}{2}$\vspace{0.2cm}}  & \parbox{0.7cm}{\vspace{0.2cm}
$+\frac{1}{2}$\vspace{0.2cm}} & \parbox{0.7cm}{\vspace{0.2cm}$+\frac{1}{2}$\vspace{0.2cm}} & 
\parbox{0.7cm}{\vspace{0.2cm}$-\frac{1}{2}$\vspace{0.2cm}} & \parbox{0.7cm}{\vspace{0.2cm}
$-\frac{1}{2}$\vspace{0.2cm}} & \parbox{0.7cm}{\vspace{0.2cm}$-\frac{1}{2}$\vspace{0.2cm}}\\\hline 
 & $g-1$ & $y_i$ & \parbox{0.7cm}{\vspace{0.2cm}$-\frac{1}{2}$\vspace{0.2cm}} & 
\parbox{0.7cm}{\vspace{0.2cm}$-\frac{1}{2}$\vspace{0.2cm}} & \parbox{0.7cm}
{\vspace{0.2cm}$+\frac{1}{2}$\vspace{0.2cm}} & \parbox{0.7cm}{\vspace{0.2cm}
$+\frac{1}{2}$\vspace{0.2cm}} & \parbox{0.7cm}{\vspace{0.2cm}$+\frac{1}{2}$\vspace{0.2cm}}
& \parbox{0.7cm}{\vspace{0.2cm}$-\frac{1}{2}$\vspace{0.2cm}} & \parbox{0.7cm}
{\vspace{0.2cm}$-\frac{1}{2}$\vspace{0.2cm}} & \parbox{0.7cm}{\vspace{0.2cm}
$-\frac{1}{2}$\vspace{0.2cm}} & \parbox{0.7cm}{\vspace{0.2cm}$-\frac{1}{2}$
\vspace{0.2cm}} & \parbox{0.7cm}{\vspace{0.2cm}$-\frac{1}{2}$\vspace{0.2cm}} \\\hline 
PCO & $g+1$ & $\{s_3\}$ & 0 & 0 &\parbox{0.7cm}{\vspace{0.2cm}$-1$
\vspace{0.2cm}} & 0 & 0 & 0 & 0 &\parbox{0.7cm}{\vspace{0.2cm}$+1$\vspace{0.2cm}} & 0 & 0\\
\hline 
 & $g-1$ & $\{s_4\}$ & 0 & 0 & 0 &\parbox{0.7cm}{\vspace{0.2cm}$-1$
\vspace{0.2cm}} & 0 & 0 & 0 & 0 &\parbox{0.7cm}{\vspace{0.2cm}$+1$\vspace{0.2cm}} & 0 \\\hline 
 & $g-1$ & $\{s_5\}$ & 0 & 0 & 0 & 0 &\parbox{0.7cm}{\vspace{0.2cm}$-1$
\vspace{0.2cm}} & 0 & 0 & 0 & 0 &\parbox{0.7cm}{\vspace{0.2cm}$+1$\vspace{0.2cm}}\\\hline 
\end{tabular}
\end{center}
${}$\\[10pt]
where the graviphotons at $z_1$ and $z_2$ are Kaluza-Klein graviphotons and the 
graviton at $z_4$ ($z_3$) a (anti-)self-dual graviton from the Weyl multiplet.

Similarly, the helicity setup of (\ref{exception2}) corresponds to the insertion of 4 KK graviphotons at the 
positions $z_i$:
\begin{center}
\begin{tabular}{|c|c|c||c|c||c||c|c||c|c||c||c|c|}\hline
\textbf{ins.} & \textbf{nr.} & \textbf{pos.} & \parbox{0.4cm}{\vspace{0.3cm}$\phi_1$
\vspace{-0.2cm}} & \parbox{0.4cm}{\vspace{0.2cm}$\phi_2$\vspace{0.2cm}} & 
\parbox{0.4cm}{\vspace{0.2cm}$\phi_3$\vspace{0.2cm}} & \parbox{0.4cm}{\vspace{0.2cm}$
\phi_4$\vspace{0.2cm}} & \parbox{0.4cm}{\vspace{0.2cm}$\phi_5$\vspace{0.2cm}} & 
\parbox{0.4cm}{\vspace{0.2cm}$\tilde{\phi_1}$\vspace{0.2cm}} & \parbox{0.4cm}
{\vspace{0.2cm}$\tilde{\phi_2}$\vspace{0.2cm}} & \parbox{0.4cm}{\vspace{0.2cm}
$\tilde{\phi_3}$\vspace{0.2cm}} & \parbox{0.4cm}{\vspace{0.3cm}$\tilde{\phi_4}$
\vspace{-0.1cm}} & \parbox{0.4cm}{\vspace{0.2cm}$\tilde{\phi_5}$\vspace{0.2cm}}\\\hline\hline
graviph. & $1$ & $z_1$ & \parbox{0.7cm}{\vspace{0.2cm}$+1$\vspace{0.2cm}} & 
\parbox{0.7cm}{\vspace{0.2cm}$+1$\vspace{0.2cm}} & 0 & 0 & 0 & 0 & \parbox{0.7cm}
{\vspace{0.2cm}$+1$\vspace{0.2cm}} & \parbox{0.7cm}{\vspace{0.2cm}$+1$
\vspace{0.2cm}} & 0 & 0 \\\hline 
 & $1$ & $z_2$ & \parbox{0.7cm}{\vspace{0.2cm}$+1$\vspace{0.2cm}} & \parbox{0.7cm}
{\vspace{0.2cm}$-1$\vspace{0.2cm}} & 0 & 0 & 0 & 0 & \parbox{0.7cm}{\vspace{0.2cm}$-1$
\vspace{0.2cm}} & \parbox{0.7cm}{\vspace{0.2cm}$+1$\vspace{0.2cm}} & 0 & 0 \\\hline 
 & $1$ & $z_3$ & \parbox{0.7cm}{\vspace{0.2cm}$-1$\vspace{0.2cm}} & \parbox{0.7cm}
{\vspace{0.2cm}$+1$\vspace{0.2cm}} & 0 & 0  & 0 & 0 & \parbox{0.7cm}{\vspace{0.2cm}$+1$
\vspace{0.2cm}} & \parbox{0.7cm}{\vspace{0.2cm}$-1$\vspace{0.2cm}} & 0 & 0  \\\hline 
 & $1$ & $z_4$ & \parbox{0.7cm}{\vspace{0.2cm}$-1$\vspace{0.2cm}} & \parbox{0.7cm}
{\vspace{0.2cm}$-1$\vspace{0.2cm}} & 0 & 0 & 0 & \parbox{0.7cm}{\vspace{0.2cm}$-1$
\vspace{0.2cm}} & \parbox{0.7cm}{\vspace{0.2cm}$-1$\vspace{0.2cm}} & 0 & 0 & 0 \\\hline 
graviph. & $g-1$ & $x_i$ & \parbox{0.7cm}{\vspace{0.2cm}$+\frac{1}{2}$\vspace{0.2cm}} & 
\parbox{0.7cm}{\vspace{0.2cm}$+\frac{1}{2}$\vspace{0.2cm}} & \parbox{0.7cm}{\vspace{0.2cm}
$+\frac{1}{2}$\vspace{0.2cm}} & \parbox{0.7cm}{\vspace{0.2cm}$+\frac{1}{2}$\vspace{0.2cm}} 
& \parbox{0.7cm}{\vspace{0.2cm}$+\frac{1}{2}$\vspace{0.2cm}}  & \parbox{0.7cm}{\vspace{0.2cm}
$+\frac{1}{2}$\vspace{0.2cm}} & \parbox{0.7cm}{\vspace{0.2cm}$+\frac{1}{2}$\vspace{0.2cm}} 
& \parbox{0.7cm}{\vspace{0.2cm}$-\frac{1}{2}$\vspace{0.2cm}} & \parbox{0.7cm}
{\vspace{0.2cm}$-\frac{1}{2}$\vspace{0.2cm}} & \parbox{0.7cm}{\vspace{0.2cm}
$-\frac{1}{2}$\vspace{0.2cm}}\\\hline 
 & $g-1$ & $y_i$ & \parbox{0.7cm}{\vspace{0.2cm}$-\frac{1}{2}$\vspace{0.2cm}} & 
\parbox{0.7cm}{\vspace{0.2cm}$-\frac{1}{2}$\vspace{0.2cm}} & \parbox{0.7cm}
{\vspace{0.2cm}$+\frac{1}{2}$\vspace{0.2cm}} & \parbox{0.7cm}{\vspace{0.2cm}
$+\frac{1}{2}$\vspace{0.2cm}} & \parbox{0.7cm}{\vspace{0.2cm}$+\frac{1}{2}$
\vspace{0.2cm}}& \parbox{0.7cm}{\vspace{0.2cm}$-\frac{1}{2}$\vspace{0.2cm}} 
& \parbox{0.7cm}{\vspace{0.2cm}$-\frac{1}{2}$\vspace{0.2cm}} & \parbox{0.7cm}
{\vspace{0.2cm}$-\frac{1}{2}$\vspace{0.2cm}} & \parbox{0.7cm}{\vspace{0.2cm}
$-\frac{1}{2}$\vspace{0.2cm}} & \parbox{0.7cm}{\vspace{0.2cm}$-\frac{1}{2}$
\vspace{0.2cm}} \\\hline 
PCO & $g+1$ & $\{s_3\}$ & 0 & 0 &\parbox{0.7cm}{\vspace{0.2cm}$-1$
\vspace{0.2cm}} & 0 & 0 & 0 & 0 &\parbox{0.7cm}{\vspace{0.2cm}$+1$
\vspace{0.2cm}} & 0 & 0\\\hline 
 & $g-1$ & $\{s_4\}$ & 0 & 0 & 0 &\parbox{0.7cm}{\vspace{0.2cm}$-1$
\vspace{0.2cm}} & 0 & 0 & 0 & 0 &\parbox{0.7cm}{\vspace{0.2cm}$+1$\vspace{0.2cm}} 
& 0 \\\hline 
 & $g-1$ & $\{s_5\}$ & 0 & 0 & 0 & 0 &\parbox{0.7cm}{\vspace{0.2cm}$-1$
\vspace{0.2cm}} & 0 & 0 & 0 & 0 &\parbox{0.7cm}{\vspace{0.2cm}$+1$\vspace{0.2cm}}\\\hline 
\end{tabular}
\end{center}
${}$\\[10pt] 
In fact, one can show that these amplitudes lead also to the same result (\ref{result1}). 


\end{document}